\input harvmac.tex

\input epsf.tex


\def\figin{\epsfcheck\figin}\def\figins{\epsfcheck\figins}
\def\epsfcheck{\ifx\epsfbox\UnDeFiNeD
\message{(NO epsf.tex, FIGURES WILL BE IGNORED)}
\gdef\figin##1{\vskip2in}\gdef\figins##1{\hskip.5in}
\else\message{(FIGURES WILL BE INCLUDED)}%
\gdef\figin##1{##1}\gdef\figins##1{##1}\fi}
\def\DefWarn#1{}
\def\frac#1#2{{#1\over#2}}
\def\figinsert{\goodbreak\midinsert}
\def\ifig#1#2#3{\DefWarn#1\xdef#1{fig.~\the\figno}
\writedef{#1\leftbracket fig.\noexpand~\the\figno}%
\figinsert\figin{\centerline{#3}}\medskip\centerline{\vbox{\baselineskip12pt
\advance\hsize by -1truein\noindent\footnotefont{\bf
Fig.~\the\figno:} #2}}
\bigskip\endinsert\global\advance\figno by1}



\lref\ArkaniHamedSI{
  N.~Arkani-Hamed, F.~Cachazo, C.~Cheung and J.~Kaplan,
  arXiv:0903.2110 [hep-th].
}
\lref\BrandhuberDA{
  A.~Brandhuber, P.~Heslop, V.~V.~Khoze and G.~Travaglini,
  arXiv:0910.4898 [hep-th].
}
\lref\KoberleSG{
  R.~Koberle and J.~A.~Swieca,
  Phys.\ Lett.\  B {\bf 86}, 209 (1979).
}

\lref\dornrecent{
  H.~Dorn,
  arXiv:0910.0934 [hep-th].
}

\lref\lastjevicki{
  A.~Jevicki and K.~Jin,
  arXiv:0911.1107 [hep-th].
}

\lref\hodgesspinor{
  A.~Hodges,
  arXiv:0905.1473 [hep-th].
}

\lref\AldayHR{
  L.~F.~Alday and J.~M.~Maldacena,
  JHEP {\bf 0706}, 064 (2007)
  [arXiv:0705.0303 [hep-th]].
}
\lref\DrummondBM{
  J.~M.~Drummond, J.~Henn, G.~P.~Korchemsky and E.~Sokatchev,
  Phys.\ Lett.\  B {\bf 662}, 456 (2008)
  [arXiv:0712.4138 [hep-th]].
}

\lref\AldayYW{
  L.~F.~Alday and R.~Roiban,
  Phys.\ Rept.\  {\bf 468}, 153 (2008)
  [arXiv:0807.1889 [hep-th]].
}

\lref\DrummondAUA{
  J.~M.~Drummond, G.~P.~Korchemsky and E.~Sokatchev,
  Nucl.\ Phys.\  B {\bf 795}, 385 (2008)
  [arXiv:0707.0243 [hep-th]].
}

\lref\BrandhuberYX{
  A.~Brandhuber, P.~Heslop and G.~Travaglini,
  Nucl.\ Phys.\  B {\bf 794}, 231 (2008)
  [arXiv:0707.1153 [hep-th]].
}

\lref\AldayMF{
  L.~F.~Alday and J.~M.~Maldacena,
  JHEP {\bf 0711}, 019 (2007)
  [arXiv:0708.0672 [hep-th]].
}

\lref\BonoraAY{
  L.~Bonora, C.~P.~Constantinidis, L.~A.~Ferreira and E.~E.~Leite,
  J.\ Phys.\ A  {\bf 36}, 7193 (2003)
  [arXiv:hep-th/0208175].
}

\lref\DrummondAU{
  J.~M.~Drummond, J.~Henn, G.~P.~Korchemsky and E.~Sokatchev,
  Nucl.\ Phys.\  B {\bf 826}, 337 (2010)
  [arXiv:0712.1223 [hep-th]].
}
\lref\AldayYN{
  L.~F.~Alday and J.~Maldacena,
  arXiv:0904.0663 [hep-th].
}
\lref\BDS{
  Z.~Bern, L.~J.~Dixon and V.~A.~Smirnov,
  Phys.\ Rev.\  D {\bf 72}, 085001 (2005)
  [arXiv:hep-th/0505205].
}

\lref\GromovBC{
  N.~Gromov, V.~Kazakov, A.~Kozak and P.~Vieira,
  arXiv:0902.4458 [hep-th].
}
\lref\ArutyunovUX{
  G.~Arutyunov and S.~Frolov,
  JHEP {\bf 0911}, 019 (2009)
  [arXiv:0907.2647 [hep-th]].
}
\lref\GrigorievJQ{
  M.~Grigoriev and A.~A.~Tseytlin,
  Int.\ J.\ Mod.\ Phys.\  A {\bf 23}, 2107 (2008)
  [arXiv:0806.2623 [hep-th]].
}
\lref\JevickiUZ{
  A.~Jevicki and K.~Jin,
  JHEP {\bf 0906}, 064 (2009)
  [arXiv:0903.3389 [hep-th]].
}

\lref\McCoyCD{
  B.~M.~McCoy, C.~A.~Tracy and T.~T.~Wu,
  J.\ Math.\ Phys.\  {\bf 18}, 1058 (1977).
}

\lref\ZamolodchikovUW{
  A.~B.~Zamolodchikov,
  Nucl.\ Phys.\  B {\bf 432}, 427 (1994)
  [arXiv:hep-th/9409108].
}

\lref\brandhuber{
  A.~Brandhuber, P.~Heslop and G.~Travaglini,
  Nucl.\ Phys.\  B {\bf 794}, 231 (2008)
  [arXiv:0707.1153 [hep-th]].
}

\lref\DeVegaXC{
  H.~J.~De Vega and N.~G.~Sanchez,
  Phys.\ Rev.\  D {\bf 47}, 3394 (1993).
}

\lref\GMNone{
  D.~Gaiotto, G.~W.~Moore and A.~Neitzke,
  arXiv:0807.4723 [hep-th].
}

\lref\GMNtwo{
   D.~Gaiotto, G.~W.~Moore and A.~Neitzke,
  arXiv:0907.3987 [hep-th].
}

\lref\HitchinVP{
  N.~J.~Hitchin,
  Proc.\ Lond.\ Math.\ Soc.\  {\bf 55}, 59 (1987).
}

\lref\DrummondAU{
  J.~M.~Drummond, J.~Henn, G.~P.~Korchemsky and E.~Sokatchev,
  arXiv:0712.1223 [hep-th].
}

\lref\DeVegaXC{
  H.~J.~De Vega and N.~G.~Sanchez,
  Phys.\ Rev.\  D {\bf 47}, 3394 (1993).
}

\lref\JevickiAA{
  A.~Jevicki, K.~Jin, C.~Kalousios and A.~Volovich,
  JHEP {\bf 0803}, 032 (2008)
  [arXiv:0712.1193 [hep-th]].
}

\lref\MaldacenaIM{
  J.~M.~Maldacena,
  Phys.\ Rev.\ Lett.\  {\bf 80}, 4859 (1998)
  [arXiv:hep-th/9803002].
}

\lref\ReyIK{
  S.~J.~Rey and J.~T.~Yee,
  Eur.\ Phys.\ J.\  C {\bf 22}, 379 (2001)
  [arXiv:hep-th/9803001].
}

\lref\AharonyUG{
  O.~Aharony, O.~Bergman, D.~L.~Jafferis and J.~Maldacena,
  JHEP {\bf 0810}, 091 (2008)
  [arXiv:0806.1218 [hep-th]].
}

\lref\BerensteinIJ{
  D.~E.~Berenstein, R.~Corrado, W.~Fischler and J.~M.~Maldacena,
  Phys.\ Rev.\  D {\bf 59}, 105023 (1999)
  [arXiv:hep-th/9809188].
}

\lref\KorchemskayaJE{
  I.~A.~Korchemskaya and G.~P.~Korchemsky,
  Phys.\ Lett.\  B {\bf 287}, 169 (1992).
}

\lref\BassettoXD{
  A.~Bassetto, I.~A.~Korchemskaya, G.~P.~Korchemsky and G.~Nardelli,
  Nucl.\ Phys.\  B {\bf 408}, 62 (1993)
  [arXiv:hep-ph/9303314].
}

\lref\BernAP{
  Z.~Bern, L.~J.~Dixon, D.~A.~Kosower, R.~Roiban, M.~Spradlin, C.~Vergu and A.~Volovich,
  Phys.\ Rev.\  D {\bf 78}, 045007 (2008)
  [arXiv:0803.1465 [hep-th]].
}
\lref\DrummondAQ{
  J.~M.~Drummond, J.~Henn, G.~P.~Korchemsky and E.~Sokatchev,
  arXiv:0803.1466 [hep-th].
}
\lref\KazakovQF{
  V.~A.~Kazakov, A.~Marshakov, J.~A.~Minahan and K.~Zarembo,
  JHEP {\bf 0405}, 024 (2004)
  [arXiv:hep-th/0402207].
}
\lref\PohlmeyerNB{
  K.~Pohlmeyer,
  Commun.\ Math.\ Phys.\  {\bf 46}, 207 (1976).
}

\lref\CecottiRM{
  S.~Cecotti and C.~Vafa,
  Commun.\ Math.\ Phys.\  {\bf 158}, 569 (1993)
  [arXiv:hep-th/9211097].
}

\lref\BershadskyQY{
  M.~Bershadsky, C.~Vafa and V.~Sadov,
  Nucl.\ Phys.\  B {\bf 463}, 420 (1996)
  [arXiv:hep-th/9511222].
}

\lref\BeisertEZ{
  N.~Beisert, B.~Eden and M.~Staudacher,
  J.\ Stat.\ Mech.\  {\bf 0701}, P021 (2007)
  [arXiv:hep-th/0610251].
}

\lref\MiramontesWT{
  J.~L.~Miramontes,
  JHEP {\bf 0810}, 087 (2008)
  [arXiv:0808.3365 [hep-th]].
}

\lref\LuKB{
  H.~Lu, M.~J.~Perry, C.~N.~Pope and E.~Sezgin,
  arXiv:0812.2218 [hep-th].
}

\lref\FreyhultPZ{
  L.~Freyhult, A.~Rej and M.~Staudacher,
  J.\ Stat.\ Mech.\  {\bf 0807}, P07015 (2008)
  [arXiv:0712.2743 [hep-th]].
}

\lref\AnastasiouKN{
  C.~Anastasiou, A.~Brandhuber, P.~Heslop, V.~V.~Khoze, B.~Spence and G.~Travaglini,
  arXiv:0902.2245 [hep-th].
}

\lref\BernIZ{
  Z.~Bern, L.~J.~Dixon and V.~A.~Smirnov,
  Phys.\ Rev.\  D {\bf 72}, 085001 (2005)
  [arXiv:hep-th/0505205].
}

\lref\KomargodskiWA{
  Z.~Komargodski,
  JHEP {\bf 0805}, 019 (2008)
  [arXiv:0801.3274 [hep-th]].
}

\lref\AldayCG{
  L.~F.~Alday,
  Fortsch.\ Phys.\  {\bf 56}, 816 (2008)
  [arXiv:0804.0951 [hep-th]].
}

\lref\ZamolodchikovET{
  A.~B.~Zamolodchikov,
  Phys.\ Lett.\  B {\bf 253}, 391 (1991).
}

\lref\AldayHE{
  L.~F.~Alday and J.~Maldacena,
  JHEP {\bf 0711}, 068 (2007)
  [arXiv:0710.1060 [hep-th]].
}

\lref\BarbashovQZ{
  B.~M.~Barbashov, V.~V.~Nesterenko and A.~M.~Chervyakov,
  Commun.\ Math.\ Phys.\  {\bf 84}, 471 (1982).
}

\lref\Bonora{ FIX THIS }
\lref\KruczenskiGT{
  M.~Kruczenski,
  Phys.\ Rev.\ Lett.\  {\bf 93}, 161602 (2004)
  [arXiv:hep-th/0311203].
}

\lref\AldayGA{
  L.~F.~Alday and J.~Maldacena,
  arXiv:0903.4707 [hep-th].
}

\lref\douglasplateau{
Douglas, J.  ``Solution of the problem of Plateau". Trans. Amer. Math. Soc. {\bf 33} 263–321 (1931).
}

\lref\BeisertRY{
  N.~Beisert,
  Phys.\ Rept.\  {\bf 405}, 1 (2005)
  [arXiv:hep-th/0407277].
}

\lref\BenaWD{
  I.~Bena, J.~Polchinski and R.~Roiban,
  Phys.\ Rev.\  D {\bf 69}, 046002 (2004)
  [arXiv:hep-th/0305116].
}
\lref\MandalFS{
  G.~Mandal, N.~V.~Suryanarayana and S.~R.~Wadia,
  Phys.\ Lett.\  B {\bf 543}, 81 (2002)
  [arXiv:hep-th/0206103].
}

\lref\ArutyunovVX{
  G.~Arutyunov, S.~Frolov and M.~Staudacher,
  JHEP {\bf 0410}, 016 (2004)
  [arXiv:hep-th/0406256].
}

\lref\FendleyVE{
  P.~Fendley and K.~A.~Intriligator,
  Nucl.\ Phys.\  B {\bf 372}, 533 (1992)
  [arXiv:hep-th/9111014].
}

\lref\FendleyXN{
  P.~Fendley,
  Nucl.\ Phys.\  B {\bf 374}, 667 (1992)
  [arXiv:hep-th/9109021].
}

\lref\ZamolodchikovCF{
  A.~B.~Zamolodchikov,
  Nucl.\ Phys.\  B {\bf 342}, 695 (1990).
}

\lref\KirillovZZ{
  A.~N.~Kirillov and N.~Y.~Reshetikhin,
  J.\ Phys.\ A  {\bf 20}, 1565 (1987).
}


\lref\dtpoles{
  P.~Dorey and R.~Tateo,
  Nucl.\ Phys.\  B {\bf 482}, 639 (1996)
  [arXiv:hep-th/9607167].
}
\lref\zampoles{
  V.~V.~Bazhanov, S.~L.~Lukyanov and A.~B.~Zamolodchikov,
  Nucl.\ Phys.\  B {\bf 489}, 487 (1997)
  [arXiv:hep-th/9607099].
}

\lref\DoreyPT{
  P.~Dorey and R.~Tateo,
  J.\ Phys.\ A  {\bf 32}, L419 (1999)
  [arXiv:hep-th/9812211].
}
\lref\DoreyUK{
  P.~Dorey and R.~Tateo,
  Nucl.\ Phys.\  B {\bf 563}, 573 (1999)
  [Erratum-ibid.\  B {\bf 603}, 581 (2001)]
  [arXiv:hep-th/9906219].
}
\lref\DoreyZX{
  P.~Dorey, C.~Dunning and R.~Tateo,
  J.\ Phys.\ A  {\bf 40}, R205 (2007)
  [arXiv:hep-th/0703066].
}

\lref\BombardelliNS{
  D.~Bombardelli, D.~Fioravanti and R.~Tateo,
  J.\ Phys.\ A  {\bf 42}, 375401 (2009)
  [arXiv:0902.3930 [hep-th]].
}

\lref\DelDucaAU{
  V.~Del Duca, C.~Duhr and V.~A.~Smirnov,
  arXiv:0911.5332 [hep-ph].
}

\lref\MartinsHW{
  M.~J.~Martins,
  Phys.\ Rev.\ Lett.\  {\bf 67}, 419 (1991).
}

\Title{\vbox{\baselineskip12pt \hbox{} \hbox{
} }} {\vbox{\centerline{
  }
\centerline{Thermodynamic Bubble Ansatz} }}
\bigskip
\centerline{Luis F. Alday, Davide Gaiotto and Juan Maldacena}
\bigskip
\centerline{ \it  School of Natural Sciences, Institute for
Advanced Study} \centerline{\it Princeton, NJ 08540, USA}

\vskip .3in \noindent

Motivated by the computation of scattering amplitudes at strong coupling,
we consider minimal area surfaces in $AdS_5$ which end on a null polygonal contour
at the boundary.
We map the classical problem of finding the surface into an  $SU(4)$ Hitchin system.
The polygon with six edges is the first non-trivial example.
For this case, we write an integral equation which determines the area
as a function of the shape of the polygon.
  The equations are identical to those  of the Thermodynamics
Bethe Ansatz. Moreover, the area is given by the
  free energy of this TBA system. The high temperature limit of the TBA system can be
  exactly solved.
  It leads to an explicit expression  for a special class of hexagonal contours.


 \Date{ }


\newsec{Introduction}

In this paper we consider a geometrical problem involving minimal
surfaces in $AdS$. We prescribe a  polygonal contour on the $AdS$
boundary and we consider a minimal surface which ends on this
contour. We devise a method for computing the area of the surface as
a function of the boundary contour.

It is a generalization of the problem of finding the shape of soap
bubbles, or Plateau problem \douglasplateau , to $AdS$ space. In the 19th century the
study of surface tension was interesting for what it could reveal
about interfaces and the size of atoms. For us,  the study of this
 $AdS$ ``soap bubbles'' is interesting  for what it could eventually teach us about
their constituents,     gluons,  and their scattering for    all values of the coupling.
While this is our motivation, the computations in this paper will be restricted
to classical geometry (or strong coupling).

\subsec{Gauge theory motivation }

Our main motivation for focusing on  this problem comes from the
study of Wilson loops or scattering amplitudes in ${\cal N}=4 $
super Yang Mills theory. These observables are interesting in
general gauge theories and contain a wealth of dynamical
information. Since planar ${\cal N}=4$ super Yang Mills theory is
integrable \refs{\BeisertRY,\BenaWD}, one hopes to be able to compute them at all values of
the coupling. At strong coupling they  can be computed in terms of
classical strings in $AdS_5$ \AldayHR .  The classical equations of motion for
a string say that the area should be extremized. Thus, we end up
studying minimal surfaces in $AdS_5$ space.
 This problem becomes tractable because it is classically integrable \refs{\MandalFS,\BenaWD}.
 In fact, we will use its classical
 integrability to find  equations that determine the area.

 We hope that our classical  analysis will be a useful starting point
 for solving the problem at all values of the coupling.
 In fact, the final classical problem has a structure that looks like  the $Y$ system for
 the Thermodynamic Bethe Ansatz.

 For the computation of   operator dimensions   the solutions to the
 classical problem \refs{\ArutyunovVX,\KazakovQF} were very useful for eventually determining the full
 quantum  solution
  \refs{\BeisertEZ,\BombardelliNS,\GromovBC,\ArutyunovUX}.
   We hope that the same will be true for the computation of scattering
  amplitudes.

\ifig\polygon{We have a polygonal loop with null sides on the boundary of $AdS$. This
polygon is   denoted here by the jagged lines.
We consider a minimal surface in the bulk that
is  ending on the polygon.}
{\epsfxsize1.5in\epsfbox{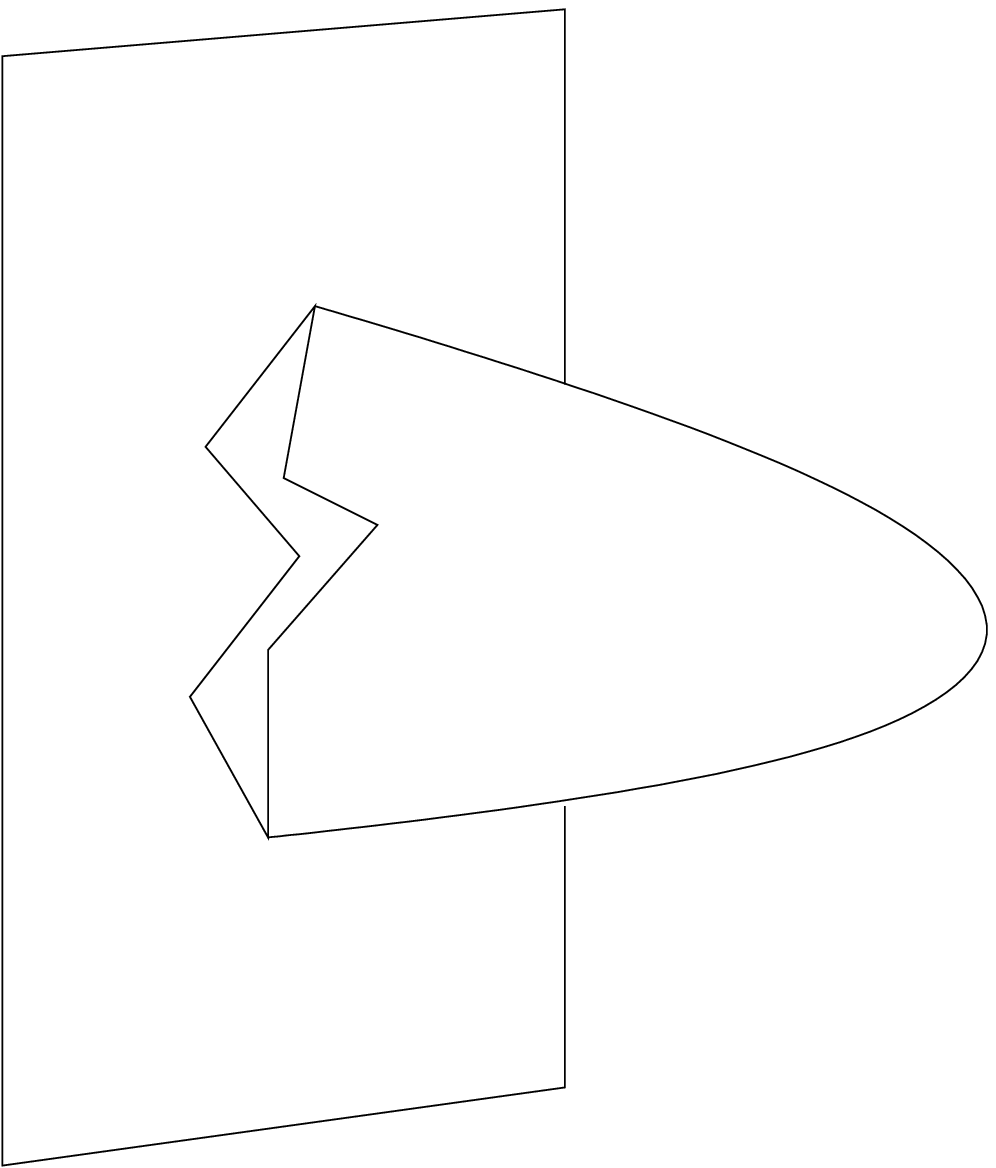}}

 We consider surfaces that  end at the AdS boundary on a special contour which is made out of null
 segments, see  \polygon . We are given a sequence of $n$ points on the boundary, such that consecutive points are at a null
 distance, $x_{i,i+1}^2 =0$. This defines a polygon with null sides. Conformal transformations act on this polygon and
 can change the positions of the points. One can define conformal invariant cross ratios which do not change
 under conformal transformations. These characterize the polygons up to conformal transformations.
 We will choose the points in such a way that the distance between all points which are not consecutive is
 spacelike. In this case, the minimal surface that ends on the contour is spacelike. It is a spacelike surface
 embedded in the Lorentzian signature $AdS_5$ space.   Its area is infinite.
 This infinity is due  to the  IR divergencies of amplitudes or the UV divergencies of Wilson loops. One can
 define a renormalized area which is finite and conformal invariant.
  Our goal is to compute the renormalized area as a function of
  the cross ratios. In the context of scattering amplitudes, this quantity is called ``remainder function''.
   See \refs{\BernAP,\DrummondBM} for a perturbative, two loop evaluation of this function for six particles.

 \subsec{Outline of the method}

 The classical action for a string in $AdS$ is conformal invariant. Thus, it contains a conserved
 holomorphic stress tensor $T(z) $, $\bar T(\bar z)$ on the worldhseet.
 The Virasoro constraints are setting these to zero.
In this case,  we also have   spin four holomorphic quantities
$P(z)$, $\bar P(\bar z)$.
 These are holomorphic and non-vanishing after we impose the equations of motion and the Virasoro constraints.
 These are additional local conserved quantities associated to the integrability of the theory.
 These currents play  a crucial role in our method for solving the problem.
 Via worldsheet
 conformal transformations $ z \to w = f(z)$ we can locally set them to one $P(z) \to P(w) =1$.
 This however, cannot be done globally without introducing some singularities.
 In our case the worlsheet can be parametrized by the whole complex plane and $P(z)$  is  a polynomial.
 The degree of the polynomial determines the number of sides of the polygon. The coefficients of the
 polynomial (together with some  extra parameters we will discuss later)
 encode the shape of the polygon .

  The problem is integrable and it has a one parameter family of flat connections parametrized by a
  spectral parameter $\zeta$. This family is encoding the infinite number of conserved non-local
  charges of the theory.

  The flat connection goes to a constant at infinity and the solutions of the linear problem $(d +{\cal A})\psi =0$
  grow at infinity. The solutions of the linear problem, for $\zeta =1$,
   encode the $AdS$ coordinates and the fact that they
  go to infinity as $z \to \infty $ implies that they go to the boundary of $AdS$. They go to different points
  on the boundary of $AdS$ in different angular sectors of the $z$ plane. These sectors are called Stokes sectors.
  In any problem which has Stokes sectors we can define cross ratios, which end up being the same as the
  spacetime cross ratios when $\zeta =1$.

   The advantage of introducing the spectral parameter $\zeta \not =1$ is that the problem of determining the
   cross ratios as a function of the polynomial coefficients simplifies when $\zeta \to 0 $ or when $\zeta \to \infty $.
   In this limit we can determine the   form of the cross ratios as a function of the coefficients of the
    polynomial $P(z)$.
    In addition,
     we expect that the cross ratios are analytic functions of $ \zeta $ away from $\zeta = 0 , \infty$.
    The cross ratios as a function of $\zeta$ display a sort of
     Stokes phenomenon as $\zeta \to 0 , \infty $.
    One can define certain cross ratios which are analytic within certain angular sectors in the $\zeta-$plane and have simple asymptotics for
    $\zeta \to 0 , \infty$. When we change angular sectors these cross ratios have discontinuities which are simple
    functions of the cross ratios themselves. Patching together the different angular sectors one defines a Riemann Hilbert problem
    which consists in finding the corresponding analytic functions in each sector with the specified jumps between sectors.
    This problem can also be rewritten as an integral equation for the cross ratios as a function of $\zeta$.

    Intriguingly, this integral equation looks very much like a Thermodynamic Bethe Ansatz equation (TBA), or
    Y-system, for some particles whose physical significance is not clear to us.
    Moreover, the renormalized area can be expressed in terms of the cross ratios via a formula which is
    the same as the formula for the free energy of the Thermodynamic Bethe Ansatz.

    This paper is organized as follows. In section two we recall some general properties of strings in $AdS_5$.
    We review the Pohlmeyer reduction and the associated Hitchin system and flat connection. We explain how the
    linear problem for the flat connection lets us recover the surface.
    In section three we explain how to derive the TBA equation from the Hitchin system. First we use a WKB
    approximation to understand the behavior of the cross ratios as a function of $\zeta$ near $\zeta =0, \infty$.
    With this information, we write down an integral equation. Finally we discuss some general aspects of the
    kinematics of the six-sided polygon.

    The connection between classical differential equations and the TBA system for conformal
    quantum integrable theories
    was studied before in \refs{\DoreyPT,\DoreyUK,\DoreyZX}. In fact, our results can be viewed as the extension
    of those results for massive integrable theories.

    In section four we show that the area can be computed from the free energy of the TBA equations.
    We also discuss a family of exact solutions of the TBA system which corresponds to the large temperature
    limit, in the TBA language.

    Finally, we present our conclusions.

\newsec{Classical Strings in $AdS_5$}

\subsec{Pohlmeyer reduction and integrability} Classical strings in
$AdS$ spaces can be described by a reduced model which depends only
on physical degrees of freedom
\refs{\PohlmeyerNB,\DeVegaXC,\JevickiAA,\GrigorievJQ,\MiramontesWT,\JevickiUZ,\dornrecent}.
 We use embedding coordinates $
Y$ with
 $  Y^2 =-1$.
 The conformal gauge equations of motion
and the Virasoro constraints are
\eqn\eom{\partial \bar \partial {Y}-(\partial
{Y}.\bar{\partial} {Y}){Y}=0 ~,~~~~~~
  \partial {Y}.\partial {Y}=\bar \partial {Y}.\bar \partial {Y}=0 }
where we parametrize the (spacelike) world-sheet in terms of complex
variables $z$ and $\bar{z}$. We can form an orthonormal basis
$q=(q_0,\cdots, q_5)$ with $q_0 = Y$, $q_4 = e^{-\alpha/2} \partial
Y$, $q_5 = e^{-\alpha/2} \bar \partial Y$ and  $q_I = B_I$ where $I$
runs from $1$ to $3$.
\eqn\orth{e^{\alpha } \equiv \partial Y  . \bar \partial Y ,~~~~B_I
. Y=B_I .
\partial Y=B_I . \bar \partial Y=0,~~~B_I . B_J =h_{IJ}}
The other inner products are fixed by the Virasoro constraints
and the fact that $Y.Y=-1$. The real vectors $B_I$'s span the
normal space to the world-sheet inside $AdS_5$. We pick them to
be orthonormal. The signature of the inner product $h_{IJ}$
is  $(-1,1,1)$ for standard Wilson loops in the usual $(3,1)$
signature, $(-1,-1,1)$ for Wilson loops in $(2,2)$ signature
and $(-1,-1,-1)$ for Wilson loops in $(1,3)$ signature\foot{
These changes of signature  correspond to the standard
procedure of analytic continuation in   the external momenta.
In the field theory language, we are changing the external
momenta but we are not changing the usual Feynman prescription
for the internal momenta, in contrast to what was done in
\ArkaniHamedSI . In particular, the IR divergencies have the
same form in all signatures.}. There is a residual gauge
freedom, either $SO(1,2)$, $SO(2,1)$ or $SO(3)$ that rotate the
$B's$, which we don't fix, see \dornrecent\ for gauge fixed
equations.

$B_3$ can be set to a constant, reducing the problem to $AdS_4$.
Setting both
  $B_3$ and $B_2$ to a constant reduces the problem to $AdS_3$. Given these
vectors we can further define
\eqn\uv{v_I=B_I . \partial^2 Y,~~~\bar{v}_I= B_I . \bar
\partial^2 Y}
Note that with our conventions, as the $B's$ are real, $v$ and $\bar
v$ are complex conjugate. We can define an $SO(2,4)$ (or $SO(3,3)$,
or $SO(4,2)$, depending on the signature of spacetime) flat
connection on the worldsheet describing the parallel transport of
$q$.
\eqn\linprob{\partial q=-{\cal A}_z q,~~~~~~~ \bar \partial q=-{\cal
A}_{\bar z} q}
Explicitly,
\eqn\conne{{\cal A}_z=- \left(\matrix{0&0&0&0& e^{\alpha/2} &0 \cr 0
& 0&d_{1}^{~2}&d_{1}^{~3}&0&-v_1 e^{-\alpha/2} \cr 0
&d_{2}^{~1}&0&d_2^{~3}&0& - v_2 e^{-\alpha/2} \cr 0 &
d_3^{~1}&d_3^{~2}&0&0& - v_3 e^{-\alpha/2}  \cr 0 &v^1 e^{-\alpha/2}
& v^2 e^{-\alpha/2} &v^3 e^{-\alpha/2} &\partial \alpha/2 &0 \cr
e^{\alpha/2} &0&0&0&0&- \partial \alpha/2 } \right)}
Here $d_{IJ}=\partial B_I . B_J$ and we lower and raise indices with
$h$. ${\cal A}_{\bar z}$ is given by a similar expression obtained
by interchanging $v \rightarrow \bar v$, $d_{IJ} \rightarrow \bar
d_{IJ}$, permuting the last two rows and permuting the last two
columns. The flatness condition for the connection ${\cal A}$ is
equivalent to the following equations \dornrecent
\eqn\flatness{\eqalign{
\partial \bar \partial \alpha-e^{\alpha}-e^{-\alpha} v^I \bar v_I=0, \cr
\partial \bar v_I-d_I^{~J} \bar v_J=0,~~~~~\bar \partial v_I -\bar d_I^{~J}v_J=0, \cr
e^{-\alpha}(\bar v_I v^J-v_I \bar v^J)=f_I^{~J},~~~~~f \equiv
\partial \bar d-\bar \partial d+[d,\bar d] }}
$f$ is the field strength of the gauge field $d$. These follow from
the original equations \eom . Conversely,
  one can first solve these equations, \flatness ,  and determine the connection ${\cal A}$.
 We can then find six independent solutions of the linear problem \linprob . We can orthonormalize these solutions at some
point. Then these six solutions will remain orthornomal
throughout the worldsheet. We have  now  an orthogonal matrix
$q_I^A$, where $A$ labels the solution number and it is the
spacetime index that all vectors $q_I$ carry. The first row of
the matrix,
  $q_0 = q_0^A = Y^A$, solves the original
equations of motion and Virasoro constraints.
The area is then given by
\eqn\areaws{
A  = 2 \int d^2 z e^{\alpha }
}
This is the area in units where the radius of $AdS$ is set to one\foot{
We have that $
\int d^2 z = \int dx dy $, with $z = x + i y $.}.
Then the amplitude or Wilson loop is given by \AldayHR\
\eqn\amplarea{
{\rm Amplitude } \sim \langle W \rangle \sim e^{ - { R^2 \over 2 \pi \alpha'} A }
~; ~~~~~~~{ R^2 \over 2 \pi \alpha' } = { \sqrt{\lambda}
\over 2 \pi }
}
where we also have given the expression for the radius in terms of the 't Hooft coupling for
${\cal N}=4$ SYM. For other theories we have other relations between the radius and the microscopic
parameters, but our result will continue to be the leading approximation when   $R^2/\alpha' \gg 1$.

An immediate consequence of \flatness\   is that
\eqn\holom{\eqalign{ \bar \partial(v^I v_I)=0 \rightarrow v^I v_I=
P(z) }}
with $P(z)$ some holomorphic function. Since $q$ is a complete
basis, we can write $\partial^2 Y$ in terms its elements
\eqn\derx{\partial^2 Y=\partial \alpha \partial Y+v^I B_I
~~\rightarrow~~ P(z)=\partial^2 Y. \partial^2 Y,~~\bar P(\bar
z)=\bar \partial^2 Y. \bar \partial^2 Y}
which gives a simple expression for this holomorphic function, and
his anti-holomorphic companion, in terms of space-time quantities.
Actually, one can verify directly from \eom\ that $P$ defined as $P
= \partial^2 Y . \partial^2 Y$ is holomorphic. We can locally define
a new variable $w$ through
\eqn\wdefi{
 d w = P(z)^{1/4} dz
 }
  This
can be viewed as a worldsheet conformal transformation $z \to
w(z)$. In the new coordinates $\tilde P(w) = \left( { \partial
z \over \partial w } \right)^4 P(z) =1$. This change of
coordinates is locally well defined. However, at points where
$P$ has a zero, we have a branch cut in the $w$ plane. We can
view the $w$ plane as a sophisticated light-cone gauge choice.
This $w$ ``plane'' is useful to think about the asymptotic
structure of the solution.

The integrability of the problem becomes manifest if we decompose
${\cal A}$ in two parts, ${\cal A} = A+\Phi$, where $A$ rotates $(Y,
B_I)$ and $(\partial Y, \bar \partial Y)$ separately among
themselves, and $\Phi$ mixes them.
\eqn\connesplit{ \eqalign{ A_z= & -\left(\matrix{0&0&0&0&& \cr 0 &
0&d_{1}^{~2}&d_{1}^{~3}&& \cr 0 &d_{2}^{~1}&0&d_2^{~3}&& \cr 0 &
d_3^{~1}&d_3^{~2}&0&& \cr &&&&\partial \alpha/2 &0 \cr
&&&&0&-\partial \alpha/2 } \right)~~~~~~~~ \cr \Phi_z=&-
\left(\matrix{&&&& e^{\alpha \over 2} &0 \cr & &&&0&-v_1 e^{-{\alpha
\over 2} } \cr &&&&0& -v_2 e^{-{\alpha \over 2}} \cr &&&&0& -v_3
e^{-{\alpha \over 2} } \cr 0 &v^1e^{- {\alpha \over 2}} &v^2 e^{-{
\alpha \over 2 }} &v^3 e^{-{ \alpha \over 2 } } && \cr e^{\alpha
\over 2} &0&0&0&& } \right)}}
 where the missing elements of the
matrices are all zero. It is easy to show that $D=d+[A,~]$ and
$\Phi$ satisfy the Hitchin equations:
\eqn\hitchi{\eqalign{ D_z \Phi_{\bar z}=0,~~~D_{\bar z} \Phi_z=0,
\cr [D_z, D_{\bar z}]+[\Phi_z,\Phi_{\bar z}]=0 }}
A simple consequence is that the spectral connection
\eqn\spec{\eqalign{ \nabla_z^{\zeta} = D_z + \zeta^{-1}
\Phi_{z} ,~~~~~~~~~~~~\nabla_{\bar z}^{\zeta} = D_{\bar z} + \zeta
\Phi_{\bar z}}}
is flat for all values of the spectral parameter $\zeta$. Notice
that unless $\zeta$ is a phase, the connection lies in $SO(6, C)$.

One can solve the linear problem $\nabla^\zeta q[\zeta]=0$ to define
a whole family of solutions $q[\zeta]$. It is easy to verify that
$Y[\zeta]=q_0[\zeta]$ is a complex solution of the usual equations
of motion and Virasoro constraints. If $\zeta$ is a phase,
$Y[\zeta]$ can taken to be real.

It turns out to be useful to study the action of the connection
${\cal A}$ on spinors of $SO(2,4)$ (or $SO(3,3)$, or $SO(4,2)$). In
practice, we can pick a set of gamma matrices and use them to
rewrite ${\cal A}$ as a $4 \times 4$ matrix. With slight abuse of
notation we will use the same symbol for the $6 \times 6$ and the $4
\times 4$ connections. We can keep the decomposition of ${\cal A}$
into $A$ and $\Phi$ simple, by using a set of gamma matrices adapted
to their block-diagonal form:
\eqn\fullgauge{\Phi_z = \left(\matrix{0 & \frac{e^{-1/2
\alpha}}{\sqrt{2}}  v_I \tau^I \cr \frac{e^{1/2\alpha}}{\sqrt{2}}
{\bf 1}_2 & 0} \right) ~~~~~~~~~ A_z = \frac{1}{4} \left(\matrix{-
\partial \alpha +d_{IJ}\tau^{IJ} & 0\cr 0 & \partial \alpha
+d_{IJ}\tau^{IJ}} \right) }
Here $\tau^{I}$ are appropriate Pauli matrices for $SO(1,2)$,
$SO(2,1)$ or $SO(3)$. Furthermore, $\Phi_{\bar z} = \bar \Phi_z$ and
$A_{\bar z} = -\bar A_z$, where the overline indicates hermitian
conjugation, together with conjugation by either $\tau^1$ or
$\tau^{12}$ for the $SO(1,2)$ or $SO(2,1)$ cases.

Although $\Phi$ and $A$ satisfy Hitchin's equations, they are not
the most general solution. There is a simple constraint which can be
added to Hitchin's equations, such that all solutions satisfying the
constraint will be of the form \connesplit . If we introduce the
constant matrix
\eqn\CCC{C = \left(\matrix{0 & \sigma_2 \cr i \sigma_2 & 0} \right)}
then the solutions of $C A^T C^{-1}=-A$ and $C \Phi^T C^{-1}=i \Phi$
have the form \fullgauge. Hence classical string solutions in
$AdS_5$ correspond to solutions of Hitchin's equations which are
fixed by the $Z_4$ automorphism $A \to -C A^T C^{-1}$ and $\Phi \to
-i C \Phi^T C^{-1}$.

In terms of the vector-valued connections, we can construct a $6
\times 6$ matrix $C$ combining a reflection of $q_i$, $i=1,2,3$, and
a rotation of $\pi/2$, $q_4 \to i q_4$, $q_5 \to -i q_5$,
so that the $Z_4$ automorphism $A \to C A C^{-1}$ and
$\Phi \to i C \Phi C^{-1}$ has fixed points of the form \connesplit\
only. In particular, the symmetry gives $\nabla^{i \zeta} = C
\nabla^\zeta C^{-1}$, and $Y[\zeta]$ must be related to $Y[i \zeta]$
by a space-time rotation.

\subsec{Boundary conditions for the connection at infinity}

In this subsection we discuss the boundary conditions at infinity
for the different components of the flat connection. We start with
the simpler case of $AdS_3$ \AldayYN\ and then express the boundary conditions
in such a way that they are easy to generalize for the most general
case.

The reduction to $AdS_3$ amounts to setting $B_2$ and $B_3$ to
a constant. Hence $d_i^{~j}=0,v_2=v_3=0$ and $v_1^2=-P(z)
\rightarrow v_1=i P^{1/2}$. It turns out that in the $AdS_3$
case, $P$ is the square of another polynomial $P \propto p^2$
\AldayYN .
 Writing the four by four connections in terms of
two by two blocks we obtain
\eqn\adethreeconn{ \eqalign{ A_z= {1 \over 4} \partial \alpha
\left(\matrix{-{\bf 1}_2 & 0 \cr 0 & {\bf 1}_2 } \right),~~~
\Phi_z= \left(\matrix{0 &  \frac{e^{-  \alpha/2}}{\sqrt{2}}
P(z)^{1/2} \sigma_1 \cr
 \frac{e^{ \alpha/2}}{\sqrt{2}}{\bf 1}_2  & 0 }\right),~~~~
 A_{\bar z}=-A_z^\dagger,~~~\Phi_{\bar z}=\Phi_z^\dagger
}}
Flatness condition implies the following equation for $\alpha$
\eqn\sinhgordon{\eqalign{\partial \bar \partial
\alpha-e^\alpha+e^{-\alpha} |P(z)|=0 }}
This generalized sinh-Gordon equation can be brought to a more
standard form by defining a new variable $w$ such that
$dw=P(z)^{1/4} dz$, \wdefi ,  and making a field redefinition
\eqn\alphahat{\alpha=\hat \alpha+\frac{1}{4} \log P(z) \bar P(\bar
z)}
As argued in  \refs{\AldayGA,\AldayYN}, the appropriate
boundary conditions are that $\hat \alpha$ vanishes at
infinity, decaying exponentially. One simple way to understand
this is to notice that the four cusp solution is simply $\hat
\alpha =0$ and $P=1$. After we go to the $w$ plane, this is
indeed the asymptotic behavior at large $w$ for any polynomial
$P$.
 This
is what we expect, since near each cusp the solution should be very
similar to the four cusp solution. In the four cusp solution the $w$
plane and the $z$ plane coincide and we have four cusps
corresponding to the four quadrants of the $w$ plane. For a
polynomial of degree $N$ which goes as $P \sim z^N + \dots$ we have
that $w \sim z^{ N+4 \over 4 } $. This implies that going once
around the $z$ plane we go around the $w$ plane ${ N+4 \over 4 }$
times. This implies that the total number of quadrants in the $w$
plane is $n = N+4$. This is the total number of cusps. In
conclusion, a polynomial of degree $N$ leads to an $n=N+4$ sided
polygon.

Such boundary conditions can be understood in the following
alternative way. If we set $\hat \alpha=0$, then we can diagonalize
$\Phi$ and $A$ by an honest gauge transformation $h$ (the
exponential of an anti-hermitian matrix)
\eqn\diag{\eqalign{h^{-1}\Phi_z h=& \frac{1}{\sqrt{2}}\left(
\matrix{ P(z)^{1/4} &&& \cr & -i P(z)^{1/4} && \cr &&
 - P(z)^{1/4}& \cr &&& i P(z)^{1/4}} \right) \equiv \Phi^{diag}_z\cr
& h^{-1}A_z h+h^{-1} \partial h=0}}
The  statement that $\hat \alpha \rightarrow 0$ at infinity is
equivalent to saying that exists a gauge transformation such that
$$h^{-1} \Phi_{z} h \rightarrow \Phi_{diag},~~~h^{-1}A_z h+h^{-1} \partial h \rightarrow 0$$
at infinity. We have similar expressions for $\Phi_{\bar z}$, $A_{\bar z}$ with the same $h$.
%
%

The boundary conditions at infinity for the general case of $AdS_5$
can be stated in a similar way. After all, near each cusp we expect
the same behavior as the one for the four cusp solution. In fact,
these  are  standard boundary conditions from the point of view of
the Hitchin system. There exist a gauge transformation such that
$h^{-1} \Phi_z h \rightarrow \Phi_{diag}$ and $h^{-1} \Phi_{\bar z}
h \rightarrow \Phi^*_{diag}$ at infinity. Notice that the boundary
conditions $h^{-1} \Phi_{\bar z} h \rightarrow \Phi^*_{diag}$
involve a simple complex conjugation. We also have that $\hat \alpha \to 0$ at infinity, where
$\hat \alpha$ is given by \alphahat .

For the case of $AdS_3$ and $AdS_4$ the matrix $d_{IJ}\tau^{IJ}$ is
diagonal, and has an off-diagonal component only for the case of
$AdS_5$. Next, we consider $A_{z}$ in the same basis that
diagonalizes $\Phi_{\bar z}$ at infinity. Flatness of $\Phi_{\bar
z}$ requires $A_{z}$  to be diagonal as well. There are two further
requirements: in the original basis $A_{z}$ should be
block-diagonal, and single-valued. This imposes the following
boundary conditions
\eqn\gaugeatinfty{h^{-1} A_{z} h +h^{-1} \partial h \rightarrow
\frac{m}{z} \left(\matrix{ \sigma_3 & 0\cr 0 & \sigma_3}
\right),~~~~~~~h^{-1} A_{\bar z} h +h^{-1} \bar \partial h
\rightarrow \frac{\bar m}{\bar z} \left(\matrix{ \sigma_3 & 0\cr 0 &
\sigma_3} \right)}
where $\bar{m}$ is minus the conjugate of $m$ for $(3,1)$ or $(1,3)$ signature
and plus the conjugate of $m$ for $(2,2)$ signature.
 The constant $m$ is allowed to be non-zero only for a $P(z)$ of
 even degree $n$: in the diagonal basis, $\Phi_{diag}$ comes back to itself under
$z \to e^{2 \pi i} z$ up to conjugation by a shift matrix. In that
basis, our asymptotic choice for $A$ is invariant under that shift
only for even $n$. Note that the $Z_4$ action that restricts the
form of the general Hitchin equation to our form (discussed around
\CCC ) has a simple action in the basis that diagonalizes $\Phi$
\diag . The matrix $C$ simply performs a cyclic permutation of all
four eigenvalues. It also permutes the eigenvalues of $A$. Thus the
condition $A = - C A^T C^{-1}$ implies that $A$ has the matrix
structure   in \gaugeatinfty .

As already mentioned, for the cases of $AdS_3$ and $AdS_4$, the
gauge field is diagonal before the gauge transformation. In
particular, its structure forbids the extra term in
\gaugeatinfty\ . This term is present only for the case of
$AdS_5$. This is related to the fact that, for $AdS_{3,4}$,
once we choose a polynomial $P(z)$, there is a unique solution
satisfying the appropriate boundary conditions, but this is not
the case for $AdS_5$.

We can use these observations also to count the number of parameters
that we expect in general. In order to count the moduli of the
Hitchin equations we can go to the region where the zeros of the
polynomial are widely spaced. In this region the matrix $\Phi$ is
diagonal in most of the space. Let us start by counting the number
of parameters in the polynomial $P$. We can always set the
coefficient of the highest power to one by a rescaling of $z$. We
can also set the coefficient of the next term to zero by a
translation. Thus we have only $N-1$ complex coefficients left. As
we explained above, as long as we can set $\hat \alpha \sim 0$ we
can also diagonalize $A$. We will then have various Wilson lines for
the connection around various one cycles of the surface. In order to
have a non-zero value for the Wilson line we need a one cycle that
is consistent with the $Z_4$ projection. Instead of counting one
cycles we can count differential forms which are odd under the
generator of $Z_4$. We have a curve $x^4 = P(z)$ with the
differential $x dz$. The $Z_4$ projection on the Hitchin system maps
$x \to i x $.
 The one forms can be written as
$ \eta = q(z) { dz \over x^2 } $. For large $z$ we require that they
decay as ${ 1 \over z^2 }  d z$ or
 faster. The
one form decaying as $1/z$ was already included in the discussion
around \gaugeatinfty . For a polynomial of even degree $N$ we have
$q \sim  z^{N/2 -2} + \cdots$. We have $N/2 -1$ complex
coefficients\foot{ The fact that the $dz/z$ term counts as only one
real coefficient is related to the fact that we just have one
compact cycle associated to this one form, which is a circle around
infinity.}.
The total number of real parameters is then $ 2(N-1) + 1
+ 2 ( N/2 -1) = 3 (N-1) = 3 n - 15 $ which is the expected number of
independent cross ratios for an $n$ sided polygon (recall that $n =
N+4$).

\subsec{Behavior at infinity of the solutions of the linear problem}

Here we show that the above conditions on the connection imply
that the solutions of the linear problem are such that the
surface ends on the boundary of $AdS$ on a polygonal contour
with light-like sides.

 The above boundary conditions imply  that the behavior at infinity of the
flat section $q$ is controlled by the degree $N$ of the polynomial
$P$. The large $|z|$ region is divided into $n=N+4$ angular sectors
of equal width. In the sector $V_i$ the six components of $Y$ grow
exponentially with a specific fixed ratio, so that the worldsheet
reaches a specific point $x_i$ at the boundary, where the $i$-th
cusp resides. Alternatively, $Y \sim y_i \exp S(z)$ where $S(z)$
grows like a power of $|z|$ and $ y_i$ is a null vector, which
identifies a point on the boundary of $AdS_5$.

Here we can think of the boundary of $AdS_5$ as the set of null
vectors $ y \in R^{2,4}$ with $y^2 =0$, together with the
identification $y \sim \lambda y $. Thus, only the direction of
the null vector $y$ is important\foot{ More explicitly, we can
say $y = ( y_+, y_-, y_\mu )$, $\mu = 0,1,2,3$. $y^2 = y_+ y_-
+ y_\mu y_\mu =0$. Then $x_\mu \equiv { y_\mu \over y_+}$ are
the ordinary Poincare coordinates. Also, given two points $y$
and $\tilde y$ we have that $ 2 y. \tilde y = -  y_+ \tilde y_+
(x - \tilde x)^2 $. Thus, the inner product of two $y$ vectors
is proportional to the distance in $R^{1,3}$. If we form cross
ratios the proportionality factors cancel. Thus cross ratios
can be written in terms of ratios of products of projective $y$
vectors. }.

As we change sectors, the point $x_i$ (or $y_i$) on the boundary
jumps, in a way which is well described by Stokes theory for the
flat connection ${\cal A}$. It is useful to consider the spinor
version of the connection. Let $\psi_a$ be a set of four linearly
independent spinor flat sections satisfying
\eqn\conne{\partial \psi_a=-{\cal A}_z \psi_a,~~~~~~~ \bar
\partial \psi_a=-{\cal A}_{\bar z} \psi_a}
We can write a general solution to the vector-valued linear problem
as
\eqn\conv{
 q_I^{A} \gamma_{A \, ab} =
\psi_{\alpha a} \Gamma^{\alpha \beta}_I \psi_{\beta b}
}
Here $\gamma^A_{ab}$ are the  six space-time gamma matrices. $a,b$
are space-time spinor indices.  $\Gamma_I$ are the six gamma
matrices used to convert the connection ${\cal A}$ from the vector
to the spinor representation. $\alpha, \beta$ are internal spinor
indices. These are the indices acted upon by the connection. Both gamma matrices
are antisymmetric in their indices.

 The boundary conditions described above   imply the
existence of $n$ sectors $W_i$ (displaced by ${\pi \over n}$ with
respect to the $V_i$) in which the $\psi^a$ grow exponentially at
large $|z|$. Much like in the vector case, this determines a
direction (a ``twistor'') $\lambda_{a,i}$ in each sector, as $\psi_{
\alpha a}  \sim \lambda_{a,i}  e_{ \alpha , i } \exp \tilde S(z)$,
where $e_\alpha$ is the direction of this growing solution in the
internal space. The location of the space-time cusp associated to a
sector $V_i$ is determined by the twistors for the sectors $W_i$ and
$W_{i+1}$ overlapping with $V_i$
\eqn\cusp{y_i^A    \gamma_{A \, ab} = ( \lambda_{a,i} \lambda_{b, i+1} - \lambda_{b,i} \lambda_{a,i+1})
~ ( e_{ \alpha , i}
 \Gamma_0^{\alpha \beta }  e_{\beta, i+1})
 }
$\Gamma_0$ appears here because the component $q_0= q_0^A$ is the
solution $Y^A$. The spinors $\lambda_{a, i}$ are determining the
direction of the vector $y^A_i$. The factor in parenthesis is simply
an overall scale. Thus, we can assign a spinor $\lambda_i$ to each
side of the polygon. The vertices of the polygon are then given by
$y_i \sim \lambda_i \lambda_{i+1} $. These spinors are the same as
the momentum twistors introduced in \hodgesspinor .

\ifig\twistors{  In (a) we see the polygon labeled by the
position of the vertices $x_i$, or equivalently, the momenta
$p_i$. In (b) we see how the data defining the polygon emerges
in our case. We have a spinor of $SU(2|2)$ $\psi$ associated to
each edge. The positions of the vertices $y_i$ are bilinears of
spinors. These coincide with the momentum twistors introduced
in \hodgesspinor .   } {\epsfxsize3.5in\epsfbox{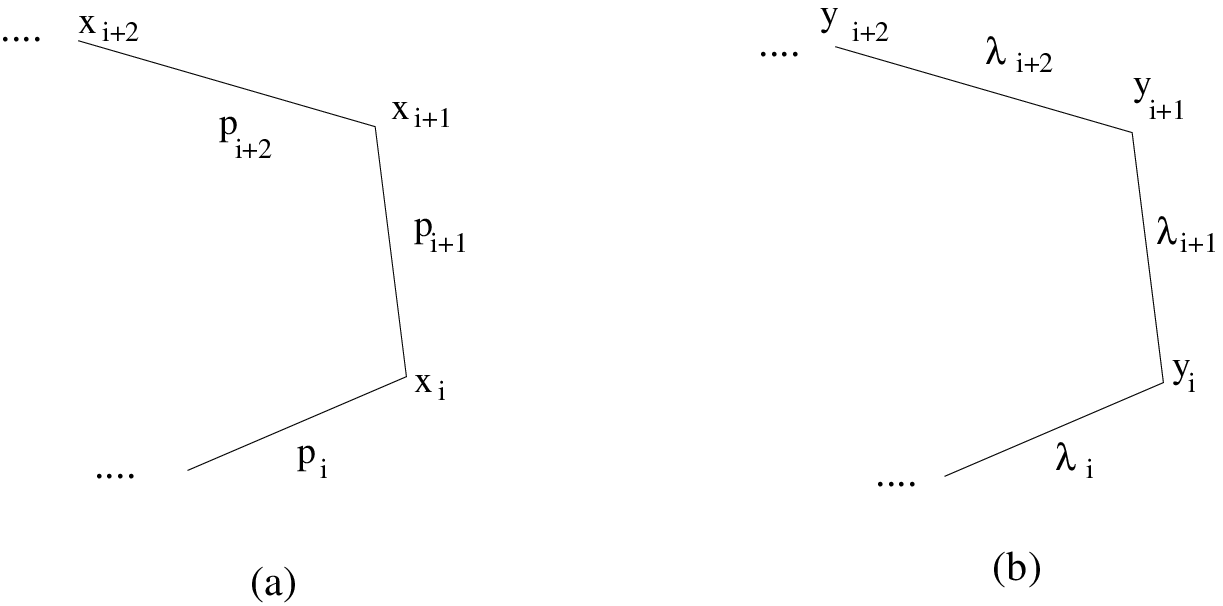}
 }
The inner product can be written as \foot{ The determinant of
 $( \lambda_1,\lambda_2,\lambda_3 \lambda_4 )$ is equal to $\epsilon^{\alpha \beta \gamma \delta}
\lambda_{\alpha, 1} \lambda_{\beta, 2} \lambda_{\gamma , 3} \lambda_{\delta , 4}$.}
\eqn\cusp{y \cdot y' = (y \cdot \gamma_{ab})(y' \cdot
\gamma_{cd})\epsilon^{abcd} = \det( \lambda_1 \lambda_2 \lambda'_1
\lambda'_2 )}
where $\lambda_{1,2}$ are the two spinors producing $y$ and
$\lambda'_{1,2}$ are the two spinors producing $y'$.
 It is then clear
that not only $y_i^2=0$, but also $ y_i \cdot y_{i+1} =0= (x_i -
x_{i+1})^2 $, hence the sides of the boundary polygon are null, as
desired.

\subsec{Linear problem at infinity and Stokes phenomena}

In this subsection we describe the nature of the Stokes phenomenon
at large $z$. We also express the Stokes data in terms of the
spacetime cross ratios.

 Once we have understood the asymptotic form
of the connection, it is straightforward to understand the
asymptotic form of the solutions $\psi_a$ of the linear problem
\conne\ . As already mentioned, we can perform a gauge
transformation in such a way that both $\Phi_z$ and $\Phi_{\bar z}$
become diagonal at infinity. In such basis, a generic solution takes
the asymptotic form
\eqn\inffield{ \psi \approx e_1 z^{m} \bar z^{\bar m}
e^{\frac{1}{\sqrt{2}}(w+\bar w)}+e_2 z^{-m} \bar z^{-{\bar m}}
e^{-\frac{i}{\sqrt{2}}(w-\bar w)}+e_3 z^{m} \bar z^{\bar m}
e^{-\frac{1}{\sqrt{2}}(w+\bar w)}+e_4 z^{-m} \bar z^{-\bar m}
e^{\frac{i}{\sqrt{2}}(w-\bar w)} }
where $w=\int P(z)^{1/4} dz \sim z^{{n \over 4}}$,
$\bar{w}=\int \bar{P}(\bar{z})^{1/4} d\bar z \sim \bar z^{{n
\over 4}}$. This asymptotic solution is a rather reliable
approximation of the exact solution at large $|z|$, as long as
one restricts the range of $\arg z$ appropriately. The failure
of the approximation is governed by Stokes phenomena. Consider
the behavior of a generic approximate solution as one varies
$\arg z$ at large $|z|$: the four exponential terms will take
turns at controlling the asymptotic behavior, so that $\psi$
will be proportional to, say, $e_1$ whenever $Re \, w$ is
positive and bigger than $|Im \, w|$ so that $(w+\bar w)$ is
the largest exponent. A generic exact solution will have a
similar behavior, and in each sector $$W_i : {2 \pi \over n}i -
{3 \pi \over n} < \arg z < {2 \pi \over n}i - {\pi \over n}$$
it will point in some direction $\lambda_i$, but
$\lambda_{i+4}$ does not have to coincide with $\lambda_i$. We
will denote as $S_i$ the appropriate largest exponent.

 \ifig\stokesone{ The light lines  show the four  exponents of the WKB approximation for large $|z|$ as a function
 of the argument of $z$. The darker line follows the behavior of a typical solution. A typical
 solution is the superposition of all four solutions so that the largest exponent dominates. }
{\epsfxsize2in\epsfbox{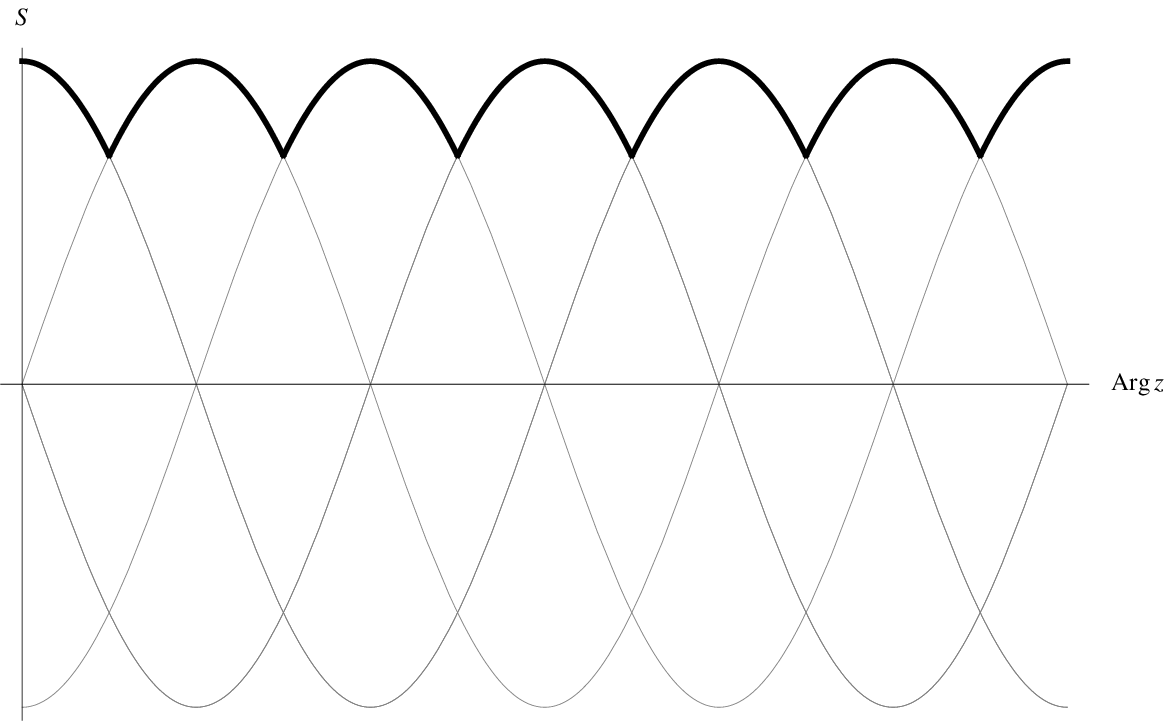}  }

\ifig\stokestwo{ The dark line shows the behavior of a "small
solution" as we change the argument of $z$. The solution is
controlled by the smallest exponent in the darkened angular
sector. The solution is still controlled by the same exponent
as we move away from the darkened region, until the exponent
grows to a maximum. Subsequently, it behaves as the generic
solution, see \stokesone . }
{\epsfxsize2in\epsfbox{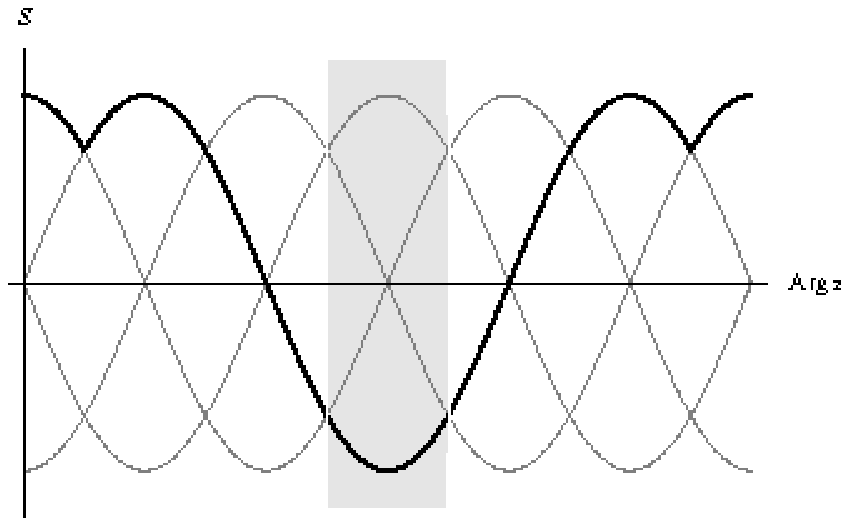}  }

There is a simple, powerful fact: if we pick a ray towards $|z|
\to \infty$ away from the boundaries of the $W_i$, we can
identify a unique (up to rescaling) exact solution by the
boundary condition which requires the fastest possible decay at
infinity. This small solution $s_i$ is the same for all rays
inside $W_i$. More generally, one can identify uniquely
$k$-dimensional subspaces of solutions whose growth is
controlled by the $k$ smallest exponentials. That way, it is
easy to see that $s_i$ will still behave as $\exp S_i$ if
continued beyond $W_i$, through $W_{i-1}$ and $W_{i-2}$ or
through $W_{i+1}$ and $W_{i+2}$, as $S_i$ grows all the way to
be the largest exponent see \stokestwo\ . Beyond that, the
behavior of $s_i$ is unconstrained.

One can generically take the set $(s_i, s_{i+1}, s_{i+2},
s_{i+3})$ to be a neat basis of linearly independent solutions.
This implies that $s_{i+4}$ can be expressed in terms of a
linear combination of the previous four solutions. In the
sector where $s_{i+2}$ is the smallest solution, $s_i$ and
$s_{i+4}$ are the two largest solutions. Thus we can compare
them. We choose to normalize $s_{i+4}$ so that it is
proportional to $-s_i$ in this sector, up to corrections by
smaller solutions. Thus we can write $$s_{i+4} + s_i = a_{i+1}
s_{i+1} + b_{i+2} s_{i+2} + c_{i+3} s_{i+3}.$$ With this choice
of normalization we have that    $ \det (s_i, s_{i+1},
s_{i+2},s_{i+3})=1$.
 We should impose the
periodicity constraint $s_{i+n} = \mu_i s_i$. We can relate the
$\mu_i$ with the formal monodromies, i.e. the monodromies of
the approximate asymptotic solutions. The formal monodromy
receives contributions from both $m,\bar m$ and the coefficient
in front of $1/z$ in the expansion of $P(z)^{1/4}$, if
present.The simplest example is the pentagon. There are no
formal monodromies, and $s_1 + s_2 + s_3 + s_4 + s_5=0$ is the
unique way to solve all the constraints.

The example of interest for us is the hexagon. In this case our
relation requires $s_1$, $s_5$, $s_9$, $s_{13} = \mu_1^2 s_1$
to grow with the same exponent $S_1(z)$, but the comparison is
made after going around the $z$ plane twice, hence $\mu_1^2 =
\exp \left( S_1(e^{4 \pi i} z)- S_1(z) \right)$, and $\mu_1$
agrees up to a sign with the formal monodromy. The formal
monodromies come from $m, \bar m$ \gaugeatinfty\ only. We can
set without loss of generality $P(z) = z^2 - U$, and then
$P(z)^{1/4} = z^{1/2} -{U \over 4 z^{3/2}} + \cdots$ has an
expansion in half integer powers of $z^{-1}$ and there is no
logarithm\foot{ More explicitly, in cases where $N$ is a
multiple of four, we will get that generically we have $d w = (
z^{ N/4 } + \cdots + \tilde m/z + \cdots )dz$ which implies
that $w$ contains a term going like $w \sim \cdots +  \tilde m
\log z + \cdots $. Such a term also contributes to the formal
monodromy.} in $w$. Hence we get formal monodromies $s_7 = \mu
s_1$, $s_8 = \mu^{-1} s_2$, $s_9 = \mu s_3$, $s_{10} = \mu^{-1}
s_4$, etc. Here $\mu= \pm \exp 2 \pi i (m-\bar m)$. In $(3,1)$
signature, $\mu$ is a phase, while in $(2,2)$ signature $\mu$
is a real number.

The coefficients $b_i$ will play a particularly important role
in what follows, since for $n=6$ they contain the non-trivial
Stokes information. Consider the determinant $\det (s_i,
s_{i+1}, s_{i+3}, s_{i+4})= - b_{i+2}$. For the hexagon that is
the same as $$b_{i+2} = -\mu \mu^{-1}\det (s_{i+6}, s_{i+7},
s_{i+3}, s_{i+4})= b_{i+5}$$ It turns out that everything can
be expressed in terms of $\mu$, $b_1$, $b_2$, $b_3$, subject to
the relation\foot{ We can change $b_i \to - b_i$ and $\mu \to
-\mu$ by changing the signs of some of the $s_i$. Thus these
two possibilities describe the same physical situation. In this
paper we concentrate in the regime where the sign of all $b_i$s
are the same, which we take to be positive. Then the two signs
of $\mu$ describe two different physical situations. }\foot{The $\mu=1$ limit of this
equation was obtained in the study of the $x^4$ anharmonic oscillator in \DoreyPT .}
\eqn\relabmu{ b_1 b_2 b_3 = b_1 + b_2 + b_3 + \mu + \mu^{-1}
~~~~~~~~~~~~~~~ b_{i+3} = b_i }

The case of the hexagon in $AdS_3$ is particularly important.
Then, the solutions at even and odd sectors do not mix, and
satisfy $s_1 - s_3 + s_5 =0$ and $s_2 - s_4 + s_6=0$, so that
$b_i = 1$ identically, and $\mu=-1$, as it should. For the
$AdS_4$ case we have that $\mu = \pm 1 $.

To conclude this subsection, we would like to relate the $b_i$
to the space-time cross-ratios for the case of the hexagon. If
we are given a solution $q_I$ of the vector problem, there is a
very simple way to extract the position of the cusps.  The
bilinear $s_i^T \Gamma^I s_{i+1}$ gives the vector solution
which decays the fastest in the sector $V_i$, hence the
contraction $y_i^A = q_I^A s_i^T \Gamma^I s_{i+1}$ gives the
position of the cusp. The inner product is
$$y_i \cdot y_j = (q_I \cdot q_J) s_i^T \Gamma^I s_{i+1} s_j^T
\Gamma^J s_{j+1}= s_i^T \Gamma^I s_{i+1} s_j^T \Gamma_I s_{j+1}
=\det(s_i, s_{i+1} ,s_j,s_{j+1}).$$ Hence the cross-ratios are
\eqn\crossratios{ u_1=\frac{x_{13}^2 x_{46}^2}{x_{14}^2 x_{36}^2}={1
\over b_2 b_3},~~~u_2=\frac{x_{24}^2 x_{15}^2}{x_{25}^2 x_{14}^2}={1
\over b_1 b_3},~~~u_3=\frac{x_{35}^2 x_{26}^2}{x_{36}^2 x_{25}^2}={1
\over b_1 b_2} }

In conclusion, we have shown that the Stokes data at infinity is given by $b_1,~ b_2,~ b_3$ and $\mu$
with the constraint \relabmu .
 The spacetimes cross ratios are related to the $b_i$ via \crossratios .

\newsec{From Hitchin to TBA}

If we introduce now the spectral parameter, $\zeta$,
 in the connection,
we can investigate the behavior of the deformed solution
$Y[\zeta]$. As the conserved polynomial becomes simply
$\zeta^{-4} P(z)$, the deformed solution will have the same
number of cusps at infinity, associated with directions
$y_j[\zeta]$ and cross-ratios $u_j[\zeta]$ (or $b_j[\zeta]$)
 which are
meromorphic in $\zeta$ away from $\zeta=0, \infty$. The only
slight subtlety is that the location of the sectors
$W_j[\zeta]$ and $V_j[\zeta]$ varies as the phase of $\zeta$
varies. The sectors $W_j[\zeta]$ are defined by
$$W_j[\zeta]:~{2 \pi j \over n} + {4 \over n}\arg \zeta  - {3
\pi \over n} < \arg z <  {2 \pi j \over n} + {4 \over n}\arg
\zeta  - {\pi \over n}.$$ Hence
 the small sections $s_j[\zeta]$,
the coefficients $b_j[\zeta]$ and the   cross-ratios $u_j[\zeta]$
 do not quite come back to
themselves as $\zeta \to e^{2 \pi i} \zeta$, but rather undergo
a shift in the index $j \to j+4$, $b_j[\zeta e^{ 2 \pi i }] =
b_{j+4}[\zeta] $. Due to the $Z_4$ automorphism of $A,\Phi$,
there is actually  a symmetry $b_{j}[i\zeta] = b_{j+1}[\zeta]$.
We will not make use of this symmetry for now.  Note that, for
the hexagon case,  the relation \relabmu\ that we had for
$\zeta =1$ remains valid for $b_i[\zeta]$, with $\mu$
independent of $\zeta$. Throughout this section we will hold
$U$ and $\mu$ fixed and we will only vary $\zeta$. Of course,
we are only interested in the values at $\zeta =1$. We will
determine these values by writing an equation for the cross
ratios as a function of $\zeta $ and then we will set $\zeta
=1$. This equation comes from demanding that the functions are
analytic plus the statement that they have a known approximate
expression (displaying Stokes phenomenon) near $\zeta \sim 0,
\infty$, which can be computed via the WKB approximation.

\ifig\motionsectors{  This figure shows the sectors in the $z$ plane for the case of $n=6$.
 They could be the sectors $W_j$ or $V_j$.
In (a) we see the various sectors for a given value of the
phase of $\zeta$. (b) as we change the phase of $\zeta$ by
$\zeta \to \zeta e^{ i \varphi}$ the sectors rotate by an angle
$2 \varphi/3$. In (c) we see the result of a full rotation $
\zeta \to \zeta e^{ i 2 \pi }$. We come back to $(a)$ up to a
relabeling of sectors.  }
{\epsfxsize4in\epsfbox{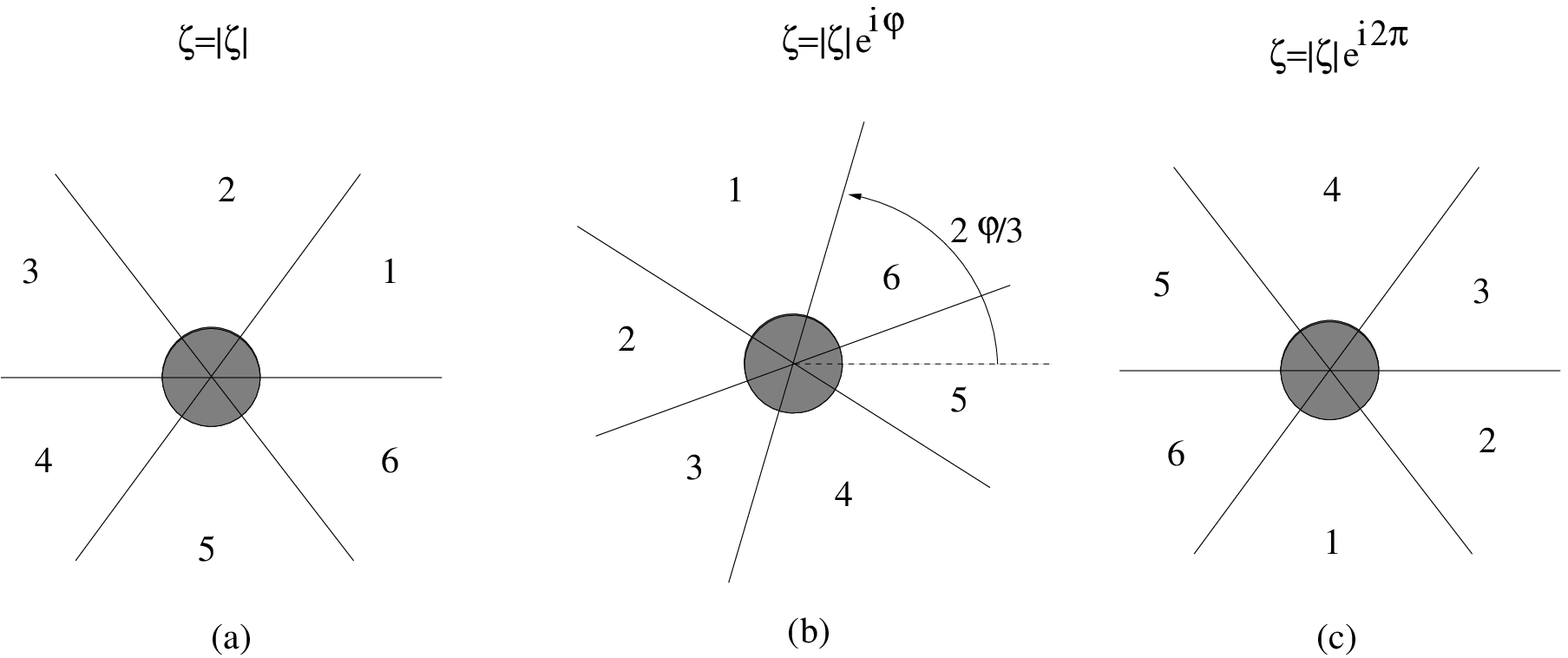}  }

 We aim to set up a Riemann-Hilbert problem, and derive
a useful set of integral equations from it. The equations take
the appearance of TBA equations. Although the resemblance is
purely formal, it is rather striking. The TBA equations do not
just compute the cross-ratios. We will show how the regularized
area of the minimal surface, which can be written as a suitably regularized
integral of
$ Tr \Phi \bar \Phi  $,
 takes the form of the free energy for the TBA equations.

For the sake of simplicity, we will give a direct derivation of
the TBA equations for the case at hand. The reader should be
aware that this is a special case of a much broader result. The
mondromy data (such as the cross-ratios) of the auxiliary
connection for any Hitchin system is always computed by a
TBA-like set of equations. Physically, this fact follows from a
relation between Hitchin systems and ${\cal N}=2$ gauge
theories \refs{\GMNone,\GMNtwo}.  In the special case of $SU(2)$
Hitchin systems, the result has been demonstrated directly, and
a crucial ingredient for the TBA equations (the spectrum) has
been determined through a simple WKB analysis \GMNtwo . For
more general higher rank systems, the simple WKB analysis is
expected to fail, and be replaced by something more complicated,
and currently unknown. Luckily for us, the special Hitchin systems
with a simple spectral curve $x^m =P(z)$, such as the one relevant for the
minimal surfaces in $AdS_5$, are still amenable to
a simple WKB analysis, although an extra layer of
analysis is required. The relation between the regularized integral of $Tr \Phi \bar \Phi$ and the
TBA free energy appears to be valid for general Hitchin
systems as well, and was not noticed in previous works.

\subsec{WKB approximation of $b_i[\zeta]$} Near $\zeta=0,
\infty$ the behavior of the connection is clearly dominated by
$\Phi$ and $\bar \Phi$ respectively, and we can use a WKB
approximation to evaluate the behavior of the cross-ratios. The
basic idea behind the WKB approximation is to go to a (complex)
gauge where, say, both $\Phi$ and $\bar A$ have been
diagonalized. If one ignores the higher order corrections as
$\zeta \to 0$, the $i$-th component of a flat section should
behave like $\exp {1 \over \zeta} \int x_a dz$, where $x_a$ are
the eigenvalues of $\Phi$. The higher order corrections mix the
different components of the flat section, and disrupt this
simple exponential behavior. The component with the highest
exponential growth (largest real part of ${x dz  \over \zeta}$)
will in general contaminate all the others, but its leading
exponential growth will be unaffected. The simple WKB
approximation computes the small $\zeta$ exponential behavior
of a flat section transported along a path for which the real
part of ${x dz \over \zeta}$ for a given eigenvalue remains the
largest. This is called  a ``WKB path'': the condition
guarantees that the flat section grows with the largest
exponent and is well described by the WKB approximation.

In the case at hand, we want to compare the ``small'' flat
vector-valued sections $s_i^T \Gamma^I s_{i+1}$ at the rays $r_i$
between the sectors $W_i$ and $W_{i+1}$. At sufficiently large $z$,
the small section is,
by definition,   the flat section which grows the fastest along
a line which goes towards smaller $z$ along that ray $r_i$. If
we can find a WKB path which flows all the way to some other
ray $r_j$ at large $z$, we can evaluate the small $\zeta$
asymptotics of the inner product $\det ( s_i, s_{i+1},
s_j, s_{j+1})$ as $\exp {1 \over \zeta} \int_{p_i}^{p_j}
x dz$. $p_{i,j}$ are some reference points near $z=\infty$ in
$r_{i,j}$. Varying $p_{i}, ~p_j$ changes the (arbitrary) choice of
normalization for the small sections. Remember that the
eigenvalue $x$ for the vector-valued connection will be the sum
of two consecutive roots $P(z)^{1/4}$. If we have appropriate
WKB paths available for all the pieces of a cross-ratio,
and we can evaluate the asymptotic
behavior of each piece and combine them. The spurious
dependence on the $p_i$ drops off from the cross-ratio, and the
various contour integrals join into a single integral of
$P(z)^{1/4}$ along an appropriate contour on the Riemann
surface $x^4 = P(z)$.

The eigenvalues $x_a$ for the vector valued $\Phi$ are of the general
form $x$, $-i
x$, $-x$, $ i x$, $0$, $0$. The real part of $\zeta^{-1} x
dz$ is the largest if the phase of $\zeta^{-1} x dz$ sits
inside an open sector of width $\pi/2$ centered around the
positive real axis. Beyond that sector, either $\zeta^{-1} (-i x) dz$
or $\zeta^{-1} (i x) dz$
will have a larger real part.
  This is the condition
which must be satisfied by a WKB path. It is particularly
useful to consider WKB paths of steepest ascent, such that the
phase $\varphi$ of $x dz$ is constant. Such a path satisfies
the WKB condition for all values of $\zeta$ inside a sector of
width $\pi/2$ centered around the ray of phase  $\varphi$ in
the $\zeta$ plane. Correspondingly, any estimate of a
cross-ratio asymptotics done with WKB paths of constant phase
$\varphi$ will hold along all rays inside the corresponding
sector in the $\zeta$ plane. Notice that the WKB condition is
satisfied in the sector both with respect to $\zeta^{-1} x dz$
and to $\zeta \bar x \bar dz$, so the same calculation controls
the asymptotics both at small and at large $|\zeta|$ for a
given phase of $\zeta$.

The WKB paths of constant phase $\varphi$ are straight lines in the $w$ plane.
 At sufficiently large $z$, away from the zeros of $P(z)$, there are always WKB
  paths joining rays $r_{i}$ and $r_{i+2}$, which give estimate of the
  reference inner products $( s_i \wedge s_{i+1} \wedge s_{i+2} \wedge s_{i+3})$.
   The straight lines in the $w$ plane can be grouped in continuous  families
   which asymptote to the same rays at infinity.
Different families are separated by the special lines which end on the zeros of $P(z)$.
It is useful to draw a picture in the $z$ plane, with one WKB path from each family for a given choice of $\varphi$
(and all possible choices of a  fourth root of $P(z)$).

 \ifig\wkbone{ This figures shows WKB lines in the $z$ plane
 with a constant phase. The doted lines end at the zeros and they
 divide various regions. WKB lines that cross at 90 degrees correspond to two different eigenvalues that differ by
 an $i$ factor. Solid lines show lines that we will use to evaluate $b_3$ and $b_1$ in this case. The
 figure on the right shows the effect of increasing the value of $\varphi$. Note that we have WKB lines joining
 cusps 12 with 45 and 23 with 56 but we do not have WKB line joining 16 with 34.   }
{\epsfxsize2in\epsfbox{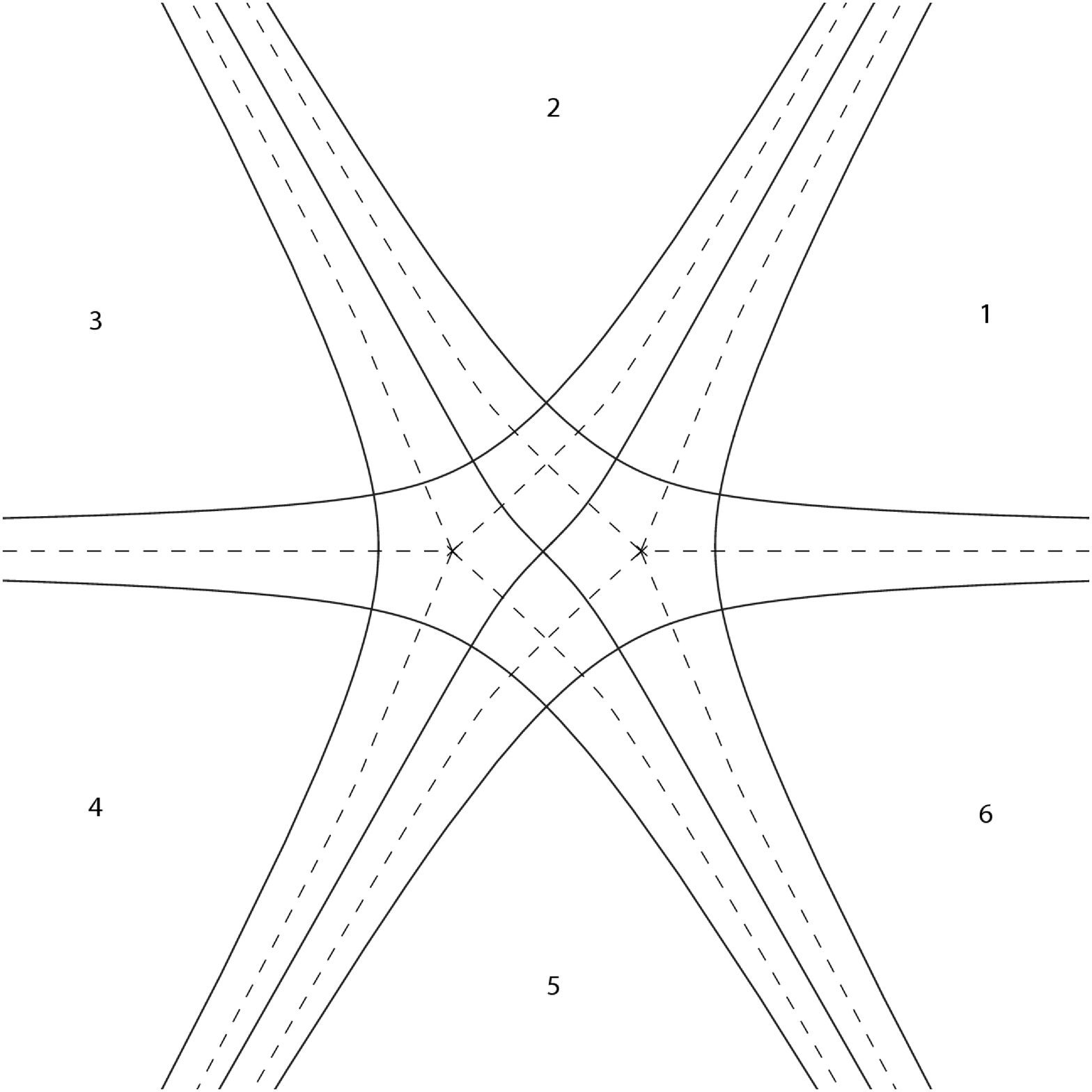}  ~~~~~ ~~~~ \epsfxsize2in \epsfbox{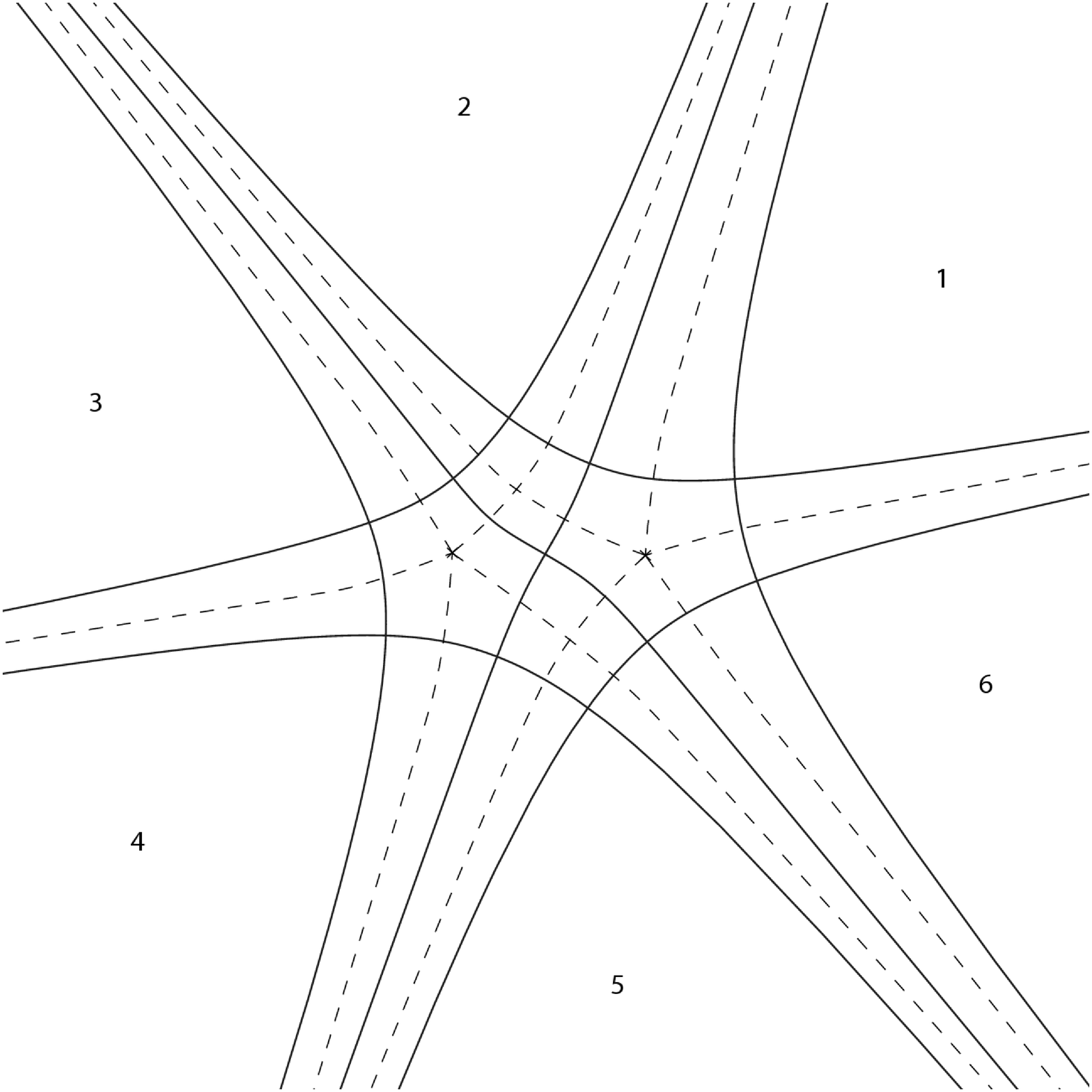}
 }
 \ifig\wkbtwo{  These are the same WKB lines as in the previous figure but for increasing values of $\varphi$.
 The figure on the right represents a critical value of the phase where the line that was joining the cusp 23 with
 61 no longer exits. Instead we have WKB lines that end at zeros and a new WKB line that joins the two zeros. In
 this case we can only use the WKB approximation to compute the cross ratio $b_1$ involving cusps 12 with 45.    }
{\epsfxsize2in\epsfbox{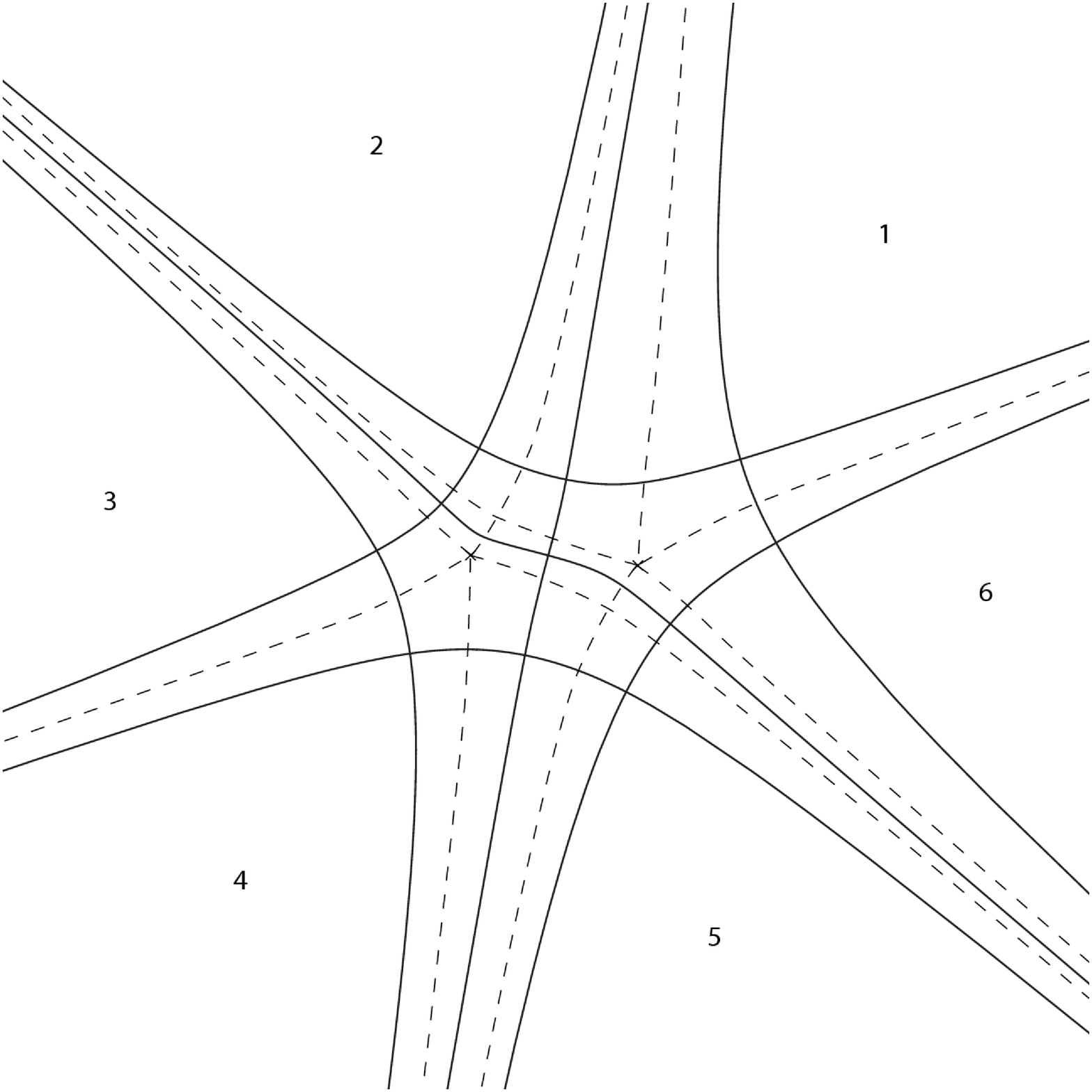} ~~~~~~~~~~ \epsfxsize2in\epsfbox{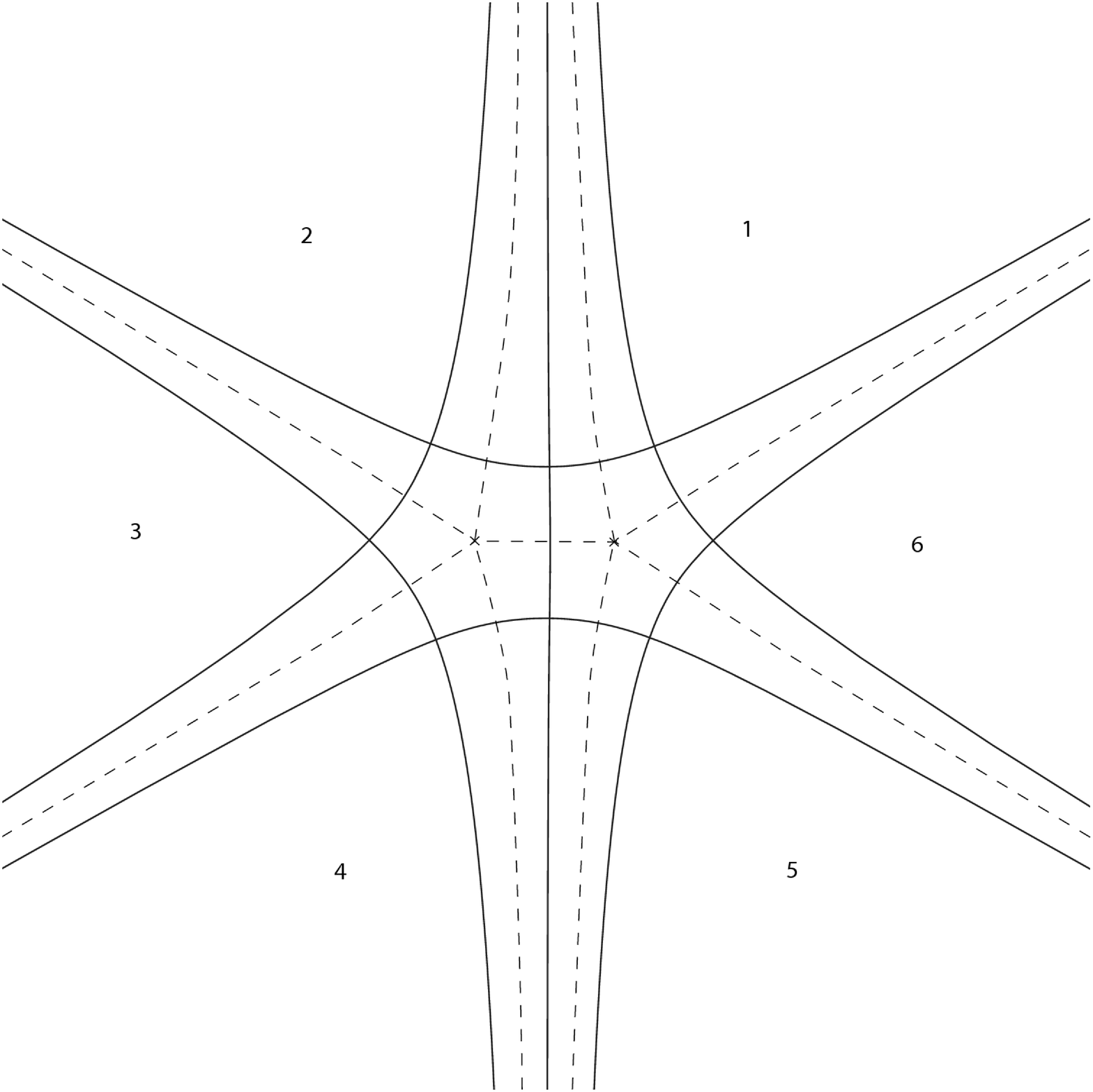}
 }
\ifig\wkbthree{  This figure shows the WKB lines for a phase
which is bigger than the critical phase. In this case we have
WKB lines joining the cusps 12 with 45 and also joining 16 with
34. In this case we can compute $b_2$ and $b_3$ via the WKB
approximation.} {\epsfxsize1.5in\epsfbox{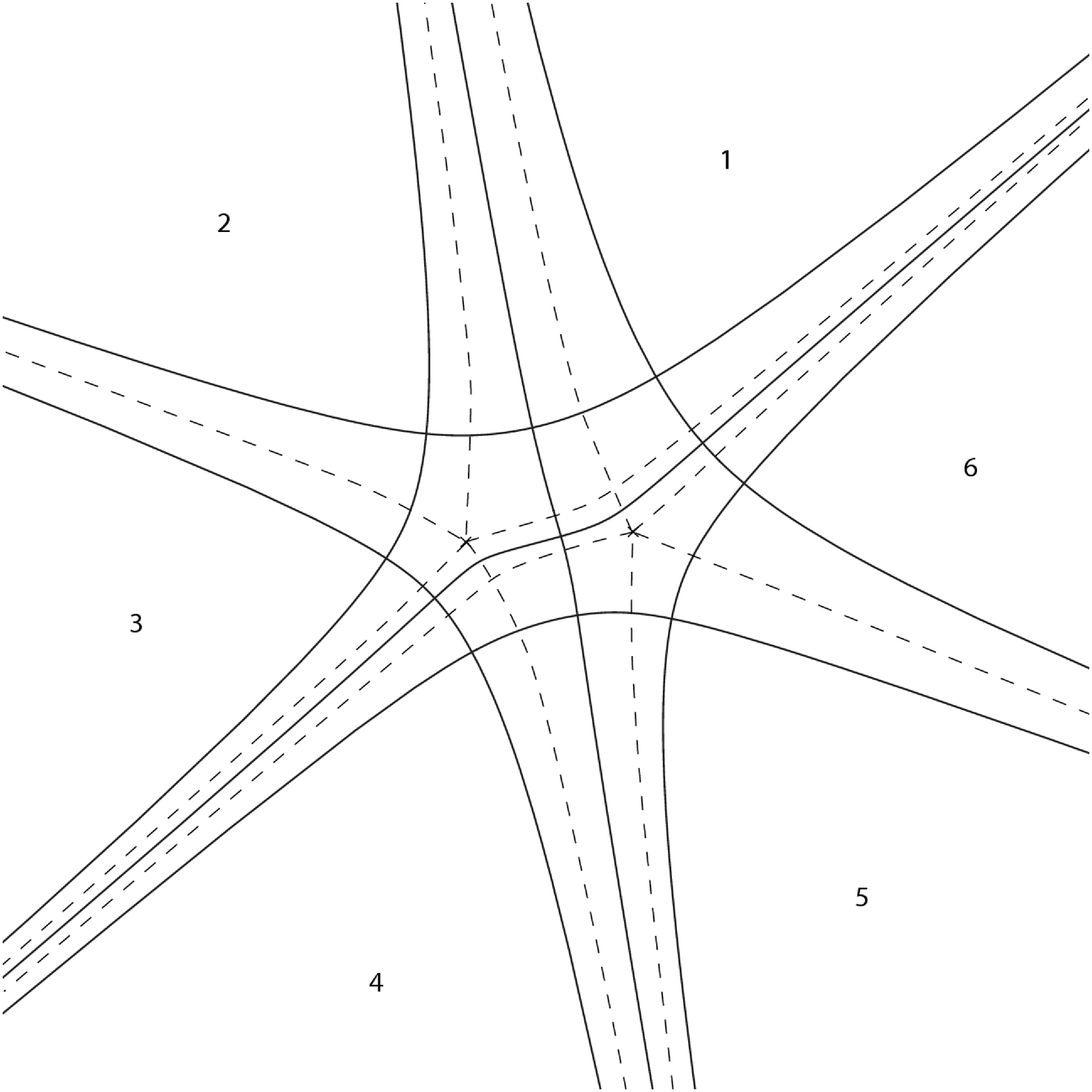}
 }

We see that only two out of three opposite pairs of rays are
joined by a WKB path, which allows the estimate of $( s_i
\wedge s_{i+1} \wedge s_{i+3} \wedge s_{i+4})$. See \wkbone .
As one varies
the value of $\varphi$ continuously, the configuration of paths
of constant phase also varies continuously. Jumps happen at
special values of $\varphi$, where the special WKB paths ending
on a zero of $P(z)$ collide with the other zero, see \wkbtwo .
Clearly, the special values of $\varphi$ coincide with the phase
of the integral of $x dz$ from a zero to the other along that
path. Since $x$ is the sum of two eigenvalues of the spinor connection,
we have  $x= { (1+i) \over \sqrt{2}  } (z^2-U)^{1/4}$, see \diag .
 We can rescale $z \to
U^{1/2} t$ to get $x dz = U^{3/4}   (1-t^2)^{1/4} dt$. The path
runs on real $t$ between $t=-1$ and $t=1$, the special values
$\varphi_i$ are the phases of the four roots of $U^{3/4}$. At
each $\varphi_i$ a single pair of opposite rays (which we can
identify without loss of generality with $r_i$ and $r_{i+3}$)
are joined by a WKB path of constant phase. The path exists in
the whole range $\varphi_{i+1} < \varphi < \varphi_{i-1}$, i.e.
$\varphi_{i}-\pi/2 < \varphi < \varphi_{i}+\pi/2$. Hence the
product $( s_i \wedge s_{i+1} \wedge s_{i+3} \wedge s_{i+4})$
has a simple WKB estimate for a range of $$\varphi_{i}- 3\pi/4
< \arg \zeta < \varphi_{i}+3\pi/4.$$


We can use this information to estimate the behavior of
$b_i[\zeta]$ as $\zeta \to 0$ and $\zeta \to \infty$ inside
this sector. We find  $\exp Z_i/\zeta$ at small $|\zeta|$ and $\exp \bar
Z_i \zeta$ at large $|\zeta|$ on the whole sector $\varphi_{i}-
3\pi/4 < \arg \zeta < \varphi_{i}+3\pi/4$ for an appropriate
function $Z_i[U]$.  Patiently assembling the WKB phase
integrals for all the pieces of the cross-ratio, $Z_i$  is
\eqn\periodZ{
Z = U^{3/4}     \int_{ -1}^1 dt
(1-t^2)^{1/4} dt = { \sqrt{   \pi } \Gamma({1 \over 4 } ) \over 3 \Gamma({3 \over 4 } ) } U^{3/4}
}
The different $Z_j$ corresponds to the four roots of $U^{3/4}$.
The specific root of $U^{3/4}$ is exactly the one with phase
$\varphi_j$, hence $b_j$ is exponentially large in the half
plane  $\varphi_{j}-\pi/2 < \varphi < \varphi_{j}+\pi/2$. The
asymptotic behavior of $b_j$ changes beyond the lines at $
\varphi_{j}\pm 3\pi/4$. We can check this explicitly using the
relation \relabmu . For example, we can write $$b_j = (\mu
+\mu^{-1}+b_{j-1} +b_{j-2})/( b_{j-1} b_{j-2}-1).$$ As we cross
the ray at $ \varphi_{j}+ 3\pi/4$, $b_{j-1}$ and $b_{j-2}$
switch dominance, and the asymptotic value of the ratio passes
from $b_j \sim b_{j-2}^{-1} \sim \exp[ Z_j/ \zeta]$ to $b_j
\sim b_{j-1}^{-1} \sim \exp[-Z_{j-1}/ \zeta] \sim \exp[ - i
Z_j/\zeta] $. On the other side, beyond the ray at $
\varphi_{j}- 3\pi/4$ we get $b_j \sim b_{j+1}^{-1} \sim \exp[
-Z_{j+1}/ \zeta] \sim \exp[ i Z_j/\zeta]$.

 \ifig\regionszeta{ We are displaying the $\zeta$ plane for a particular value of the phase of $U$,
 chosen so that along the positive real
 axis, $b_1[\zeta]$ becomes largest. Along the positive imaginary axis,   $b_3[\zeta]$ is the largest, and
 so on. Due to to the fact that the $b_j[\zeta e^{2 \pi i} ] = b_{j+4}[\zeta] = b_{j+1}[\zeta] $ we have
 a branch cut, denoted by the doted line where we switch the indexing of the $b_j$. We have indicated
 what each $b_j$ changes into. As we change the phase of $U$ the lines where $b_i$ are the biggest rotate
 rigidly.}
{\epsfxsize2in\epsfbox{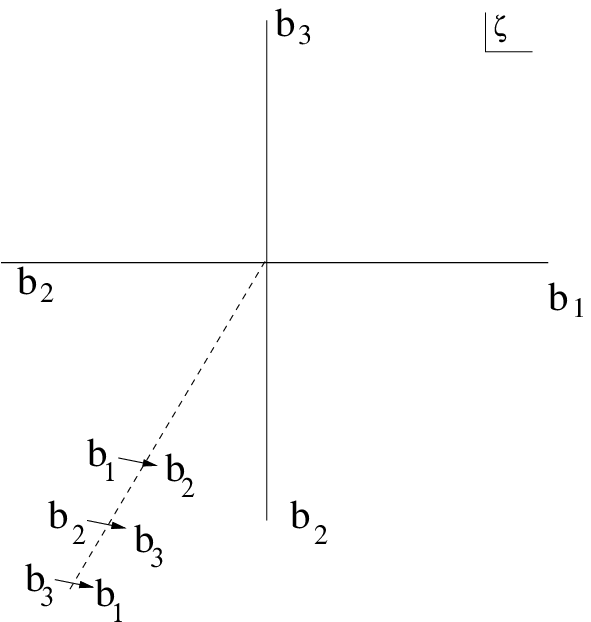}}

This is all the information we will need from the WKB analysis.

\subsec{A Riemann-Hilbert problem}

In this subsection we set up a Riemann-Hilbert problem. For this we need to combine the
fact that the $b_j[\zeta]$ are meromorphic in $\zeta$ (except for $\zeta =0, \infty$) with
the fact that they have different asymptotic behaviors as $\zeta \to 0, \infty$ depending on
the phase of $\zeta$. In other words, the $b_i[\zeta]$ display Stokes phenomenon, but now in
  $\zeta$.

In order to set up this Riemann-Hilbert problem, we would like
to identify some functions ${\cal X}_a$ of the $b_i$ whose
asymptotic behavior at small and large $\zeta$ is uniform on
the whole $\zeta$ plane. More concretely, we would like to
define the ${\cal X}_a$ in such a way that they behave as
${\cal X}_a \sim e^{Z_a/\zeta}$ as $\zeta \to 0$, for all
phases of $\zeta$. Here $Z_a$ denotes the four eigenvalues of
the connection which go like $Z, -i Z, -Z, i Z$.
 Of course, due to the
Stokes phenomenon in $b_i$, the
 price to pay is that there are    discontinuities in the ${\cal X}_a[\zeta]$ along
 some  rays in the $\zeta$ plane. The Riemann-Hilbert problem is captured by the data of
such discontinuities. As we will show explicitly below, such
discontinuities can be extracted from the various  WKB
approximations discussed above. One may set up many slightly
different  Riemann-Hilbert  problems, based on slightly
different definitions of the ${\cal X}_a$. There is a canonical
choice though, inspired by the connection to ${\cal N}=2$
theories in four dimensions \GMNone , which gives
discontinuities well suited to formulate the final integral
equations we are aiming for.

It is instructive to first consider a non-ideal set of functions $\Upsilon_a$,
allowing for discontinuities at the rays $\varphi_i$ in order to keep the uniform asymptotics
 $\Upsilon_a \sim \exp Z_a/\zeta$ uniformly true everywhere. For instance, let us define
\eqn\upsildef{ \eqalign{
\Upsilon_1[\zeta] &=  b_1[\zeta] ~~~~~{\rm for} ~~~~  - { \pi \over 2 }  < \arg \zeta -\varphi_1< { \pi \over 2}
\cr
 \Upsilon_1[\zeta] & =   1/b_2[\zeta] ~~~~~{\rm for} ~~~~   {  \pi \over 2 }  < \arg \zeta -\varphi_1 < { 3 \pi \over 2 }
 }}
 where $\varphi_1$ is the phase of $Z_1$. We have
 similar expressions for $\Upsilon_2  $
 where we change $b_j \to b_{j+1}$ and $\varphi_1 \to \varphi_2 = \varphi_1 -\pi/2$ in the above expression.
 We then
 define $\Upsilon_{a+2} = \Upsilon_{a}^{-1}$.
 Now, we would like to show that $\Upsilon_a$ has uniform asymptotics. Let us start with $\Upsilon_1 $.
 For simplicity, let us set the phase of $Z_1$ to zero, we will restore it later.
 In the upper half $\zeta$ plane $b_3$ is very large and we can use \relabmu\ to write
 $ b_1 b_2 -1 = b_3^{-1} ( b_1 + b_2 + \mu + 1/\mu ) $.
 %
 %
As $b_3$ is exponentially larger at small $\zeta$ than the factor in parenthesis
around the boundary  $\arg \zeta = \varphi_1 + \pi/2$  (see \regionszeta ),
we see that
$ b_2 \sim 1/b_1 $. This shows that $1/b_2 \sim \exp[ Z_1/\zeta]$ near this line. But since the asymptotics of
$b_2$ is already given by the WKB approximation, we see that this continues to be true throughout the second sector in
\upsildef .
  The same is true for all the $\Upsilon_a$.

We aim to derive integral equations for the logarithms $\log \Upsilon_a$ from their discontinuities.
At this stage, the discontinuities  are rather ugly-looking.
For example,   at $\arg \zeta = \varphi_{1} + \pi/2$,
$\Upsilon_1$ goes from $b_1$ to $b_2^{-1}$, and the ratio of the two values is
a complicated function. This can be easily computed by using the relation \relabmu\ to write
 $$ \Upsilon_1(\zeta^+)/\Upsilon_1(\zeta^-) =
  b_1/b_2^{-1} = b_1 b_{2}  =   1+ b_{3}^{-1} (b_{1} + b_2  + \mu+\mu^{-1}).$$ Here $\zeta^\pm$ denotes arguments for
  $\zeta$ which are slightly bigger or smaller than the discontinuity line. It
cannot quite be written as a combination of $\Upsilon_a$ which
is continuous on this line.
 It is always convenient to write the discontinuities in terms of functions where are continuous on that line
 \GMNone .

\ifig\xdefinition{ Here we see the explicit definitions of
${\cal X}_1$ and ${\cal X}_3$
 in each angular sector of the $\zeta$ plane. (We have
set the phase of $Z_1$ to zero for simplicity). Of course the
$b_i$ are discontinuous only at the branch cut. The fact that
we define ${\cal X}_a$ differently in different sectors is the
source of the discontinuities. }
{\epsfxsize1.5in\epsfbox{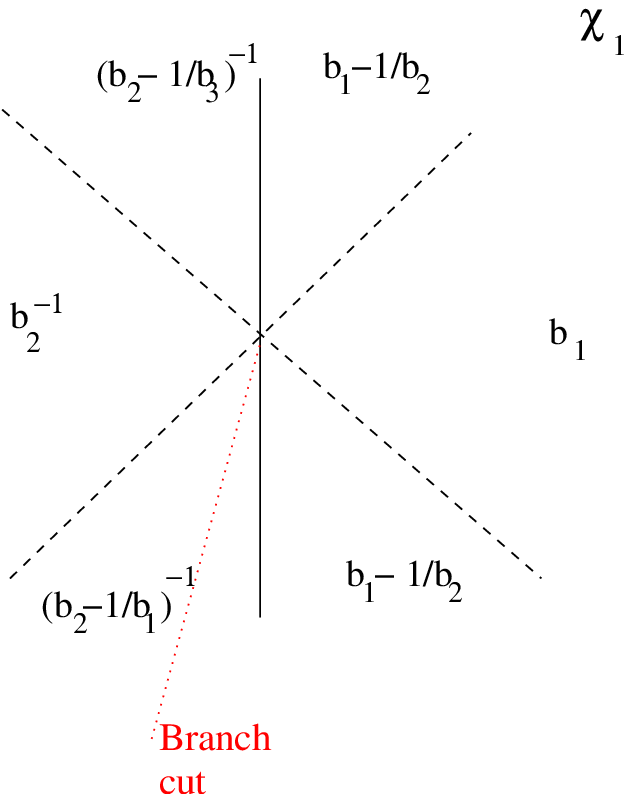} ~~~~~~~~~~~~~~
\epsfxsize1.7in\epsfbox{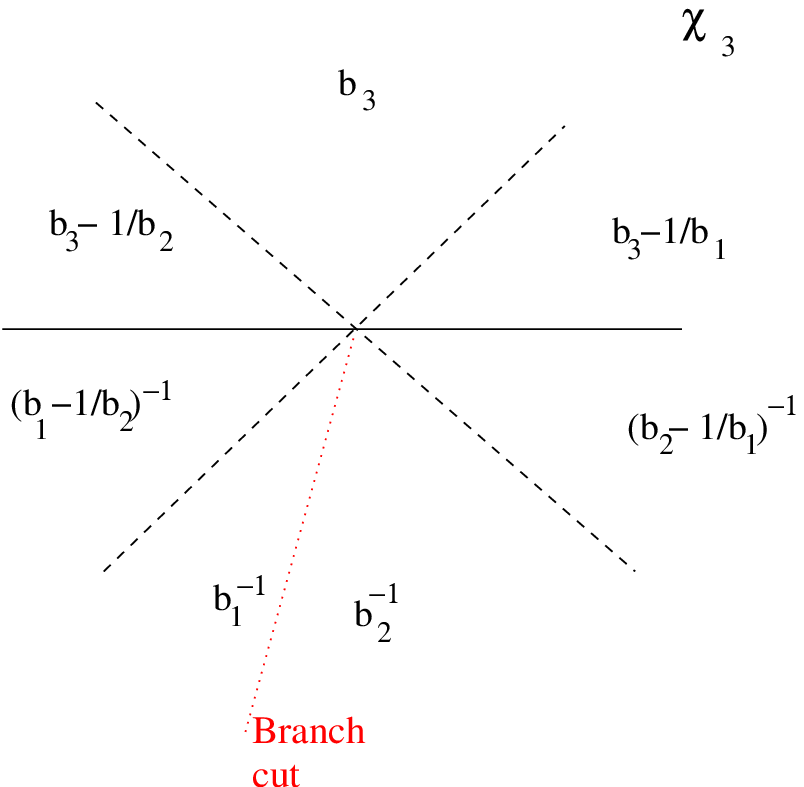} }

%

 \ifig\discontchi{ We see how ${\cal X}_1$ and ${\cal X}_3$ change as we cross the lines of discontinuity.
 We have defined $\Delta_a = (1 + \mu/{\cal X}_a)(1+ { 1 \over \mu {\cal X}_a} ) $, $\Delta_{a,a-1} = (1 + { 1 \over {\cal X}_a {\cal X}_{a-1} }) $.
 These are defined so that, for example, across the top vertical line we have ${\cal X}_1^+ = {\cal X}_1^- \Delta_3^{-1} $ where
 ${\cal X}_1^+$ denotes the value of ${\cal X}_1$ just after crossing the line (and ${\cal X}_1^-$ just
  before) if we move in an anti-clockwise
 fashion.   }
{\epsfxsize1.5in\epsfbox{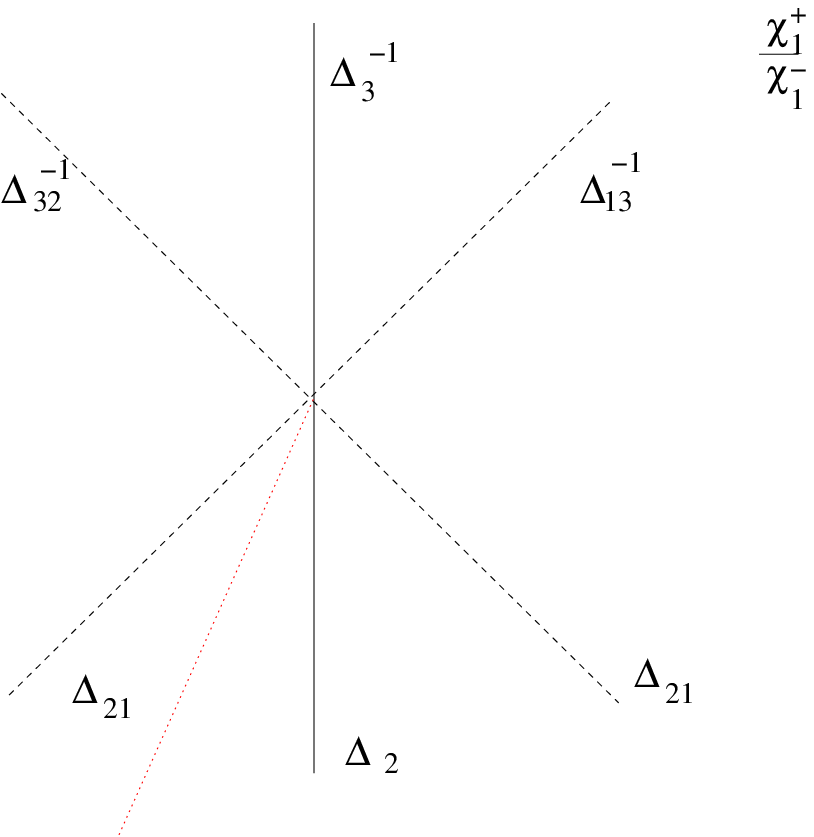} ~~~~~~~~~  \epsfxsize1.5in\epsfbox{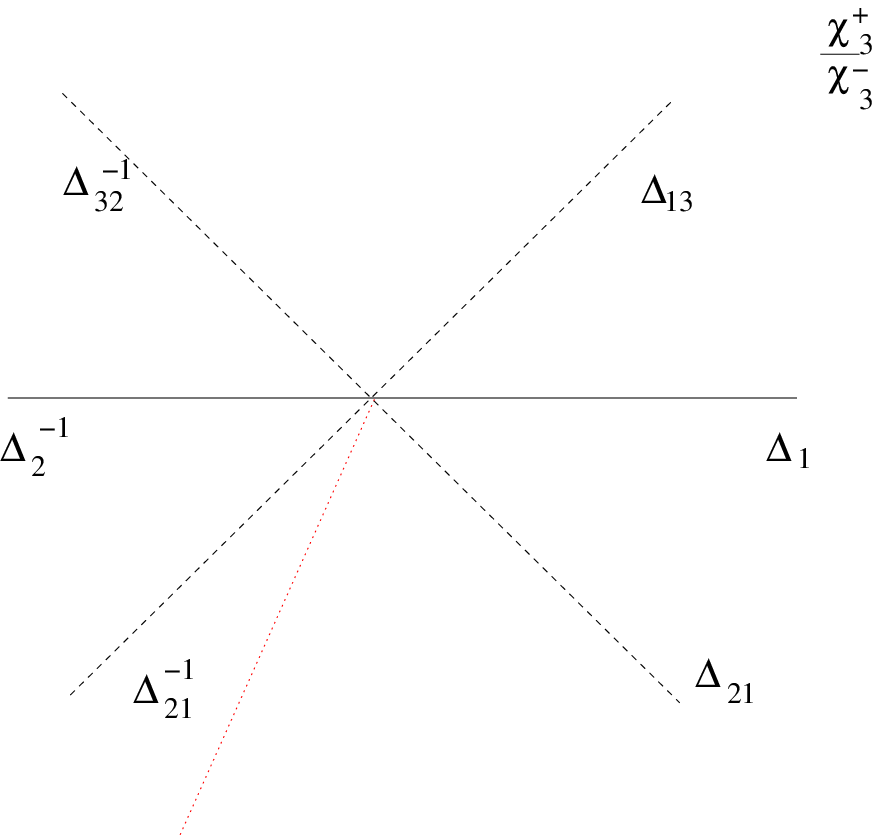}  }

 We now define new variables ${\cal X}_a$ which have simpler discontinuities, although more discontinuity lines.
Let us define ${\cal X}_a(\zeta)$ to coincide with $b_a$ only
for  $\varphi_{a}-\pi/4 < \arg \zeta < \varphi_{a}+\pi/4$. To
define it in the remaining regions we define ${\cal X}_a {\cal
X}_{a-2} = 1$ and also require ${\cal X}_a {\cal X}_{a-1} = b_a
b_{a-1}-1$ for $\varphi_{a} < \arg \zeta < \varphi_{a-1}$. The
important property of ${\cal X}_a {\cal X}_{a-1} $ is that its
analytic continuation has a simple WKB expression in a sector
of angle slightly bigger than $\pi$. This defines each ${\cal
X}_a$ on all the quadrants. In \xdefinition\ we see the
resulting definition of ${\cal X}_1$, ${\cal X}_3$. Compared to
the variables $\Upsilon_a$ we have now introduced extra
discontinuities. The correction is mild, but important. There
are now eight lines of discontinuity. At $\varphi_a$,  ${\cal
X}_a$ is continuous. For ${\cal X}_{a-1}$ we have
  the definitions ${\cal X}_{a-1} = b_{a-1}-1/b_a$ and
${\cal X}_{a+1} = b_{a+1}-1/b_a$ on the two sides of the
$\varphi_a$ line. We find
$${\cal X}_{a-1}(\zeta^+)/{\cal X}_{a-1}(\zeta^-) = {\cal X}_{a-1}(\zeta^+)
{\cal X}_{a+1}(\zeta^-) =
 (1 + \mu {\cal X}_a^{-1})(1 +  \mu^{-1} {\cal X}_a^{-1} ),$$ where we used \relabmu .
Hence the discontinuity of $\log {\cal X}_{a-1}$ is a function
of the continuous variable ${\cal X}_a$: $\log (1 + \mu {\cal
X}_a^{-1})(1 +  \mu^{-1} {\cal X}_a^{-1} )$. At $\varphi_a+
\pi/4$, ${\cal X}_a {\cal X}_{a-1}$ is continuous by design.
For ${\cal X}_a$,
 we need to compare the definitions
${\cal X}_a(\zeta^-) = b_a$ and ${\cal X}_a(\zeta^+) = b_a -
1/b_{a-1} $. We find $${\cal X}_a(\zeta^+)/{\cal X}_a(\zeta^-)
= 1 - { 1 \over b_a b_{a-1}}  = ( 1 + { 1 \over  {\cal X}_a
{\cal X}_{a-1}} )^{-1}.$$ Hence the discontinuity of $\log
x_{a}$ is a function of the continuous variable ${\cal X}_a
{\cal X}_{a-1}$ only: $- \log ( 1+ { 1 \over {\cal X}_a {\cal
X}_{a-1} } )$. The resulting discontinuities are summarized in
\discontchi .

We have our final Riemann-Hilbert problem: two functions $\log
{\cal X}_a[\zeta]$, with uniform asymptotic behavior at large
and small $\zeta$, and discontinuities which are functions of
the ${\cal X}_a$ themselves. The discontinuities take the
canonical form expected from the connection to ${\cal N}=2$
theories
 in four dimensions. Indeed, both the asymptotic behavior and the discontinuities of the ${\cal X}_a(\zeta)$
  are identical to the
ones for the cross-ratios of an $SU(2)$ Hitchin system, associated with the $N=4$ Argyres-Douglas theory
in the notation of \GMNtwo . The spectral curve for that system takes the form $x^2 = z^4 + U$,
which is identical to the curve for the hexagon if one takes $z\leftrightarrow x$. The transformation
 also relates the canonical differentials $x dz$, as $x dz + z dx$ is a total derivative.
 This relation guarantees the coincidence of the periods $Z_i$, but it does not imply the
  identity of the discontinuities. That is a non-trivial, and rather  unexpected fact.

\subsec{An integral equation}

Let us first assume that the functions ${\cal X}_a[\zeta]$ have
no poles or zeros in their region of definition. Along various
lines the ${\cal X}_a$ will have discontinuities that follow in
a simple way from the above definitions. In \discontchi\ we
have summarized the information about the discontinuities both
for ${\cal X}_1$ and ${\cal X}_3$.

Functions with such discontinuities can be obtained as the solution of integral equations of the
form
\eqn\gmneq{ \eqalign{ \log {\cal X}_a=& \frac{Z_a}{\zeta}+\bar
Z_a \zeta+ \cr
 &   ~~~~ + \sum_b n_{ab} K_{\ell_b}* \log \left((1+{ \mu \over {\cal X}_b}
  )(1+\frac{1}{\mu {\cal X}_b}) \right)+\sum_b m_{ab} K_{\tilde
\ell_b }* \log{(1+{ 1 \over {\cal X}_b {\cal X}_{b-1} } )} \cr
& ~~~  K_{\ell} * f= \frac{1}{4\pi i}\int_\ell \frac{d
\zeta'}{\zeta'}\frac{\zeta'+\zeta}{\zeta'-\zeta}f(\zeta')~, \cr
&~~~~~~\ell_j: \frac{Z_b}{\zeta'} \in  R^+,~~~~~~~~~~~~~\tilde
\ell_b: \frac{Z_b+Z_{b-1}}{\zeta'} = (1 +i ) { Z_b \over \zeta}
\in R^+
 }}
The Kernels are making sure that when $\zeta$ crosses the contour we pick a contribution which is precisely the
one we expect for the corresponding discontinuity.
The integers $n_{ab}$, $m_{ab}$ can be read off from figure \discontchi . Namely, for $a=1$ or $a=3$ they
 are   the exponents
of the $Y$s in  \discontchi . They
have the values
\eqn\inte{ \eqalign{n_{aa}=0 ~~~ n_{a, a-1} = -1~~~n_{a,a-2}=0
~~~ n_{a, a-3} = 1 \cr m_{aa}= -1 ~~~ m_{a, a-1} = -1
 ~~~m_{a,a-2}= 1 ~~~ m_{a, a-3} = 1}}

This equation can be further simplified by using the fact that
${\cal X}_{a-1}[i\zeta] = {\cal X}_a[\zeta]$ which is due to
the $Z_4$ symmetry of our Hitchin problem. This implies that
the values of ${\cal X}_a$ that are appearing along the rays
$\ell_a$ are all the same function. Let us set the phase of
$Z_1$ to zero for simplicity. We can then introduce two
functions $\epsilon$ and $\tilde \epsilon$ which parameterize
the values of the ${\cal X}_a$ and ${\cal X}_a {\cal X}_{a-1}$
at all the rays.
  Namely
  \eqn\definfu{
   e^{\epsilon(\theta)}  \equiv {\cal X}_1[ \zeta = e^{\theta}]~,~~~~~~~~~~~~~~~~
   e^{ \tilde \epsilon(\theta) } \equiv {\cal X}_1 {\cal X}_{3} [ \zeta = e^{i \pi/4} e^{\theta}]
   }
The functions ${\cal X}_1$ and ${\cal X}_3$ on the full $\zeta$
plane can be determined by the two functions of a real variable
$\epsilon $ and $\tilde \epsilon$ through the right hand side
of \gmneq. The integral equations for $\epsilon$ and $\tilde
\epsilon$ follow then simply by evaluating the left hand side
along the rays used to define $\epsilon $ and $\tilde \epsilon
$. We thus get \eqn\foldedequ{ \eqalign{ \epsilon(\theta) = & 2
|Z| \cosh \theta  + {\sqrt{2} \over \pi }  \int { d \theta'  }
{ \cosh (\theta-\theta')
 \over \cosh 2 (\theta-\theta') }
\log( 1 + e^{ -\tilde \epsilon} ) +
\cr &~~~~~~~~~
+ { 1 \over 2 \pi } \int  d\theta' {  1 \over  \cosh (\theta-\theta')} \log(1 +   \mu e^{ - \epsilon} )(1 + { e^{-\epsilon } \over \mu  }  )
\cr
 \tilde \epsilon(\theta)   = &  2  \sqrt{2}|Z| \cosh \theta + { 1  \over \pi} \int d\theta'
 {  1 \over   \cosh (\theta-\theta')}  \log( 1 + e^{-\tilde \epsilon } ) +
 \cr
 & ~~~~~~~~~~~+ { \sqrt{2} \over \pi }
 \int d\theta'  { \cosh (\theta-\theta') \over \cosh 2 (\theta-\theta') }
  \log(1 + { \mu e^{-\epsilon } } )(1 + { e^{-\epsilon}  \over \mu   }  )
 }}

 These are completely explicit integral equations for two functions. The integrals run between minus
 infinity and infinity. The functions $\epsilon,~\tilde \epsilon$ inside the integrals are evaluated at $\theta'$.
 These turn out to coincide with the TBA equations of the $A_3$ (or $Z_4$ symmetric)   integrable  theory
 introduced in \KoberleSG\ and further studied in \ZamolodchikovET . This
  theory has two particles of mass $m$ and one bound state with mass $\sqrt{2} m$.
 A very similar  equation was obtained in  the study of an $x^4$ anharmonic oscillator in \DoreyPT. Their
 equation can be obtained by the replacemnt
  $|Z|\cosh \theta \to e^\theta $ which is related to the UV, or conformal, limit of the $A_3$ theory.

 Once we find a solution to \foldedequ\  we can
 then write the cross ratios, $b_i$, which we can define through the right hand side of \gmneq\ and from these
 we can construct the $u_i$. Denoting $Z = |Z|e^{i \varphi}$ then note that $b_a[ Z , \zeta=1] =
 b_a[ |Z|, \zeta = e^{-i\varphi}]$. We can pick a definite octant, say $ -\pi/4 < \varphi < 0$. We can then compute
 all three $b_i$ on this octant from the two expressions
 \eqn\uthreed{ \eqalign{
   \log( { 1 \over u_2 } -1 )  = & \log {\cal X}_1 {\cal X}_3 = 2 \sqrt{2} \cos \hat \varphi  |Z| + {1 \over \pi }
   \int d\theta' {1  \over   \cosh( \theta' + i \hat \varphi ) } \log( 1 + e^{-\tilde \epsilon} ) +
   \cr
 &  +  { \sqrt{2 } \over \pi } \int d\theta' { \cosh(\theta'+ i  \hat \varphi) \over \cosh 2(\theta' + i \hat \varphi) }
    \log(1 + { e^{-\epsilon } \mu } )(1 + { e^{-\epsilon } \over \mu   }  )
\cr
 \log b_1 = & \log {\cal X}_1 =  2 |Z| \cos \varphi  + {\sqrt{2} \over \pi }
  \int { d \theta'  } { \cosh (i \varphi + \theta') \over \cosh 2 (i \varphi + \theta') }
\log( 1 +  e^{-\tilde \epsilon} ) +
\cr &~~~~~~~~~
+ { 1 \over 2 \pi } \int  d\theta' {  1 \over  \cosh (i \varphi + \theta')}
\log(1 + {  e^{-  \epsilon}\mu  } )(1 + {  e^{-  \epsilon} \over \mu   }  )
  }}
Where $\hat \varphi = \varphi + \pi/4$. From these and \relabmu\ we can determine
\eqn\otherbs{ \eqalign{
b_3 = &  { 1 \over b_1  u_2}
\cr
b_2 = & { b_1 + b_3 + \mu + { 1 \over \mu } \over b_1 b_3 -1}
}}
 Once we have defined the $b_i$ on this octant, we can use the symmetries of the problem to map them to other
 octants. In fact this single octant is a complete solution, since we can always permute the $u_i$ so that we are
 on this octant.

Notice that the equation we wrote is almost indifferent to the
signature.  The only difference is that in $(3,1)$ and $(1,3)$
signature $\mu$ is a phase, while in $(2,2)$ signature $\mu$ is
real. The main question one may ask is if the condition on the
absence of zeros and poles in the ${\cal X}_a$ is
self-consistent. There is a regime where the self-consistency
is manifest, namely $|U|\gg 1$. In that case the WKB
approximation is valid essentially for all $\zeta$, and ${\cal
X}_a \sim \exp \left[ \frac{Z_a}{\zeta}+\bar Z_a \zeta \right]$
are such that the logarithms on the right hand side of \gmneq\
are exponentially small along the contours of integration. In
the set up analyzed in \GMNone\ this corresponded to a certain
``large radius'' limit. Thus the problem becomes simple for
$|Z| \gg  1 $ and it is a good starting point for a numerical
solution of the equations. It corresponds to the low
temperature limit of the TBA system.

Let us see what large $|U|$, or large $|Z|$, corresponds to in our case.
Let us assume that we are in a regime where $b_1$ and $b_3$ are both
large and have a good WKB approximation. Writing $Z = |Z| e^{ i \varphi}$, this
corresponds to $ -\pi/2 < \varphi < 0$. We then have the expressions
\eqn\collinear{ \eqalign{
& \log b_1 \sim 2 |Z|\cos\varphi ~,~~~~~~ \log b_3 = 2 |Z| \cos( \varphi + {\pi \over 2 } )
\cr
&b_2 \sim  { 1 \over b_1 } + { 1 \over b_3 }  \to  u_1+ u_3 \sim 1 ~,~~~~~~~~{ u_1 \over u_3} = { b_3 \over b_1 }
\cr
& u_2 \sim { 1 \over b_1 b_3 }  \ll 1
}}

We see that if the phase $\varphi$ is generic then we have  $(u_1,u_2,u_3) = (1,0,0) $ or $(0,0,1)$.
On the other hand if the phase $\varphi$ approaches $-{\pi/4}$ as $|U| \to \infty$, then we can have a finite
value of $u_1/u_3$ and we get a segment $(1-u_3, 0, u_3)$, $0 < u_3 < 1$ which joints the two points mentioned above.
In other angular sectors  we   get other segments
with $u_i \to u_{i+1}$.

Thus, the large $|U|$, or large $|Z|$, limit corresponds to cross ratios that are on this triangle. We can move away
from this triangle by lowering the value of $|U|$. $U$, together with $\mu$ span the three dimensional space
of cross ratios $u_i$. Within this space there will be solutions that can be interpreted as living in various
signatures. In the next subsection we describe more precisely the region in the space of cross ratios which is
spanned by polygons living in space with various signatures.

\subsec{Kinematics, signature and the space of cross ratios}

In this section we   discuss the kinematics
relevant for the hexagonal Wilson loop (or scattering of six
particles), in several signatures. We consider an hexagon with
cusps at points $x_i$ in four dimensions, such that the
distance between two consecutive points is light-like. One can
construct three independent cross-ratios
\eqn\crossratios{
u_1=\frac{x_{13}^2 x_{46}^2}{x_{14}^2 x_{36}^2},~~~
u_2=\frac{x_{24}^2 x_{15}^2}{x_{25}^2 x_{14}^2},~~~
u_3=\frac{x_{35}^2 x_{26}^2}{x_{36}^2 x_{25}^2}
}
we would like to understand the range of such cross-ratios in
different signatures, namely, $(3,1)$, $(2,2)$ and $(1,3)$. It
is convenient go to a conformal frame where we have sent three
points, for instance $x_{6,1,5}$, to infinity. Furthermore, we set
the position of $x_3$ as the origin. To be more specific, we
denote four dimensional coordinates by $x=(x^+,x^-,x_\perp)$
and choose
\def\vecp{{ \vec p \,}}
\def\vecq{{ \vec q \,}}
\eqn\sixpoints{\eqalign{ x_3=&(0,0,0),~~~~~~x_2=(-1,x_2^-,\vec p),~~~~~~~~~x_4=(x_4^+,-1,\vec q)\cr
x_1=&x_2+(0,\Lambda,0),~~~x_5=x_4+(\Lambda,0,0),~~~x_6=(\Lambda,\Lambda,0)+....
}}
Since $x_2$ and $x_4$ are light-like vectors, we require
$x_2^-=-{\vec p \, }^2,~~x_4^+=-{\vec q \,}^2$. Here,  $\vecp$ and $\vecq$ are two vectors living in the
transverse space. This transverse space has various signatures. They can be used
to write the cross ratios.
\eqn\crossperp{\eqalign{u_1=\frac{1}{1-\vecq^2},~~~
u_2=\frac{1 + \vecq^2 \vecp^2  -2 \vec q . \vecp
 }{(1-\vec q^2)(1- \vecp^2)},~~~u_3=\frac{1}{1-\vecp^2},
}}
The regular hexagon embedded in $AdS_3$ corresponds to
$\vec q = \vec p=0$, with $u_i=1$. We are interested in
solutions continuously connected with the regular hexagon, for
which $u_i>0$. A negative sign for some    $u_i$  means that some of the distances between
cusps are becoming   timelike. In this paper we will discuss only the case where the $u_i \geq 0$ where distances
are not yet timelike.

 \ifig\uplane{ This is   the three dimensional space of cross ratios parameterized by $(u_1,u_2u_3)$.
 The lines belonging to the edges of the triangle correspond to  the large $|U|$ limit.
 These edges    correspond to various collinear limits.
  The vertices of the triangles correspond to soft limits.  }
{\epsfxsize2.0in\epsfbox{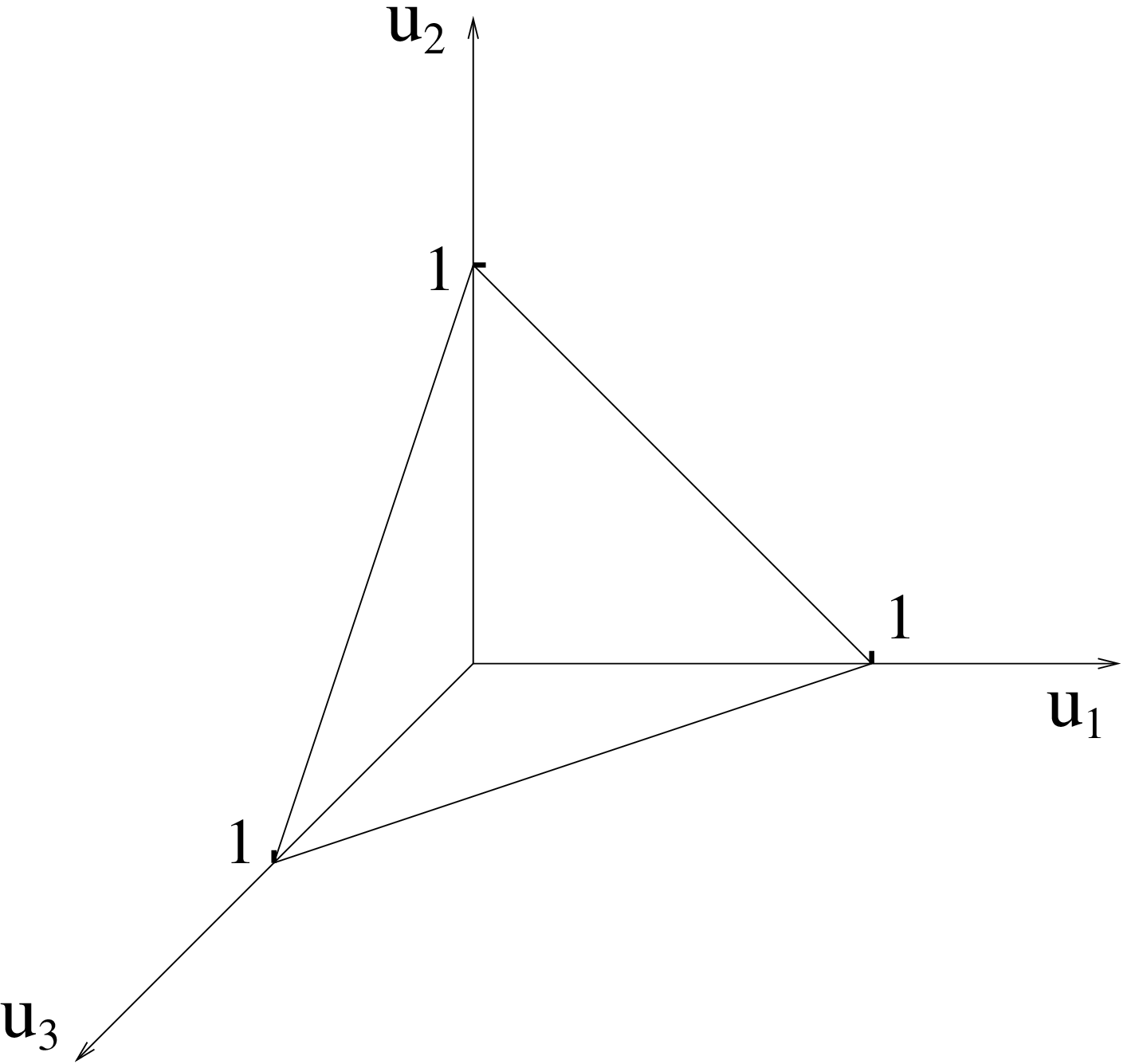}}

Let us discuss in more detail the physical meaning of the   large $|U|$
lines, for example $u_2 =0$, $u_1 + u_3 = 1$. The second
condition sets $\vecq^2\vecp^2=1$. Since, $u_i$ are positive, we see
that
$u_1$, $u_3$  are less than one. This means that
 $\vecq^2$ and $\vecp^2$ are
  negative. The   condition $u_2=0$ requires $\vecp \cdot \vecq=1$,
 which implies that $\vecp$ and $\vecq$
are collinear and point in opposite directions. We see then
that the four dimensional vectors $x_2$ and $x_4$ are collinear
as well. We conclude that the large $|U|$  regime is
actually the collinear limit. In fact, the fraction $z= { k^\mu_{23} \over k^\mu_{23} + k^\mu_{34} } =  u_{1}$,
where the $k_{i,i+1} = x_{i+1} - x_i$ are the spacetime momenta.
Thus, as we move along the segment $(1-u_3,0,u_3)$ we change this fraction which  is a free parameter in the
collinear limit.
Note that for $z$ to be different from   zero or infinity we
need to fine-tune the phase of $|U|$ as we move to large
values of $|U|$. However, in this regime the area, $A_{free}$ is not sensitive to this phase. In fact, we
will see that our remainder function remains constant in this limit.
Taking the large $|U|$ limit along a generic direction gives us a simple soft limit which simply eliminates
one of the momenta. Let's say that $u_3, u_1 \to 0$ in this limit, then we conclude that $\vecq^2 \to \infty$, which
after a rescaling , is making $x_2 \sim x_3  $. Thus,  we are sending the momentum $k_{23} \to 0$.

Note that the expressions \crossperp\ are valid for any
signature under consideration. Of course, the sign of
$\vecq^2$  depends on the signature. For instance, for
$(3,1)$ signature, $\vecq^2 \ge 0$ and hence $u_i>1$, while
for $(1,3)$ signature, $\vecq^2<1$ and $u_i<1$.
However, the converse is not true. The reason is simple, with
$(2,2)$ signature we can have either sign for $\vec q$.

To get more information we look at the sign of
 $\vecp^2 \vecq^2 - (\vecp.\vecq)^2$,  which is positive in $(3,1)$ and $(1,3)$, but is
negative in $(2,2)$. This   can be
translated into a condition on $\mu$.  Combining \relabmu\ with \crossratios\  we
can write $\mu$ as
\eqn\muvsu{\eqalign{\mu+\mu^{-1}=\frac{1-u_1-u_2-u_3}{\sqrt{u_1
u_2 u_3}}}}
Using the above formulas we can write
\eqn\muvsu{\eqalign{ - \frac{1}{4}(\mu-\mu^{-1})^2=\frac{u_1
u_3}{u_2} \left(   \vecp^2 \vecq^2- (\vecp.\vecq)^2  \right)}}

 \ifig\uspace{  Here we see the $AdS_4$ surface corresponding to $\mu=\pm 1$ in the space of cross
 ratios parametrized by $(u_1,u_2,u_3)$. Points on this surface corresponds to cross ratios that arise
  when we have a polygonal contour in a three dimensional space. The surface pinches at $u_i=1$.
  The region
 inside the ``bag'' in the $u_i <1$ region correspond to configurations in
  $(1,3)$ signature. The region outside corresponds to surfaces in  $(2,2)$ signature. Finally the
  region inside the surface for $u_i > 1$ corresponds to ordinary $(3,1)$ signature. We show the same surface
  from two different viewing directions.  }
{\epsfxsize4.0in\epsfbox{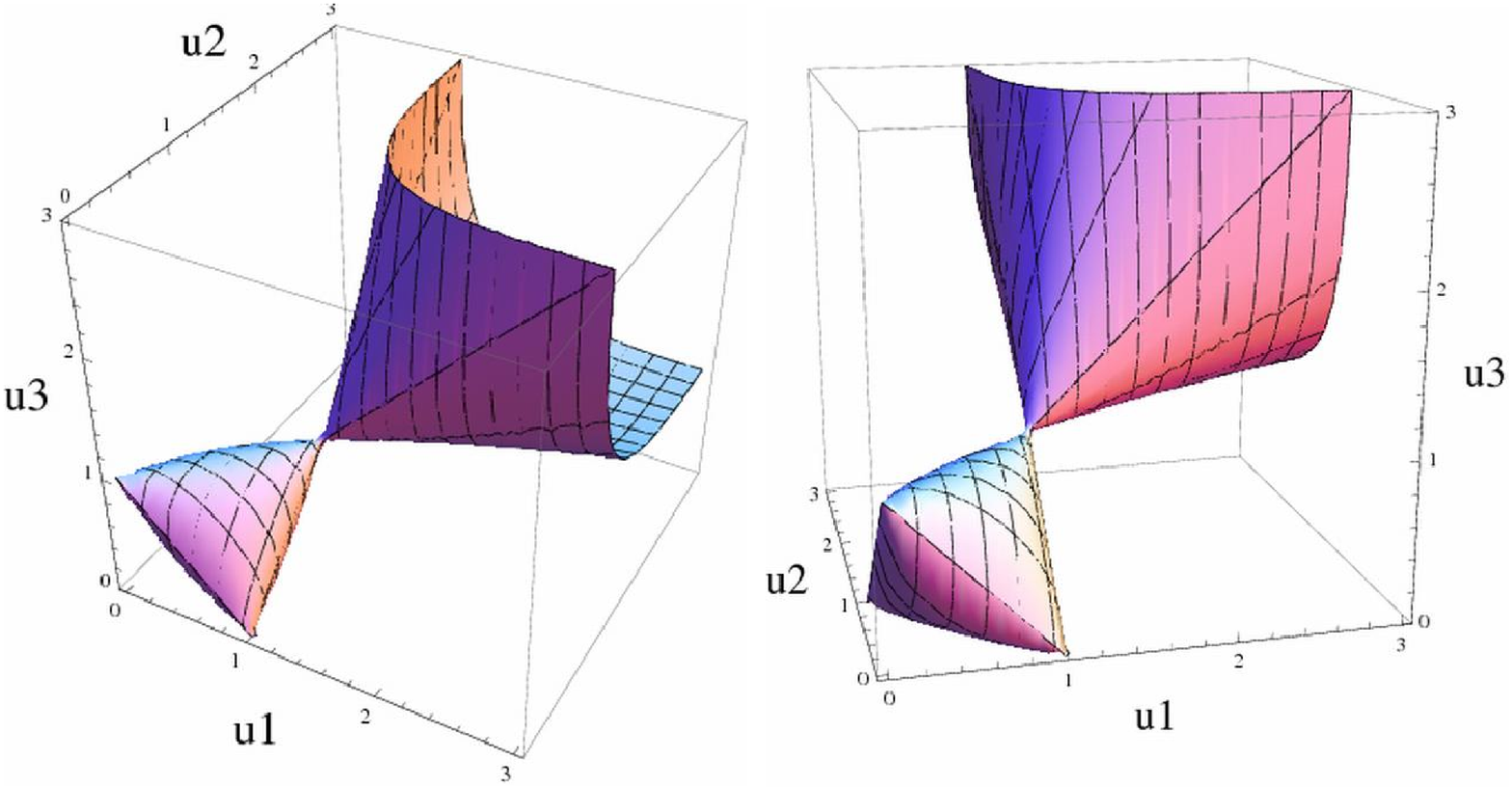}}

We conclude that  $\mu$ is a phase for $(3,1)$ and $(1,3)$ but
$\mu$ is real for $(2,2)$. Thus the boundary of the $(2,2)$
signature lies at $\mu = \pm 1$. These values coincide with the
ones corresponding to surfaces that can be embedded in $AdS_4$.
This is not surprising, we need to make one spacetime
coordinate vanish if we are going to change its signature.  We
plot this surface in  \uspace . The surface pinches at
$u_1=u_2=u_3=1$. This corresponds to the regular hexagon
embedded in $AdS_3$, all three signatures meet here. The
$AdS_4$ surface has two sheets emanating from the large $|U|$
triangle. One sheet has $\mu=-1$, it starts at the large $|U|$
triangle, goes through the pinch at $u_i=1$ and then continues
through to $u_i
> 1$. The other has $\mu =1$ and it stays at $u_i < 1$. It has
the topology of a disk and it contains an interesting polygon
with $u_1=u_2=u_3 ={ 1 \over 4 }$. This is another kind of
regular hexagon which has a simple picture in $(2,2)$ signature
and it is discussed further in appendix B. These two branches
together define the full $AdS_4$ surface. The region inside the
surface for $u_i<1$ has $(1,3)$ signature, the region outside
has $(2,2)$ signature and the region inside for $u_i>1$
corresponds to ordinary $(3,1)$ signature.

It is interesting to look at the surfaces of constant $\mu$
which end on the large radius triangle. We have already discussed the $\mu =\pm 1$ surfaces.
If $\mu$ is a phase
 the surface has the topology of a disk  which sits in
the $(1,3)$ signature region (one space three times). We expect that as we solve the
integral equation for decreasing values of $|U|$ starting from
the large radius region, we will get a nice map from the $U$
plane covering the disk once. It turns out
 that $U=0$ corresponds to the line $u_1=u_2=u_3$.
We move along the line by changing $\mu$.  In section five we find the
exact solution corresponding to this case.

 If we take $\mu$
to be real and negative, the surface  has the topology of a
cylinder in the region $u_i>0$, and has some pieces in the
regions with some $u_i$ negative. We expect that as one moves
towards small $|U|$, it will cover the cylinder, and then at
some point one $b_a$ will become negative, making two $u_i$
negative. In order to reach the $(3,1)$ signature region, a
possible strategy is to solve the integral equation starting
from the large radius region, with $\mu$ real and slightly
smaller than $-1$, follow it until we get to the $u_i>1$
region, and then cross into the $3,1$ signature by changing
$\mu$ back to a phase. Some interesting phenomena can occur in
the process, if the functions ${\cal X}_a[\zeta]$ acquire
zeroes or poles. This is a common (even desirable at times)
occurrence in solving TBA-like equations \refs{\zampoles,\dtpoles}. In appendix C we
discuss this a bit further.

\newsec{Area and TBA free energy}

In this section we will show how to compute a certain regularized area as the free energy of
the TBA system for the hexagon case.

The area is given by $ 2 \int e^{\alpha}$ which is divergent. This should be regularized in a physical way.
For the time being let us just define a regularized version by subtracting the behavior of $e^{\alpha}$ far away.
We have seen that $\hat \alpha$ \alphahat\ goes to zero at infinity.
This means that for large $z$, $e^{\alpha } \sim |P|^{1/2}$.
Furthermore, in the case of the hexagon, we also have that $|P|^{1/2} \sim |z|$ at infinity.
Thus we find it convenient to introduce the following definitions
\eqn\derif{\eqalign{
A_{reg} = &  A_{periods} +A_{free} = 2 \int d^2 ( e^{\alpha } - |z| )
\cr
A_{periods} \equiv  & 2 \int d^2z ( |P(z)|^{1/2} - |z| ) = |Z|^2
\cr
A_{free} \equiv & 2 \int d^2 z ( e^{\alpha} - |P|^{1/2} )
}}
where we have evaluated explicitly $A_{periods}$ in terms of \periodZ . Note that this is a convergent integral in
our case since $P(z) = z^2 -U$.

It is convenient to write $A_{reg}$ in terms of the ``Higgs" fields of the Hitchin system using
$$Tr \Phi_z \Phi_{\bar z} = e^\alpha - v^I \bar v_I e^{-\alpha} = 2 e^\alpha - \partial \bar \partial \alpha $$
Hence,  up to a total derivative which integrates to $\pi N$, we
can define (the trace is in the spinor representation)
$$A_{free} = \int d^2z
 \left[ Tr \Phi_z \Phi_{\bar z} -2  |z| \right] d^2z$$

The formal expression $\int  Tr \Phi_z \Phi_{\bar z}$
has a nice interpretation in the context of the Hitchin system. Recall
there is a natural Poisson bracket on functionals of $A,\Phi$
\eqn\poisson{\{U,V\} = \frac{i}{2} \int d^2z Tr\left[
\frac{\delta U}{\delta\Phi_z}
  \frac{\delta V}{\delta\Phi_{\bar z}}+\frac{\delta U}{\delta
A_z}   \frac{\delta V}{\delta A_{\bar z} } -( z \leftrightarrow \bar z ) \right]  }
and the flow generated by $\int  Tr \Phi_z \Phi_{\bar z}$
rotates $\Phi_z$ and $\Phi_{\bar z}$ in opposite directions as
$\delta \Phi_z = \frac{i}{2} \Phi_z$, $\delta \Phi_{\bar z} =
-\frac{i}{2} \Phi_{\bar z}$

The Poisson bracket is gauge invariant, and it actually
descends to a nice Poisson bracket on functionals of the flat
sections of the spectral connection, such as the $b_i$. On such
functionals, it reduces to
\eqn\poisson{\{U,V\} = i \int d^2z Tr\left[ \frac{\delta
U}{\delta {\cal A}_z}   \frac{\delta V}{\delta {\cal A}_{\bar
z} } -( z \leftrightarrow \bar z ) \right]  }

It is straightforward, but tedious, to compute the Poisson
bracket $\{b_i, b_{i-1}\} = b_i b_{i-1} -1$ along the lines of
appendix B of \GMNtwo .
 There is a  self-consistency check, one can  see that $\{b_i[\zeta], b_{i-1}[\zeta]\}$ had to be
some polynomial in the $b_i[\zeta]$ with constant coefficients
independent of $\zeta$. The right hand side $b_i b_{i-1} -1$ is
the only possibility consistent with the asymptotics of $b_i$
at small $\zeta$ along all the rays in the $\zeta$ plane. Also,
$\{b_i, \mu\}=0$. Another consistency check is that this Poisson bracket
is consistent with the relation \relabmu . Taking the Poisson bracket of
the two sides of \relabmu\ with $b_1$ we get
  $$\{b_1, b_1 b_2 b_3\} = b_1 b_2 (b_1b_3-1) - b_1(b_1b_2-1) b_3   = b_1 b_3 - b_1 b_2 = \{b_1, b_1 + b_2 + b_3 +
\mu + \mu^{-1}\}.$$ Finally, the neatest result: $\{{\cal X}_a,
{\cal X}_{a-1}\} = {\cal X}_a {\cal X}_{a-1}$ identically in
all sectors. In particular, the discontinuities in the ${\cal
X}_a[\zeta]$ preserve the Poisson bracket, and the
corresponding symplectic form
$$\omega = d\log {\cal X}_1 d\log {\cal X}_3 = d \log {\cal X}_1 d\log({\cal X}_1 {\cal X}_3) ~.$$
The asymptotic behavior of ${\cal X}_a$ at $\zeta=0, \infty$
shows that $\omega$ holomorphic in $\zeta$ everywhere, and
hence is $\zeta$ independent.

A rotation of $\Phi_z$ and $\Phi_{\bar z}$ in opposite
directions can be absorbed by a phase rotation of $\zeta$. Can
we identify a function $F$ which can generate this phase
rotation directly? It should satisfy the equation \eqn\fordf{
 dF = \frac{i}{2} \left[  d\log {\cal X}_1~  \zeta \partial_\zeta \log {\cal X}_3 - d\log {\cal X}_3 ~ \zeta \partial_\zeta
\log {\cal X}_1 \right] ~.}
 $F$ , as $\omega$, will be $\zeta$ independent.
This allows us to compute it at, say, very small $\zeta$. From
the integral equation, we can extract the behavior of $\log
x_a$ at small $\zeta$ \eqn\formeq{ \eqalign{ \log {\cal X}_a
\sim & \frac{Z_a}{\zeta}+  K_a + \zeta \left(\bar Z_a  +
K'_a\right)
}}
%
%
Here $K_a$ and $K'_a$ are simply the value at $\zeta=0$ and the first derivative at $\zeta=0$ of the
convolution terms in the integral equation. Then
$$
\zeta \partial_\zeta \log {\cal X}_a \sim -\frac{Z_a}{\zeta}+\zeta (\bar Z_a+ K'_a )
$$
and $dF$ can be written as $dF = dF_1 + dF_2$ where
\eqn\fonec{
 dF_1 =
  \frac{i}{2} d(Z_1 \bar Z_3 - \bar Z_1 Z_3) = d |Z|^2
 }
 where we used that $Z_1 = Z$ and $Z_3 = i Z$.
This is the dominant contribution in the large $|Z|$ limit.
 The second term is
 \eqn\seondfr{
 dF_2 = \frac{i}{2} d\left(Z_1  K'_3  - Z_3 K'_1\right)
 }
  $F_2$ can be rearranged a
little, to the form \eqn\resultff{ F_2 =  \frac{1}{ 4 \pi
}\int_{\ell_a} \frac{d \zeta'}{\zeta'}\frac{Z_a}{\zeta'} \log
\left((1+ { \mu \over {\cal X}_a}
 )(1+\frac{1}{\mu {\cal X}_a}) \right) + \frac{1}{4\pi  }\int_{\tilde \ell_a} \frac{d
\zeta'}{\zeta'}\frac{Z_a+Z_{a-1}}{\zeta'} \log{(1+{ 1 \over
{\cal X}_a {\cal X}_{a-1}}) } } A similar calculation at large
$\zeta$ produces a similar result, but with
$\frac{Z_a}{\zeta'}$ replaced by $\bar Z_a \zeta'$. The two
expressions must be identical, and we can take the average
\eqn\tbafree{ \eqalign{ F_2 = &   \frac{1}{8\pi  }\int_{\ell_a}
\frac{d \zeta'}{\zeta'}\left(\frac{Z_a}{\zeta'} +\bar Z_a
\zeta'\right) \log \left((1+ { \mu \over {\cal X}_a}
 )(1+\frac{1}{\mu {\cal X}_a}) \right) +
\cr
  & +
\frac{1}{8\pi  }\int_{\tilde \ell_a} \frac{d
\zeta'}{\zeta'}\left(\frac{Z_a+Z_{a-1}}{\zeta'} + (\bar Z_a
+\bar Z_{a-1}) \zeta'\right) \log{(1+{ 1 \over {\cal X}_a {\cal
X}_{a-1}}) } }}

Notice that   $F_2$ goes to zero in the large
$|Z|$ limit. In this limit the dominant piece is $F_1$, which reproduces $A_{periods}$.
Note that there could have been a $\mu$ dependent integration constant in $F$. However, this
is not the case because we know from the properties of the solutions that $A_{free}$ as define in
\derif\ goes to a constant for large $|Z|$.
Thus the conclusion is that, up to a constant, we have that
$A_{\rm free}= F_2$.
Finally, let us write $A_{\rm free}$ in terms of the solutions
of the integral equation \uthreed\

\eqn\finalafree{ A_{\rm
free} = {  1   \over 2 \pi }  \int_{-\infty}^\infty d\theta  2
|Z| \cosh \theta
  \log{ ( 1 + {e^{-\epsilon } \mu   } )(1 + { e^{-\epsilon } \over \mu  } ) }
+  2 \sqrt{2}  |Z| \cosh \theta \log{ (1 +e^{- \tilde \epsilon
} ) } }
This has precisely the form of the free energy (up to
an overall sign) of the TBA system in \foldedequ .

In order to get the full physical answer of the problem we need
to introduce a more physical regulator. The physical regulator
consists in cutting off the integral when the radial $AdS$
coordinate gets too close to the boundary. If $z$ is the radial
$AdS_5$ coordinate we say that $z > \epsilon_c$. Here
$\epsilon_c$ is a short distance
 UV cutoff from
the Wilson loop perspective or an IR scale from the amplitude point of view. Thus we are cutting off the $z$ integral
in a way that depends on the solution.
However, this dependence on the solution can be expressed in terms of the distances between the cusps (it depends
on distances, not only on cross ratios).
We explain this in detail in the appendix.
We find that
\eqn\definit{
A = A_{\rm div} + A_{\rm BDS-like} + A_{\rm periods} + A_{\rm free} + {\rm constant}
 }
 where $A_{\rm div}$ contains the $\epsilon_c$ dependent divergent terms. $A_{\rm BDS-like}$ is finite but it depends
 on physical distances and not just the cross ratios.
  Its explicit form is given in appendix A. It obeys the anomalous conformal symmetry Ward
 identities derived in \DrummondAU . It has become conventional to define a remainder function
 by subtracting  the specific function $A_{BDS}$ given in \BDS ,
  which has the functional form of the one loop result. Thus one defines a remainder function
   via
 \eqn\definr{\eqalign{
 R \equiv & -( A - A_{div} - A_{BDS}) = -( A_{BDS-like} - A_{BDS} ) -  A_{\rm periods} - A_{\rm free} + {\rm constant}
\cr
R  = &    \sum_{i=1}^3
\left( { 1 \over 8} \log^2 u_i + { 1 \over 4 } Li_2(1-u_i) \right)
 -  |Z|^2 - A_{\rm free} + {\rm constant}
}}

In the large $|Z|$ limit, which is the collinear limit, one can check using \collinear\ that the first term in
$R$,
involving $u_i$, cancels the $ |Z|^2$ term, up to a constant.  Thus the remainder function goes to
a constant in the large $|Z|$ limit, since $A_{free} \to 0$ in this limit.
As argued in \BDS , the $A_{BDS}$ term gives the expected result in the collinear limit. Thus, the remainder
function should be finite. This is a simple check of our results.

Let us summarize the final result.
The objective is to find the remainder function $R$ as a function of the spacetime cross ratios $u_i$.
For this purpose, we compute $\mu$ from the $u_i$ via \muvsu\ and we introduce an auxiliary
 complex parameter $Z$.
We then solve the integral equations \foldedequ\ and determine the functions $\epsilon(\theta), ~\tilde \epsilon(\theta)$.
These functions depend on the parameters $|Z|$ and $\mu$. We then solve for the spacetime
cross ratios $u_i$ (or equivalently $b_i $ \crossratios ) using \uthreed , \otherbs . {\it In principle}, one
would be able to invert that relation and
compute $Z$ and $\mu$ in terms of the spacetime cross ratios $u_i$.
Finally, one can compute $A_{free}$ as a function of $|Z|$ and $\mu $ using \finalafree\ and insert the result
in \definr . In practice one can numerically compute the functions $\epsilon,~\tilde \epsilon$, the $u_i$ and
the area as a function of $|Z|$ and $\mu$.

\newsec{Analitic solution of the integral equation}

In the following we will compute the area from the integral
equation in the particular limit $U \rightarrow 0$. This
corresponds to the special kinematical configurations with
$u_1=u_2=u_3=u$.  From the point of view of the TBA equations
this limit corresponds to a high temperature/conformal limit
and has been analyzed for instance in \refs{\FendleyXN,
\FendleyVE} (see also \refs{\ZamolodchikovCF,\KirillovZZ,\MartinsHW}) to
which we refer the reader for the details. We start by
recalling the general result in these references and then we
focus on the specific case at hand. Consider the standard form
of TBA equations
\eqn\TBAstandard{\eqalign{ \epsilon_a(t) = m_a \cosh t-\sum_b
\int \frac{dt'}{2\pi} K_{ab}(t-t')\log(1+\lambda_b
e^{-\epsilon_b(t')})}}
where the indices $a,b$ run over different species, $m_a$
denote their masses and $\lambda_a$ their chemical potential.
Then, the free  energy of the system can be written as
\eqn\groundenergy{F=-\sum_a \frac{m_a}{2\pi} \int dt \cosh t
\log(1+\lambda_a e^{-\epsilon_a(t)})}
In turns our that the free energy can be exactly computed in
this limit \refs{\FendleyXN,\FendleyVE
,\ZamolodchikovCF,\KirillovZZ}
\eqn\uvlimit{F(m_a \rightarrow 0) = -\frac{1}{\pi} \sum {\cal
L}_{\lambda_a}(x_a),~~~{\cal L}_\lambda(x)=\frac{1}{2}
\left(\log x \log(1+\frac{\lambda}{x})-2Li_2\left(
-\frac{\lambda}{x}\right) \right) }
The parameters $x_a$ are the solution of the system of
equations
\eqn\xaeqn{x_a=\prod_b \left(1+\frac{\lambda_b}{x_b}
\right)^{N_{ab}},~~~N_{ab}=-\frac{1}{2\pi}\int_{-\infty}^\infty
dt K_{ab}(t) }
Note that the final result does not depend much on the details
of the Kernels (only their integrals)

\subsec{Particular case at hand}

The integral equations \foldedequ\ are exactly in the standard
form \TBAstandard\ , so that we can apply the above results
straightforwardly. We have three species of particles, that we
call $\epsilon_1$, $\epsilon_2$ and $\tilde \epsilon$ (with
$\epsilon_1=\epsilon_2=\epsilon$). The mass of each excitation
is \foot{Quite interestingly, note that the mass of these three
excitations agree (up to the proportionality factor $|Z|$) with
the masses of the three fluctuations around the  classical string
 solution describing a single cusp (or four cusps) in  $AdS_5$, see \AldayMF\ .}
\eqn\masses{m_{\tilde \epsilon} =\sqrt{2} |Z|,~~~
m_{\epsilon_1}=m_{\epsilon_2}=|Z|}
The chemical potential for each species is $\lambda_{\tilde
\epsilon}=1$, $\lambda_{\epsilon_1}=\mu$ and
$\lambda_{\epsilon_2}=1/\mu$. From the particular kernels in our equation
\foldedequ\
  we can easily compute $N_{ab}$.  We find
\eqn\Nab{\eqalign{N_{\tilde \epsilon \tilde
\epsilon}=1,~~~N_{\tilde \epsilon \epsilon_i}=N_{\epsilon_i
\tilde \epsilon}=1,~~~N_{\epsilon_i \epsilon_j}=\frac{1}{2}}}
Pluggin this in \xaeqn ,  we can solve for $x_{\tilde \epsilon}$ and
$x_\epsilon=x_{\epsilon_1}=x_{\epsilon_2}$, we find
\eqn\xasol{e^{\tilde \epsilon} = x_{\tilde
\epsilon}=1+\mu^{2/3}+\mu^{-2/3},~~~~~~~~~~~~e^\epsilon = x_\epsilon=\mu^{1/3}+\mu^{-1/3}}
The free energy is simply
\eqn\freenergy{F=-\frac{1}{\pi}\left({\cal L}_1(x_{\tilde
\epsilon})+{\cal L}_\mu (x_\epsilon)+{\cal
L}_{1/\mu}(x_\epsilon) \right)}
Using the explicit expression for ${\cal L}$ and identities
involving di-logarithms one can show the very simple result
\eqn\freenergysimple{-F=A_{free}=\frac{\pi}{6}-\frac{1}{3
\pi}\phi^2,~~~~~~\mu=e^{i \phi}}
Several comments are in order. First, is it possible to solve
numerically the integral equation for small values of $Z$ (for
instance $Z=0.01$). After several iterations we find numerical
results that are in perfect agreement with the analytic
solution \freenergysimple\ , see for instance the following
figure
\ifig\exactvsnumeric{Analytic (continuous line) vs. numeric
(points) solutions of the integral equation for the case
$Z=0.01$ and $\mu=e^{i\phi}$, with $\phi$ running from $0$ to
$\pi$.} {\epsfxsize2.0in\epsfbox{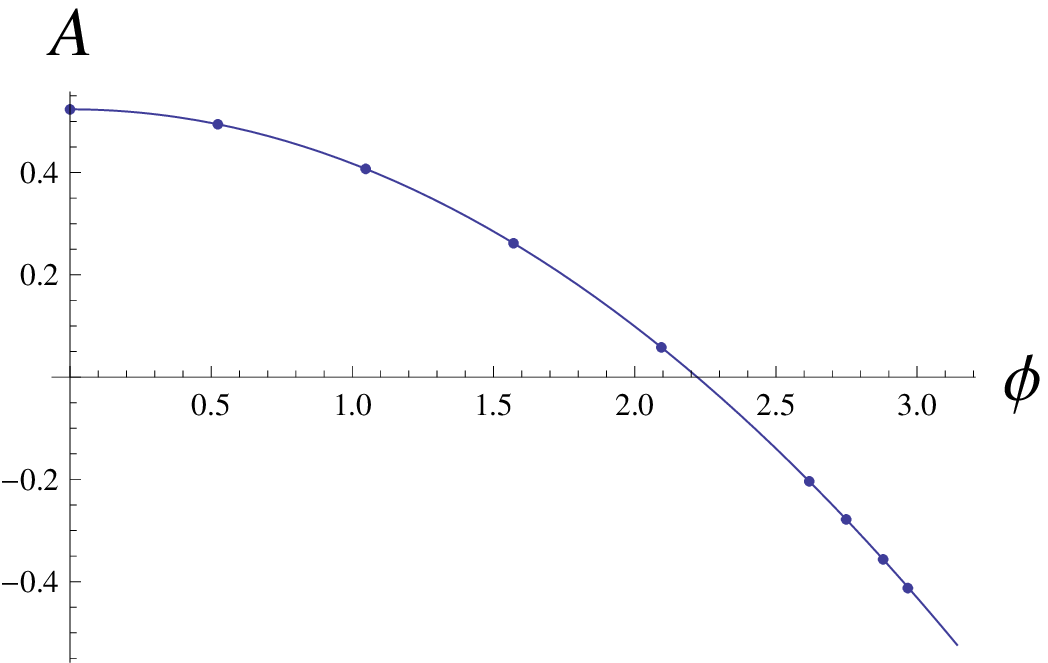}}
In particular, this shows that the numerical solutions can be
quite reliable.

Second, we see that the values of the regularized area/free
energy for $\mu=\pm 1$ is simply given by $\pm \frac{\pi}{6}$.
This values of $\mu$ correspond to $u=1/4$ and $u=1$
respectively, and the value for the free energy is in very good
agreement with what we expect from the numerical analysis of
appendix B. \foot{The substraction of $2 A_{pentagon}$ in
appendix B is needed since the free energy was defined in such
a way that is zero for very large values of $|Z|$.} Recall that $\mu = \pm 1$
corresponds to the two kinds of regular hexagons.

Third, the free energy can be expressed in terms of $u$ by
using
\eqn\uvsphase{\mu+\mu^{-1}=2\cos \phi =\frac{1-3u}{u^{3/2}}}
where $\phi$ is not necessarily real. We are interested in the
whole region $u >0$. We can cover this region by going along
the following contour in the $\mu-$plane
\ifig\muvsu{Different values of $u$ as we follow the path shown
by the figure in the $\mu-$plane. For $u>1/4$ $\mu$ is a
phase.} {\epsfxsize2.0in\epsfbox{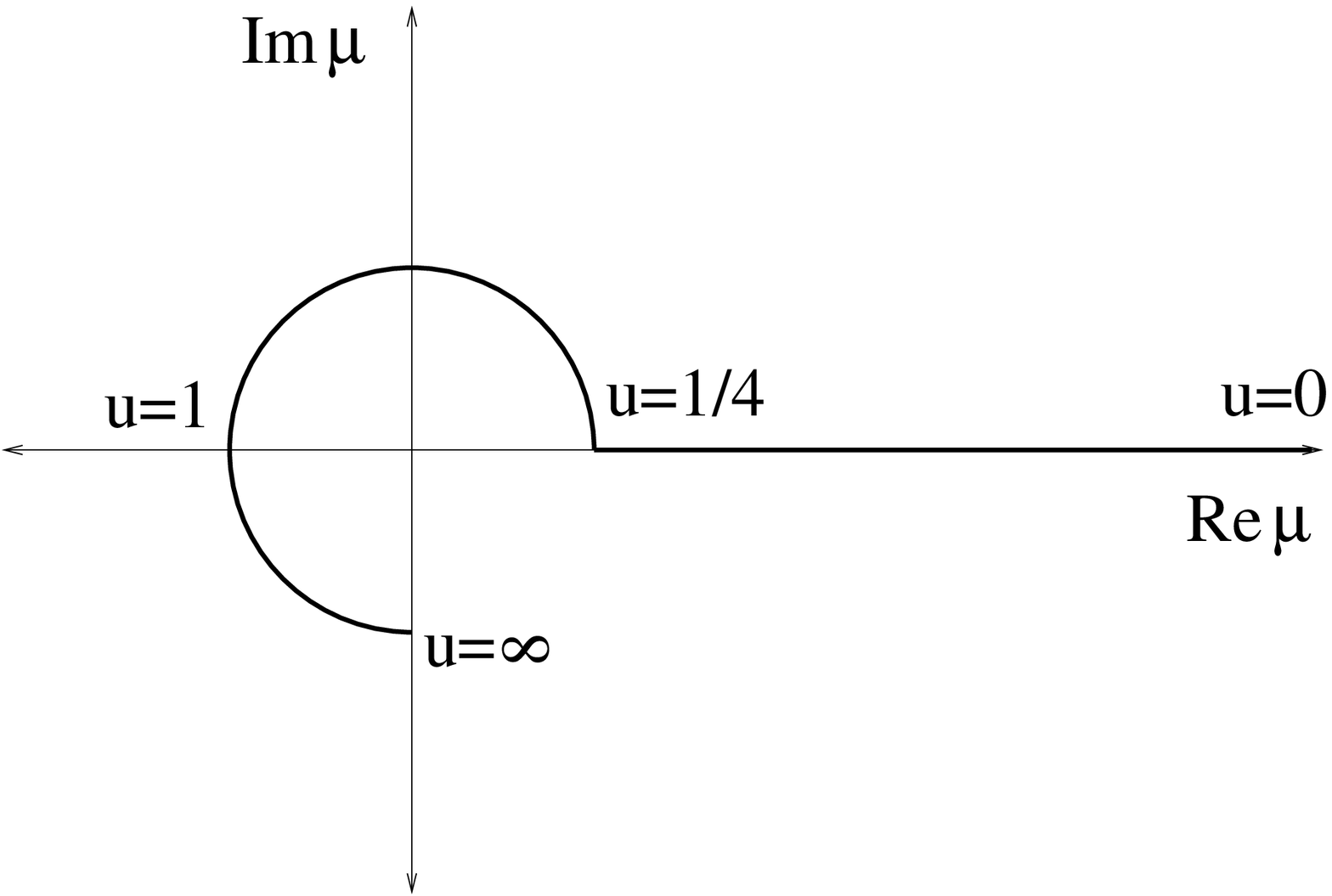}}
As $\mu$ is real and very large, $u$ is very close to zero. As
we approach $\mu=1$, $u$ grows until $u=1/4$ at $\mu=1$. For
$u>1/4$, $\mu$ is a phase, with $\mu=-1$ at $u=1$ and $u$
becoming very large as $\mu$ approaches $-i$.

The final answer for the remainder function in this regime is
then obtained by adding up all the contributions
\eqn\finalparticular{R(u,u,u)=-\frac{\pi}{6}
+\frac{1}{3\pi}\phi^2+\frac{3}{8} \left(\log^2 u+2 Li_2(1-u)
\right),~~~~~~u=\frac{1}{4 \cos^2(\phi/3)}}
This is the remainder function for the scattering of six
gluons at strong coupling in the particular kinematical
configuration in which all the cross-ratios $u_i$ coincide.
\ifig\strongvsweak{Remainder function \finalparticular\ at strong coupling for
$u_1=u_2=u_2=u$} {\epsfxsize2.0in\epsfbox{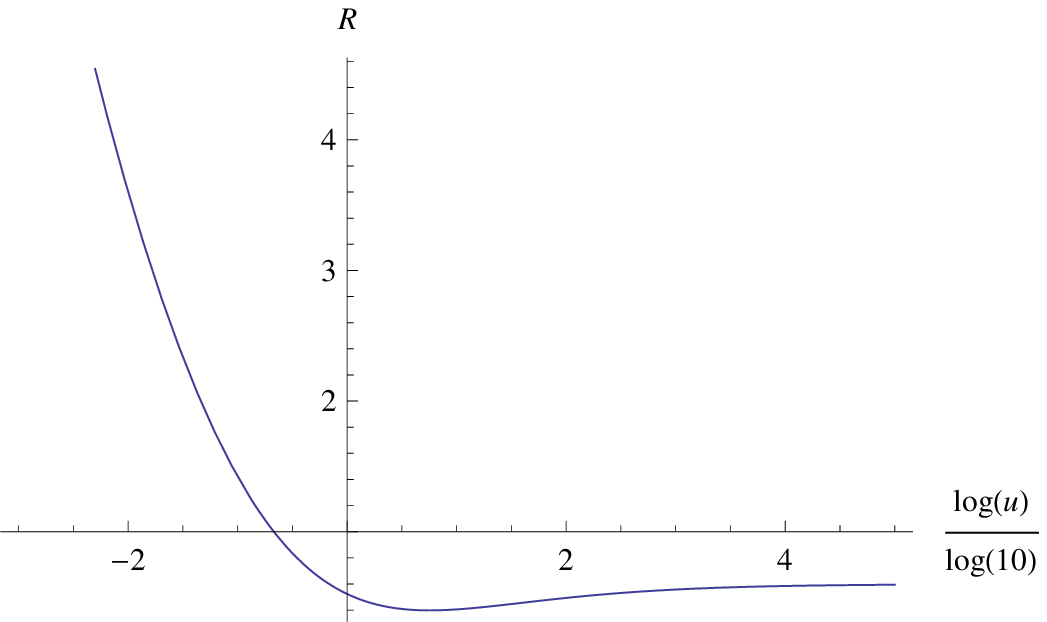}}

\subsec{A curious observation}

The remainder function at two loops for the case at hand was
extensively considered in \AnastasiouKN , based on previous
work \refs{\DrummondBM,\BernAP,\DrummondAQ}. In that paper the
remainder function was computed numerically for several values
of $u$ (in particular, $u=1/9,1/4,1,3.83$ and $u=100$). We
could try to fit their numerics by a function with some
arbitrary coefficients but the general structure of
\finalparticular\ , similar to what it was done in
\BrandhuberDA\ for the case of the octagon in $AdS_3$. More
precisely, we consider a function of the form

\eqn\finalgeneral{R_{c_1,c_2,c_3}=c_1 \left(-\frac{\pi}{6}
+\frac{1}{3\pi}\phi^2 \right) +c_2 \frac{3}{8} \left(\log^2 u+2
Li_2(1-u) \right)+c_3}
The idea is that the high temperature of the TBA equations are not too sensitive to the precise
form of the kernels, so that perhaps this functional form holds for all values of the coupling.
In addition, in our computation, the two terms arise in a somewhat independent fashion, so we have given
us the freedom to change their relative coefficients\foot{Note that we obtain that $c_1/c_2 \approx 1.07 $, which
 is close
to one, so perhaps we should not change the relative coefficients of the two terms!}.
It turns out that for certain values of $c's$, namely $c_1
\approx 12.2,~c_2 \approx 11.4,~c_3 \approx -9.1$,  we get a
quite good approximation of the two loops result, see the following figure
\ifig\strongvsweak{$R_{c_1,c_2,c_3}$ (continuous line) vs.
numerical values (points) for the remainder function at two
loops} {\epsfxsize2.0in\epsfbox{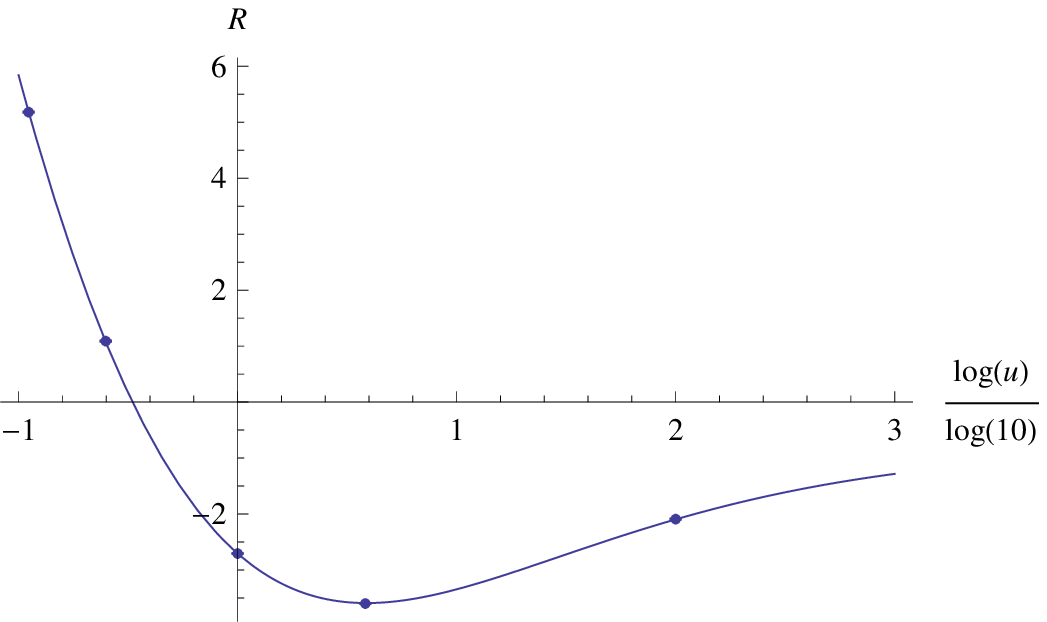}}
However, the asymptotic value for $R$ as $u \rightarrow \infty$
seems slightly higher than the value computed in \AnastasiouKN
. It would be interesting to try to understand better the two
loops result in this particular regime. In particular, it would
be interesting to find its analytical expression and compare it
with the strong coupling result.

Comparing the two loops (figure 8 of \AnastasiouKN\ ) and
strong coupling results , we see that many qualitative features
are very similar and are probably independent on the coupling.
For instance, the apparently universal behavior $R \approx h(\lambda)
\log^2 u$ for small $u$ may be simple to understand and
$h(\lambda)$ may be computable for all $\lambda$.

After the first version of this paper was published, an analytic expression for
the two loop result was found \DelDucaAU . The function looks very similar to the strong
coupling result,  but it {\it cannot }
be fitted by an ansatz of the form \finalgeneral .

\newsec{Conclusions}

In this paper we have studied minimal surfaces in $AdS_5$. These are the classical solutions for
strings moving in $AdS_5$.
 We have shown that the problem can be mapped into an $SU(4)$ Hitchin problem.
More precisely, it is a $Z_4$ projection of an $SU(2,2)$
Hitchin problem. An important observation is the existence of a
spin four holomorphic current $P(z)$ on the worldsheet. For
surfaces that end on a polygonal contour,  $P$ is a polynomial
whose degree depends on the number of sides of the polygon. In
the Hitchin language we have an essential singularity at
$z=\infty $. This singularity has its associated Stokes
phenomenon. This Stokes phenomenon is important for the
emergence of the polygon on the boundary of $AdS$. In fact, the
shape of the polygon is related to the Stokes matrices of the
solutions of the flat Hitchin connection. The shape of the
polygon is encoded in the coefficients of the polynomial as
certain phases which correspond to other moduli of the Hitchin
system.

The integrability of the Hitchin equations can be used to analyze this problem. We introduced the parameter
$\zeta$ and studied the problem as a function of $\zeta$. When $\zeta$ is a phase, we have the ordinary physical
problem, but with complex $\zeta$ it is some sort of deformation of the physical problem. The nice feature is
that for $\zeta \to 0$ or $\zeta \to \infty$ one can solve it using a simple WKB approximation.
This solution displays a Stokes-like phenomenon but now
in the $\zeta$ plane.  The Stokes ``factors" can be found explicitly by analyzing the small
$\zeta $ problem. Given that we have an analytic function of $\zeta$  (analytic away for $\zeta =, \infty$)
with given Stokes-like behavior at zero or infinity, we can recover the function. This is done by first
formulating a Riemann-Hilbert problem and then writing its associated integral equation.

Surprisingly, this integral equation \foldedequ\ has the form of a TBA equation with rather simple kernels.
 In fact,  they are
the TBA equations for an $A_3$  (or $Z_4$) theory discussed in \refs{\KoberleSG,\ZamolodchikovCF}.
The area coincides with the free energy of the TBA system.

The TBA equation can be solved rather simply by a numerical iterative process. The TBA equation depends
on two parameters $|Z|$ and $\mu$ \foldedequ . The non-trivial part of the area, which is the free energy of the TBA,
also depends on $|Z|$ and $\mu$ \finalafree . Once the solution is found we can determine the spacetime cross
ratios, $u_1,~u_2,~u_3$, in terms of $Z$ and $\mu$  \uthreed \otherbs . This leads to the final result
for the remainder function \definr .

There is a special set of cross ratios
$u_1=u_2=u_3$  for which  the problem simplifies. This corresponds to the high temperature limit of the
TBA. This limit was studied in \refs{\ZamolodchikovCF,\FendleyXN}. Using these results we could find
an explicit, and rather simple, analytic expression for the full answer \finalparticular .

For us, the connection to the TBA system  was an
``experimental'' observation. It would be nice to find a
simple argument leading more directly to this TBA equation. It
is interesting  that the TBA that we found corresponds to the
$A_3$ theory \refs{\KoberleSG,\ZamolodchikovCF}, which is
associated to $SU(4)$, the group associated to the Hitchin
equations\foot{ Note that the spectrum of particles in the
$A_3$ integrable theory  \refs{\KoberleSG,\ZamolodchikovCF},
two particles of mass $m$ and one particle of mass $\sqrt{2} m$
coincides precisely with the spectrum of particles for small
fluctuations around the string worldsheet describing a single
cusp, analyzed in \AldayMF .}.

Extending the techniques used in this paper to the case of more
gluons should be straightforward. We expect that we will end up
with a more complicated pattern of discontinuity lines in the
$\zeta$ plane. We expect that the lines will organize in groups
of four lines at right angles, but  with different groups
rotated with respect to each other. The TBA will have a similar
structure, probably with one  unknown function per group of
four lines.   But the final area is still given by the free
energy of this TBA system.

One would also like to generalize these results
to the full quantum problem.
It is natural to imagine  that perhaps the full quantum answer is also given by a TBA equation involving the same
 particles but with a different kernel which includes all quantum corrections. Or perhaps we have a similar kernel
 but a different mapping between the parameters $Z$ and $\mu$ of the integral equations with the spacetime cross
 ratios.
 A natural next step would be to extend the Pohlmeyer reduction and the relation to the Hitchin equations for
 the full  $PS(2,2|4)$ supercoset theory. This should make it clear how to compute N$^k$MHV amplitudes.

 The large temperature limit of the TBA calculation is insensitive to the precise form of the kernel. What is
 important is the total integral of the kernel \xaeqn . Thus, one can speculate that perhaps for all values of the
 coupling one has a form similar to the one we found, but perhaps with a different coefficient \finalgeneral .
 In fact,
 if one allows for general coefficients, the functional form that we found is very close.

\newsec{Acknoledgments}

We thank N. Arkani-Hamed and F. Cachazo for discussions.

This work was supported in part by   U.S.~Department of Energy
grant \#DE-FG02-90ER40542.

\appendix{A}{ Regularization of the area }

In this appendix we regularize the area using a physical cutoff which
is appropriate to regularize  the Wilson loop or the amplitude. The
idea is very similar to the one used in \AldayYN . For the sake of completeness
we give a brief description. The story is relatively simple in the case that the
number of gluons is not a multiple of four, $n \not = 4 k$.
We will treat only this case here, since we mainly deal with the $n=6$ case which certainly
obeys this condition \foot{  The case where $n=4 k$ was treated explicitly in
\AldayYN\
for $AdS_3$ kinematics. We expect similar results for generic $AdS_5$ kinematics.}.

The area can be written as \eqn\areawri{ \eqalign{ A = & A_{\rm
free} + A_{\rm periods} + A_{\rm cutoff } \cr A_{\rm free} = &
2 \int d^2 z [ e^{\alpha} - ( P \bar P)^{1/4}  ] \cr A_{\rm
periods} = & 2 \int_{\Sigma }  d^2z   ( P \bar P)^{1/4} - 2
\int_{\Sigma_0}  d^2 w = 2 \int_{\Sigma} d^2 w -2
\int_{\Sigma_0}  d^2 w \cr A_{\rm cutoff} = &2  \int_{ \Sigma_0
, Y^+ > 1/\epsilon_c } d^2 w }} Where $A_{free}$ is the
non-trivial function that is computed as the free energy of the
TBA equation \tbafree for the case of $n=6$.
 In the definition of $A_{periods}$ we have used that the polynomial $P$ defines a surface $\Sigma$
which is the $w$ plane with the appropriate cuts. We have introduced $\Sigma_0$ as a reference surface with a single
branch point at the origin and the same structure at infinity. In order to subtract the two it is very important to
regularize both computations with a cutoff in the $w$ plane, such as $|w| = \Lambda \gg 1$, but not a cutoff in the
$z$ plane. Using this cutoff for defining the difference in $A_{periods}$ we find that it can
  be explicitly evaluated in terms of periods on the Riemann surface $x^4 = P(z)$.
It has the form of $ Z_{e_i} \bar Z_{m_i} - Z_{m_i} \bar Z_{e_i} $ where $e_i$, $m_i$ is a basis of
electric and magnetic cycles (i.e. a basis of cycles which are symplectically normalized).
 For the particular case of the hexagon, $A_{\rm periods} =
  |Z |^2$, where $Z$ is defined in \periodZ , see \derif .
$A_{\rm cutoff}$ is imposing the physical cutoff which requires
that the $AdS$ radial coordinate $Y^+$ should be larger that
$1/\epsilon_c$. Here $\epsilon_c$ is a distance scale providing
a UV (IR) cutoff in the Wilson loop  (amplitude)
interpretation.

In order to evaluate $A_{\rm cutoff}$ we can take $\Lambda =0$
and replace the $w$ surface by a simple surface with all
the branch points at the origin.
Our task is now to evaluate $A_{\rm cutoff}$. In order to evaluate it we will need to use a couple of properties
of the solutions discussed in section 2.4.
Near each cusp  we have a behavior which goes roughly like
\eqn\beha{
Y_i^A \sim y_i^A \times \left( e^{ 2  Re[w] } ,~~~{\rm or}~~ e^{ 2  Im[w] }, ~~~{\rm or}~~~
 e^{ -2  Re[w] } ,~~~{\rm or}~~ e^{- 2  Im[w] }  \right)
}
with a particular   choice at each cusp (we have rotated
 the $w$ plane so simplify these expressions). The position of the cusp on the boundary is determined by the
 null vector $y_i$ ($y_i . y_i=0$).
Recall that we had a basis of small solutions $s_i$, normalized so that $det(s_i,s_{i+1},s_{i+2}, s_{i+3})=1$.
The behavior of the spacetime solution near the cusp $i$
 could be found as  $ y^A_i = q^A_I s_i^T \Gamma^I s_{i+1}$.
 Which, together with the normalization condition
 on the $s^I$ implies
 \eqn\reslim{
 y_i . y_{i+2 } = 1
 }
 Let us say that at cusp $i$ we have $Y_i \sim y_i e^{ 2 Re[w]}   $.
Then the spacetime cutoff has the form \eqn\stap{ Y^+ = { 1
\over \epsilon_c } = y_i^+ e^{ 2 Re[w]} } which gives a line in
the $w$ plane,    $2 Re[w] = - \log\epsilon_c - \log y_i^+ $.
We have a similar relation on the cusp $Y_{i+2}^+$, where the
line is now at $2 Re[w] = -(- \log[\epsilon_c] - \log y_{i+2}^+
) $. We first imagine some reference lines at $2 Re[w] = \pm
\log \epsilon_c $ which are then displaced by small distances
proportional to $\delta_i = -\log y_i^+$. It is convenient to
relate the $\delta_i$ to the spacetime distances. This is done
as follows. We know that $y_i . y_{i+2} = 1$. At the same time
we know that the spacetime distances can be expressed as
$d_{i,i+2}^2 \sim { y_i . y_{i+2} \over y_i^+ y_{i+2}^+ } $.
This implies that \eqn\deltadif{ \delta_{i} + \delta_{i+2} =
\ell_i \equiv \log d^2_{i,i+2} } where we used \reslim . This
equation, together with $\delta_{i+n} = \delta_i$ can be used
to solve for all the $\delta_i$ in terms of the logs of the
nearby distances $\ell_i$. Here it is important that $n$ is not
a multiple of four.

\ifig\areawplane{ Here we show the computation of $A_{cutoff}$. The $w$ surface $\Sigma_0$  has a
 single branch point at the origin. We show only a portion of this surface. Here $L = - \log \epsilon_c$ and
$\delta_i = - \log y_i^+$. The total area is obviously the sum of the areas of the various rectangles.
Note that $L \gg \delta_i$ as we take the cutoff away. }
{\epsfxsize2.0in\epsfbox{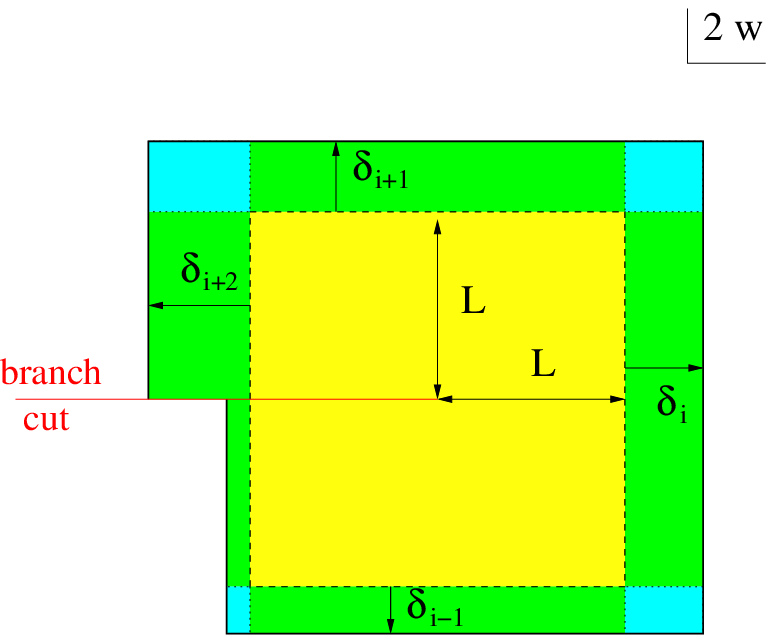}}

We can then compute the area by summing the areas of various rectangles  in \areawplane .
We get
\eqn\areacu{ \eqalign{
A_{\rm cutoff} = &{ 1 \over 2 } \left[
  n L^2 + 2 L \sum_i \delta_i + \sum_i \delta_i \delta_{i+1} \right]
\cr A_{\rm cutoff} = & A_{\rm div} + A_{\rm BDS-like} \cr
A_{\rm div} = & { 1 \over 2 } \sum_i  ( - \log \epsilon_c  + {
\delta_i + \delta_{i+2} \over 2} )^2 = { 1 \over 8 } \sum_i (
\log(\epsilon_c^2 d_{i,i+2}) )^2 \cr A_{\rm BDS-like} = & - { 1
\over 4} \delta_i^2 - { 1 \over 4} \delta_i \delta_{i+2} +  { 1
\over 2} \delta_i \delta_{i+1} }} We can rewrite this as
\eqn\areacu{ \eqalign{
 A_{\rm BDS-like}   = & -{ 1 \over 8 } \sum_{i=1}^n  \left( \ell_i^2 +
\sum_{k=0}^{2 K } \ell_i \ell_{i +1 + 2 k } (-1)^{k+1}  \right)~,~~~~~~~~ {\rm for } ~~n = 4 K + 2
\cr
= &-  { 1\over 4 }  \sum_{i=1}^n  \left( \ell_i^2 +
\sum_{k=0}^{2 K } \ell_i \ell_{i +1 + 2 k } (-1)^{k+1}  \right)~,~~~~~~~~ {\rm for } ~~n = 4 K + 1 ~, ~~4K +3
 }}
Here $A_{div}$ is the standard divergent term that is present for light-like Wilson loops
 or for amplitudes \refs{\KorchemskayaJE,\BassettoXD,\BernIZ}. On the other hand $A_{\rm BDS-like}$
 is a finite term which obeys the anomalous Ward identities
of broken conformal invariance \DrummondAU . An easy way to check this is to note that if we shift
each log of a distance $\log d^2_{i,j} \to \log d^2_{ij} + \epsilon_i + \epsilon_j$, then
\eqn\variat{
 { d A_{\rm BDS-like } \over d \epsilon_i }|_{\epsilon =0}  = { d A_{\rm BDS}
\over d \epsilon_i }|_{\epsilon=0}  =  { 1 \over 4 } \log{   d^4_{i-1,i+1} \over  d^2_{i,i+2 }
d^2_{i,i-2}  }
}
where $A_{\rm BDS}$ is the   expression written in \BDS , which is the one loop result in the weak coupling
expansion, except that here it comes with a different overall factor, due to the different value of
the cusp anomalous dimension.
When we act with the dual
conformal generators we see that we need to add with these derivatives. Thus if we have
the same derivatives as the BDS answer, then it obeys the same equations as the BDS answer.
Note that if $n$ is not a multiple of four we cannot make any cross ratios from the distances
$d_{i,i+2}$. This implies that there is a unique solution of the anomalous ward identity which
can be written only in terms of $d_{i,i+2}$. This unique solution is $A_{\rm BDS-like}$.

In order to find the difference between $A_{\rm BDS} - A_{\rm BDS-like}$ we simply start with any
non-nearby distance, $d_{i,j}$,  appearing in $A_{\rm BDS}$ and we write it in terms of the unique
cross ratio $c_{i,j}$ which involves $d_{i,j}$  and nearby distances $d_{i,i+2}$.

Thus, if one is interested in the remainder function, which is conventionally defined
in terms of what we have to add to $A_{\rm BDS}$ to get the full answer, then it can be expressed
as follows
\eqn\resuldif{\eqalign{
-A = & -A_{\rm div} - A_{\rm BDS} + R
\cr
R = & -( A_{\rm BDS-like} - A_{\rm BDS} ) -  A_{\rm periods} - A_{\rm free} + {\rm constant}
}}
where the constant is independent of the kinematics.

In the $AdS_3$ limit these formulas  become the
expressions  for  $n/2$ odd in \AldayYN\ (Formula 5.8).

In this paper we are mostly interested in the case $n=6$. In
this case, the BDS ansatz reads\foot{Note that what we are calling $A_{BDS}$ is minus the expression in
 \BDS\ (up to a factor) because of the minus sign in \amplarea . }
\eqn\formex{ A_{\rm BDS} =   \sum_{i=1}^6  \left[  { 1 \over 4} \log
\frac{x_{i,i+2}^2}{x_{i,i+3}^2} \log
\frac{x_{i+1,i+3}^2}{x_{i,i+3}^2}-\frac{1}{16} \log^2
\frac{x_{i,i+3}^2}{x_{i+1,i+4}^2}+ \frac{1}{8}Li_2
\left(1-\frac{x_{i,i+2}^2 x_{i+3,i+5}^2}{x_{i,i+3}^2
x_{i+2,i+5}^2} \right) \right] }
 We find that
\eqn\BDSminusours{A_{\rm BDS}- A_{\rm BDS-like} =  \sum_{i=1}^3
\left( { 1 \over 8} \log^2 u_i+ { 1 \over 4 } Li_2(1-u_i) \right)}
Hence, the remainder function is obtained by adding the above
difference to the regularized area discussed in the body of the
paper.

\appendix{B}{A new class of regular solutions}

In the following we will consider a generalization of the
regular polygon solutions considered in \AldayYN .  These are simple solutions for which
the area can be evaluated directly. Thus, they provide special cases which can be compared
to the general results we obtained above. We consider  solutions that possesses $Z_n$ symmetry.
In other words, we have
\eqn\Znsym{ Y^A( z e^{ i \frac{2\pi}{n}}, \bar z e^{ - i \frac{2\pi}{n}}  )=R^A_{~B}
 Y^B( z , \bar z  )}
where $R$ a rotation in space-time, $R^n=1$. These solutions are associated to regular  polygons.
 Using the expression \derx\  for the holomorphic
function $P(z)$, this implies
\eqn\Znsym{P( z e^{ i \frac{2\pi}{n}} )=e^{-4 i
\frac{2\pi}{n}}P(z )}
The simplest holomorphic function satisfying this condition is
a monic polynomial $P(z)=z^{n-4}$. For the case of $AdS_3$ we
have $n$ even. In the general case of $AdS_5$, the simplest
elementary solution would correspond to $P(z)=z$, which would
correspond to a regular pentagon.

As seen in \AldayHE ,  it is
convenient to use $(2,2)$ signature to construct the pentagon because
we can then choose all non-consecutive cusps to be spacelike separated.
 Furthermore, we will see
that it  can be actually embedded into
$AdS_4$, very much like the regular polygons studied in \AldayYN\
could be embedded in $AdS_3$.

We start by considering the embedding coordinates of $AdS_5$ in
the case of $(2,2)$ signature.
\eqn\adsfivetwotwo{ -Y_{-2}^2-Y_{-1}^2-Y_0^2+Y_1^2+Y_2^2+Y_3^2=-1}
the restriction to $AdS_4$ corresponds to setting $Y_3=0$. Note that this is
an $AdS_4$ space with two times, whose boundary has two time and one space directions.
We
can instead use ordinary $AdS_5$  Poincare coordinates, which correspond to
${1 \over z} =Y_{-2}+ Y_3 = Y_{-2} $, $(t_1,t_2)=\frac{(Y_{-1},Y_0)}{Y_{-2}}$ and
$(x_1,x_2)=\frac{(Y_1,Y_2)}{Y_{-2}}$. The regular polygons to
be considered in this appendix can be embedded into this
$AdS_4$, hence, they correspond to contours in the $(x_1,x_2)$
plane and in the $(t_1,t_2)$ plane, such that
\eqn\regularcontours{ x_1^2+x_2^2-t_1^2-t_2^2=1}
This is a surface inside $R^{2,2}$ with two time and one space directions.
This surface is the boundary of an $AdS_4$ subspace of $AdS_5$ where the surface is residing.
In addition, we require each segment to be light-like, so the
lengths in the $x-$plane and $t-$plane coincide. Finally, we
consider only polygons such that at each cusp the momentum
transfer is space-like. This can be achieved by requiring that
the angle between consecutive segments
in the $x-$plane is larger than the angle in the
$t-$plane. A prototypical example of such contour is the
regular pentagon

\ifig\regularpentagon{Regular pentagon in $(2,2)$ signature.
The lengths of each segment in the $x-$plane and $t-$plane
coincide and the angles are always smaller in the $t-$plane.}
{\epsfxsize2.5in\epsfbox{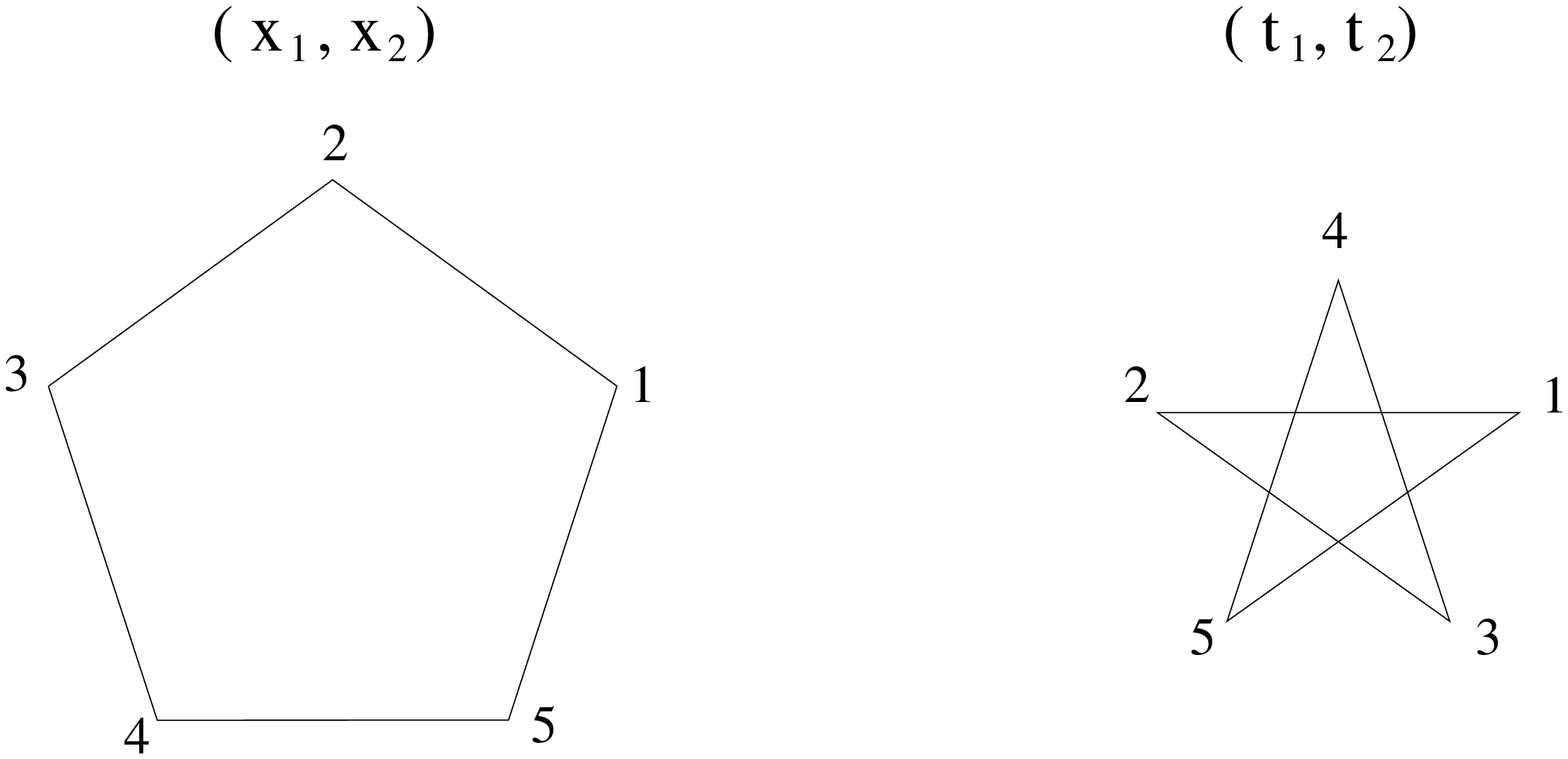}}

\subsec{Strings on $AdS_4$}

Next, let us briefly consider the reduction of strings on
$AdS_4$. In this case $B_1$ and $B_2$ are non trivial and we
have $v_1$ and $v_2$ correspondingly. From now on, let us focus
on  the case of $(3,1)$ signature and parametrize
\eqn\vparam{v_1=i P(z)^{1/2} \cos{(\beta/2)},~~~v_2=P(z)^{1/2}
\sin(\beta/2)}
where $\beta$ is real. Later on, we will then continue to
$(2,2)$ signature by simply taking $\beta \rightarrow i \beta$.
We would like to express the physical field $\beta$ purely in
terms of space-time quantities. For that, let us recall the
following expressions
\eqn\secondder{\partial^2 Y= \partial \alpha \partial Y+v^I
B_I,~~~\bar \partial^2 Y=\bar \partial \alpha \bar \partial
Y+\bar v^I B_I}
Viewing this as an expression for $v^I$, $\bar v^I$ we can compute
\eqn\betaspacetimeone{- v^I \bar v^I = |P(z)| \cos \beta=e^\alpha \partial
\alpha \bar \partial \alpha-\partial^2 Y. \bar \partial^2 Y}
In addition, with a little bit of effort, one can show
\eqn\betaspacetimetwo{e^\alpha |P(z)| \sin
\beta=\epsilon_{abcde} Y^a
\partial Y^b \bar \partial Y^c \partial^2 Y^d \bar \partial^2
Y^e}
From this relation we can immediately see that $\beta$ is real
and also that $\beta \rightarrow i \beta$ is the correct
prescription in order to continue $(3,1)$ signature into
$(2,2)$ signature.

\subsec{Boundary conditions at zero}

Let us now focus on  the case of $(2,2)$ signature. In that
case, the equations of motion and Virasoro constraints imply
the following equations for $\hat \alpha=\alpha- { 1 \over 2 }  \log
|P(z)|$ and $\beta$
\eqn\alphabetaeqs{\partial \bar \partial \hat \alpha=e^{\hat
\alpha}-e^{-\hat \alpha}\cosh \beta,~~~~\partial \bar
\partial \beta=e^{-\hat \alpha} \sinh \beta }
where we have written these equations in the $w-$plane, defined
by $dw=P(z)^{1/4}dz$.As it can be seen, $\hat \alpha$ and
$\beta$ are massive fields that decay exponentially at
infinity.

In order to understand the boundary conditions as we approach
one of the zeros of $P(z)$, we use polar coordinates $z=\rho
e^{i \varphi},\bar z=e^{-i \varphi}$ and consider the expansion
of a general solution around $\rho=0$
\eqn\gralsol{Y_{-2}=1+\sum_{\ell=1}
\Phi_\ell^{(-2)}(\varphi)\rho^\ell,~~~~~Y_{i}=\sum_{\ell=1}
\Phi_\ell^{(i)}(\varphi)\rho^\ell,~~~i=-1,0,1,2. }
By considering the equations of motion order by order in
$\rho$, the functions $\Phi_\ell^{(i)}(\varphi)$ are
determined, up to integration constants at each order. In
addition, the requirement $Y^2=-1$ and the Virasoro constraints
impose some conditions on these integration constants. When
considering the case of the regular pentagon, we have an
additional constraint, coming from the $Z_5$ symmetry of the
contours, see \regularpentagon\ . More precisely, we can
introduce complex spacetime coordinates $X = Y_1 + i Y_2$, $T =
Y_{-1} + i Y_0 $ and we then require
\eqn\reuq{
X( z    e^{ i \frac{2\pi}{n}}, \bar z    e^{ - i \frac{2\pi}{n}}) = e^{ i  { 2 \pi \over n } r_x } X(z, \bar z ) ~,~~~~~
T( z    e^{ i \frac{2\pi}{n}}, \bar z    e^{ - i \frac{2\pi}{n}}) = e^{ i   { 2 \pi \over n }r_t } T(z, \bar z )
}
Where $r_x$ and $r_t$ are two integers which parametrize the spacetime  $Z_n$ rotation matrix.
For the pentagon case, $n=5$, $r_x=1$ and $r_t =2$, see \regularpentagon .
  Once all the
constraints are worked out, we can compute the expansions of
$\alpha$ and $\beta$ around the origin. As expected, $\alpha$
is regular, while
\eqn\betabc{\cosh \beta = \frac{e^\alpha \partial \alpha \bar
\partial \alpha-\partial^2 Y. \bar \partial^2 Y}{|\partial^2 Y . \partial^2 Y|} \approx \frac{1}{\rho}+...}
Hence $\beta$ has a logarithmic divergence as we approach $\rho
\rightarrow 0$. A similar argument for $(3,1)$ signature, shows
that we should expect some winding in $\beta$ as we approach
the origin. Since $\cosh \beta  \ge 1$ and $\cos \beta  \le 1$
it is reasonable to expect such behaviors for $\beta$ from
\betabc\ \foot{More precisely, the left hand side of \betabc\
will have a general expansion of the form $$ [\cosh \beta ~~{\rm or ~~} \cos \beta ]= \frac{e^\alpha
\partial \alpha \bar
\partial \alpha-\partial^2 Y. \bar \partial^2 Y}{|\partial^2 Y . \partial^2 Y|}=\frac{c_1+c_2 z+ c_2^*z +...}{\rho} $$ For $(2,2)$ signature, as $\cosh \beta \ge 1$, we expect $c_1 \neq 0$.
On the other hand, for $(2,2)$ signature, as $\cos \beta \le 1$
we expect $c_1=0$ and $c_2 z+ c_2^*z = \cos (\varphi + \gamma)$,
resulting in non-trivial winding for $\beta$. The winding for
$\beta$ in $(3,1)$ signature is consistent with the assumption
of $\beta_{\rm Dorn}$ as defined by \dornrecent , being regular
everywhere. In \dornrecent\ $P$ is
factorized in two polynomials $P(z)=v_+(z) v_-(z)$ and the
relation between our $\beta$ and the one used in \dornrecent\
is $\beta_{\rm ours}=\beta_{\rm Dorn} - \frac{i}{2} \log \left(\frac{v_- \bar
v_+}{v_+ \bar v_-} \right)$. }.

Hence we expect the regular pentagon in $(2,2)$ signature to
correspond to two fields $\hat\alpha$ and $\beta$ with
logarithmic divergences at the origin and exponential decay at
infinity. Furthermore, it is reasonable to assume that such
solutions are rotationally symmetric. Namely $\hat \alpha$ and $\beta$ are functions
of $|w|=\rho$ only with the following boundary conditions.
\eqn\betabcwplane{\eqalign {\hat \alpha(\rho) =
-\frac{2}{5}\log \rho+c_\alpha+...,~~~\beta(\rho)=-\frac{4}{5}
\log \rho+c_\beta+...,~~~~\rho \to 0 \cr
 \hat \alpha(\rho) \approx K_0(2 \sqrt{2} \rho),~~~~\beta(\rho) \approx K_0(2 \rho),~~~~\rho \rightarrow \infty}}
The constants $c_\alpha$ and $c_\beta$ have to be chosen in
such a way that the solutions decay at infinity. We find
numerically that such solutions indeed exist and their
exponential decaying is in perfect agreement with the
expectations. Solving numerically the shooting problem, we find
$c_\alpha \approx -0.28,c_\beta \approx -0.38$. These values
are consistent with a crude version of the analysis of
\lastjevicki\ \foot{However, in \lastjevicki\ it was assumed that
$\beta$ is regular at $\rho \rightarrow 0$, while we assume a
logarithmic divergence \betabcwplane r.}. Once the numerical solution is found,
we can compute its regularized area. We find the approximate
result
\eqn\areapentagon{A_{\rm pentagon} = 2 \int (e^{\hat \alpha}-1)d^2 w
\approx 1.18}
with an estimated error of a few percent. Note that $\beta
\rightarrow -\beta$ is a symmetry of the equations and we get
another solution where $\beta \to - { 4 \over 5} \log \rho$ at
the origin. These two pentagons  differ by a spacetime parity
operation\foot{The fact that spacetime parity changes the sign
of $\beta$ can be seen from \betaspacetimetwo .}. For a more
general polynomial we can choose either boundary condition,
 $\beta^{\pm} \rightarrow \pm \frac{4}{5} \log \rho$,  at each zero  of
  $P(z)$.

\subsec{Other regular solutions}

Now we can imagine that $P(z)$ has two zeros, in each of which
$\beta$ has boundary conditions $\beta^{\pm}$. As we bring the
two zeros together there are two distinct situations:

\noindent 1.- We bring together two zeros where the boundary
conditions are $\beta^+$ and $\beta^-$. In this limit, we
expect to recover the usual regular hexagon, which can be
embedded into $AdS_3$ and for which $\beta = 0$. In terms of
contours in $(2,2)$ signature, its boundary looks like
\ifig\regularhexagon{Usual regular hexagon in $(2,2)$
signature. As it can be embedded into $AdS_3$, since one of the time
coordinates vanishes identically. The cross ratios are $u_1=u_2=u_3 =1$.}
{\epsfxsize2.5in\epsfbox{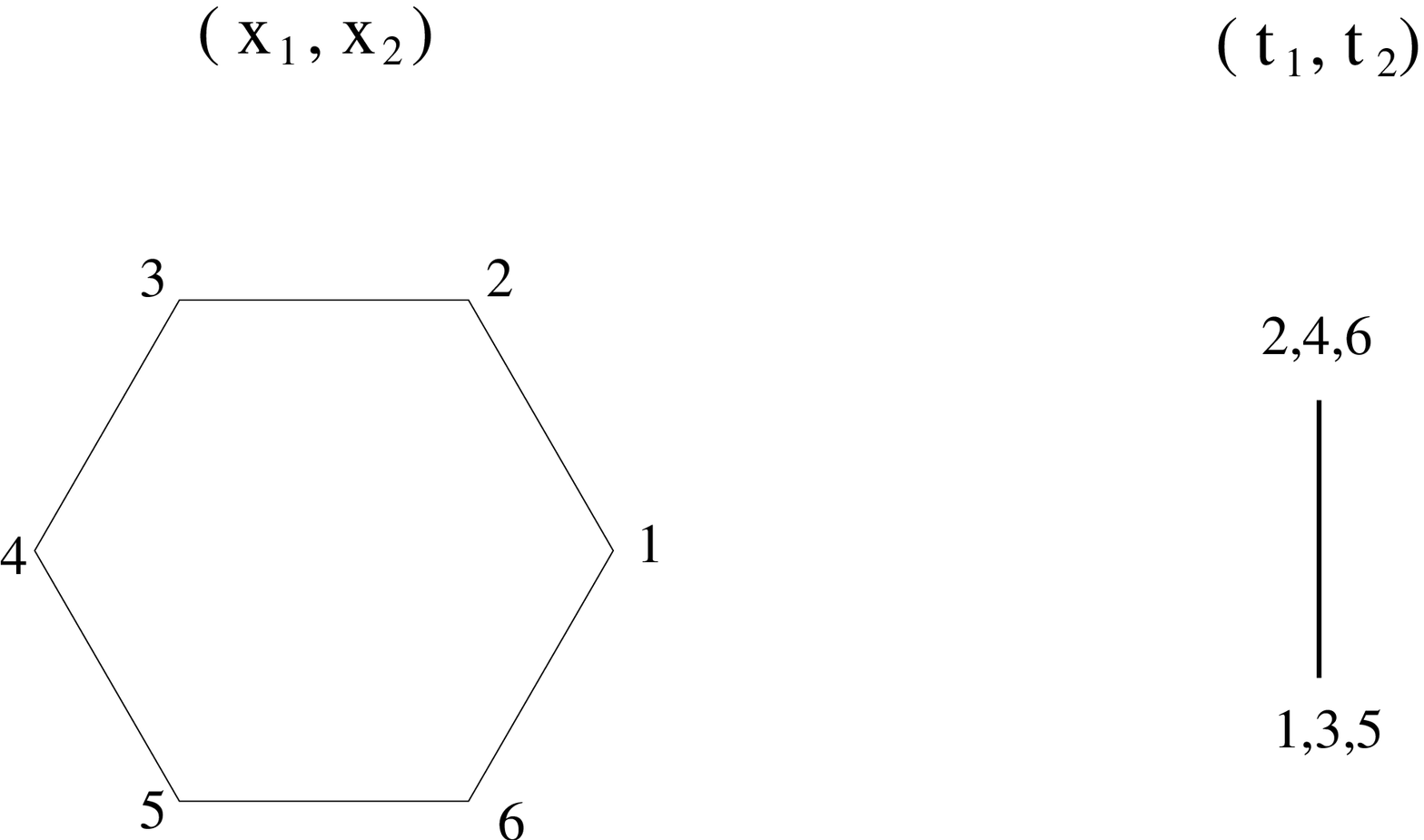}}
The regularized area corresponding to this hexagon was computed
in \AldayYN , with the result
\eqn\areafirsthexagon{A^{+-}_{\rm hexagon}  = \frac{7}{12} \pi}

\noindent 2.- We bring together two zeros where the boundary
conditions are both $\beta^+$ (or both   $\beta^-$).
In this limit, we expect to recover a different regular
hexagon, still with $Z_6$ symmetry, but which cannot be
embedded in $AdS_3$ because $\beta \not =0$. Furthermore, the field $\beta$
corresponding to this solution has a logarithmic singularity at
the origin. In terms of contours in $(2,2)$ signature, its
boundary looks like

\ifig\secondregularhexagon{Alternative regular hexagon in
$(2,2)$ signature. It still posses a $Z_6$ symmetry but cannot
be embedded into $AdS_3$. The cross ratios are $u_1=u_2=u_3 ={ 1 \over 4 } $. }
{\epsfxsize2.5in\epsfbox{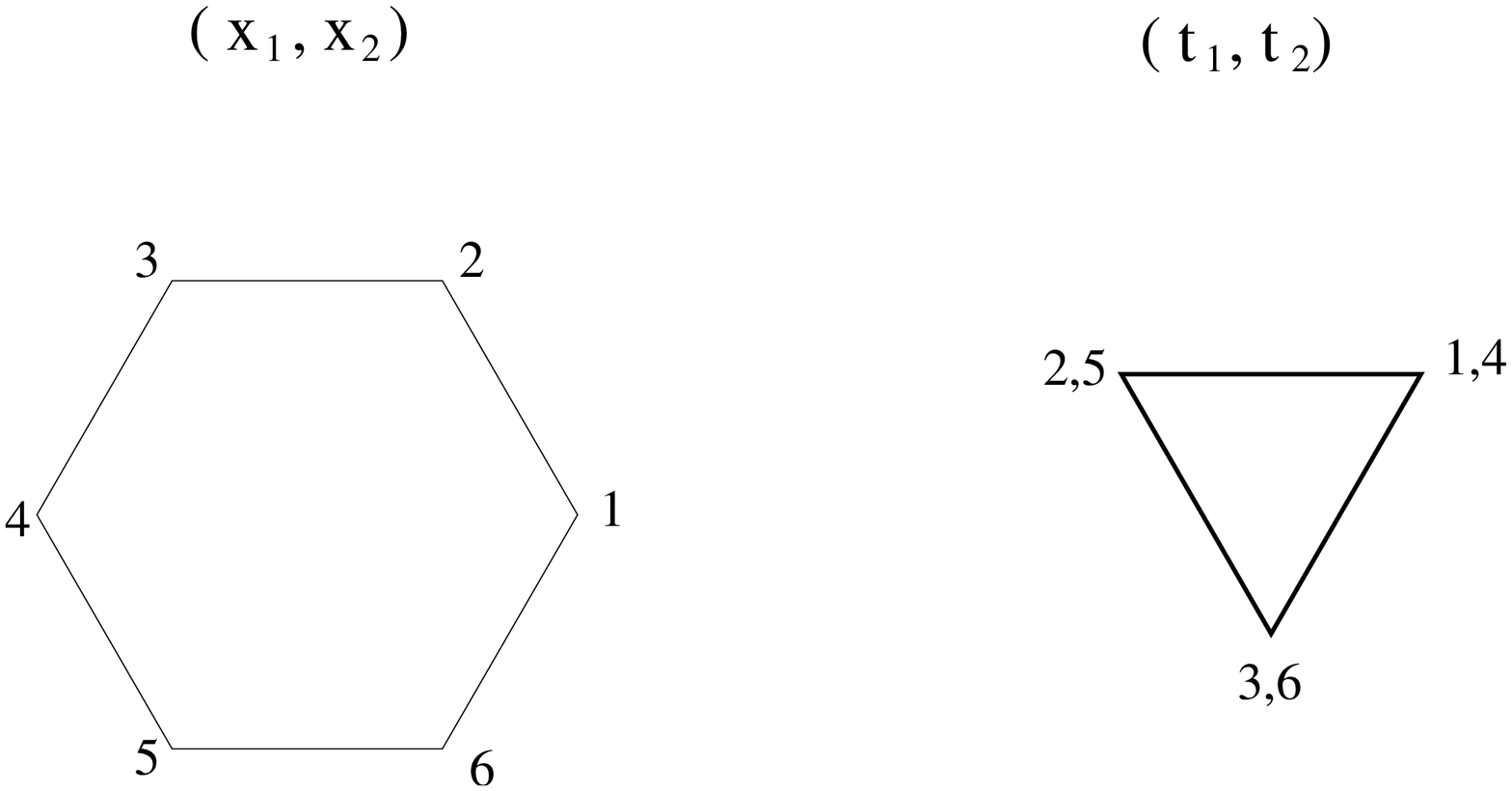}}

In the $w-$plane we expect the following boundary conditions at
the origin
\eqn\betabcwplanehexa{\eqalign {\hat \alpha(\rho) =
-\frac{2}{3}\log \rho+c_\alpha+...,~~~\beta(\rho)=-\frac{4}{3}
\log \rho+c_\beta+...,~~~~\rho \to 0}}
The condition for $\beta$ is
 deduced by looking at \betaspacetimeone\ near the origin and demanding that the right hand side is
 approximately constant near the origin.
Solving numerically the shooting problem and computing the
approximate area, we find
\eqn\areasecondhexagon{A^{++}_{\rm hexagon}  \approx 2.85}
The value of the cross-ratios is $u_i=1$ for the first, usual,
regular hexagon and $u_i=1/4$ for the second kind of regular
hexagon. It is difficult to compare these individual areas to the result
of the integral equation due to possible overall additive constants.
However we can compute  the hexagon areas
minus twice the pentagon area, which corresponds to the
difference in the regularized area when the two zeros are
brought together from a very large distance.  From our numerical
estimates we find
\eqn\testareas{A_{\rm hexagon}^{+-}-2A_{\rm pentagon} \approx -0.52 \approx - { \pi \over 6 }
,~~~~~~~A_{\rm hexagon}^{++}-2A_{\rm pentagon} \approx 0.50 \approx { \pi \over 6 } }
The values $\pm \pi/6$ are the exact results obtained using the integral equation.

For a generic number of zeros of $P(z)$ we expect a similar
situation. Given a polygon of $n$-sides, we can draw a regular
polygon such that we need to perform a rotation by $\omega=e^{i
2\pi p/n}$ in order to go from one cusp to the next one, with
$p=1,2,...$. Hence, we will have a family of inequivalent
polygons labeled by two integers $(p_x,p_t)$, drawing a regular
polygon with $p_x$ in the $x-$plane and with $p_t$ in the
$t-$plane, see \reuq . For instance, in the following figure we see
different choices $p=1,2,3,4$ for $n=9$.
\ifig\nonagons{Shape of different regular nonagons, with
$p=1,...,4$ from left to right. The full polygon in (2,2) signature is given by choosing a
pair of these with $p_x < p_t$. }
{\epsfxsize2.5in\epsfbox{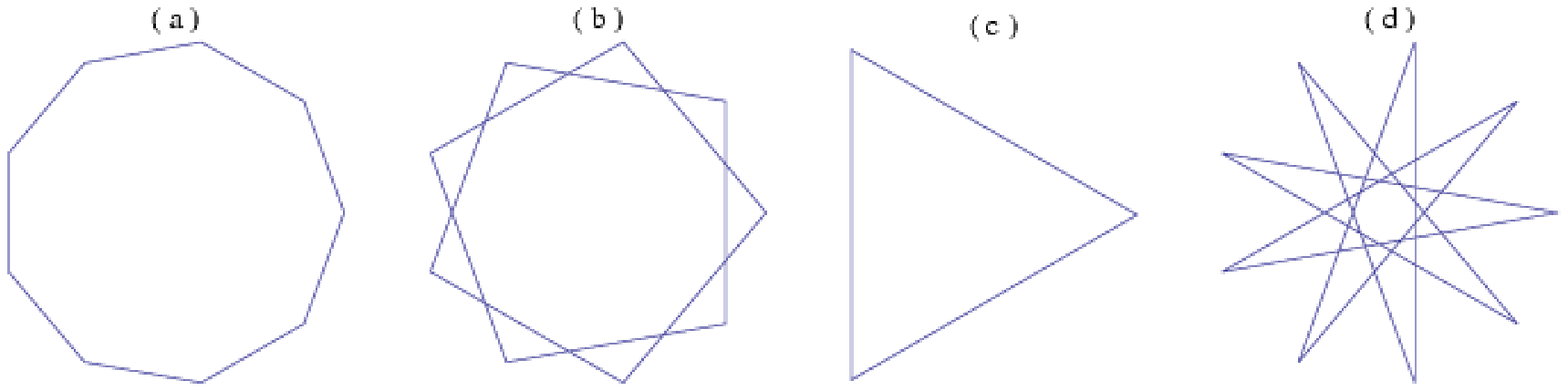}}
Furthermore, the requirement of space-like momentum transfer
can be achieved by imposing $p_x < p_t \leq [{n\over2 } ]$.
Finally, let us mention that such polygons can be embedded into
$AdS_4$ and the distance between adjacent cusps can be made
light-like. Centering the above polygons at the origin we can
rescale their relative size so that the segments in the
$x$-plane and the $t$-plane have the same length. Then the pair
of regular polygons produces a null-polygon in $R^{2,2}$.

It would be interesting to understand these solutions better.
It is likely that they correspond to radially symmetric solutions. One could
attempt to find
 the precise boundary conditions for $\beta$ in each
case.

\appendix{C}{On the appearance of poles in the TBA equation}

Although in the initial large $|Z|$ region the $x_i[\zeta]$
had no zeroes or poles in $\zeta$, they may appear at smaller $|Z|$. This
is a familiar phenomenon in TBA-like equations \refs{\dtpoles,\zampoles}. Poles and
zeroes   appear in a very specific manner. Lets' start from
a situation without poles or zeroes, and for simplicity
consider some function ${\cal X}(\zeta)$, with a discontinuity
${\cal X}^+ = {\cal X}^- (1+{\cal Y})$ across a cut $\ell$. In
general, both the ${\cal X}^+$, ${\cal X}^-$ which define
${\cal X}$ on the two sides will be nice analytic functions
across the cut. ${\cal X}^+$ is supposed to have no zeroes or
poles above the cut, but it may have either below the cut. It
ie easy to see it has to be a zero: ${\cal X}^-$ and ${\cal Y}$
are not supposed to have zeroes or poles there, all can happen
is that $1+{\cal Y}$ is zero at some point $\zeta_0$.

As we vary the parameters in the integral equation, the point
$\zeta_0$ defined by ${\cal Y}[\zeta_0]=-1$ may move across the
cut. In that case, we do not expect the zero of ${\cal X}^+$ to
suddenly disappear, and a pole of ${\cal X}^-$ to suddenly
appear if we follow our solution continuously. That would
happen, though, if we did not modify the integral equation. The
reason is clear: the integral equation is of the form $\log
{\cal X} = \cdots + K_\ell * \log(1+{\cal Y})$: the $\log$ has
a singularity at $\zeta_0$, which we are passing across the
cut.

The correct way to continuously follow a solution is to accept
that ${\cal X}^+$ should have a zero at $\zeta_0$ above the
cut. The integral equation is easily modified to account for
it, schematically as
$$\log {\cal X} = \cdots + \log (\zeta-\zeta_0) + K_\ell * \log(1+{\cal Y})$$
The extra term actually coincides with the difference between
the convolution $K_\ell * \log(1+{\cal Y})$ along paths just
above and just below $\zeta_0$. Notice that the location of the
zero is fixed by ${\cal Y}[\zeta_0]=-1$, in order for ${\cal
X}^-$ not to have the zero there.

The expression for the TBA free energy is similarly modified by
the addition of a schematic term $Z/\zeta_0 + \bar Z \zeta_0$.
In TBA these extra terms have the interpretation of an
excitation added to the ground state. In our specific setup,
the location of possible zeroes and poles is determined by
equations of the form $ u_i[\zeta_0]= \infty $ from the cuts $\tilde
\ell_{i+1}$ and equations of the form $b_i[\zeta_0]=-\mu$ or
$b_i[\zeta_0]=-\mu^{-1}$. It is easy to see that if, say,
$b_i=-\mu$ then either $b_{i+1}=-\mu^{-1}$ or
$b_{i-1}=-\mu^{-1}$, and one of the cross-ratios equals $1$. The
converse is also true. It seems possible that as we follow a
path towards $(3,1)$ signature such phenomena will occur.

\listrefs

\bye